\documentclass[journal]{IEEEtran}

\usepackage[utf8]{inputenc}
\usepackage{cite}
\usepackage{setspace} %%% for double line spacing
\usepackage{todonotes}
\usepackage{soul} %%% highlight text
\usepackage{amssymb} % for complex number signs  + checkmark 
\usepackage{booktabs}
\usepackage{multirow}

\ifCLASSINFOpdf
\else
\fi

\usepackage{amsmath}
\usepackage{bm} %bold greek symbols
\usepackage[ruled, vlined]{algorithm2e}

\makeatletter
\newcommand{\nosemic}{\renewcommand{\@endalgocfline}{\relax}}% Drop semi-colon ;
\newcommand{\dosemic}{\renewcommand{\@endalgocfline}{\algocf@endline}}% Reinstate semi-colon ;
\makeatother

\usepackage{pgfplots} % for bar diagram
\usetikzlibrary{fit}
\usepackage{filecontents} %for ploting w/ pgfplots using data.csv 
\usepackage{multicol} % for multiple figures across 2-column document
\usepackage[export]{adjustbox} % for subfloat alignment
\usepackage[position=bottom]{subfig} %for subfloat (subfigure)
\usepackage{gensymb} % for \degree
\usepackage{multirow}

\pgfplotsset{compat=1.11,
        /pgfplots/ybar legend/.style={
        /pgfplots/legend image code/.code={%
        \draw[##1,/tikz/.cd,bar width=10pt,yshift=-0.2em,bar shift=0pt]
                plot coordinates {(0cm,0.8em)};},
},
}
\usepackage{hyperref}

\usepackage{cleveref} % for redefining multiple refs to 1 footnote

\usepackage{tabstackengine}
\setstackEOL{\cr}

\usepackage[group-separator={,}, group-minimum-digits={3}]{siunitx}

\usepackage{gensymb}

\usepackage[bottom]{footmisc}

\DeclareMathOperator*{\argmax}{argmax}
\DeclareMathOperator*{\argmin}{argmin}

\usepackage{xcolor, colortbl}
\definecolor{Gray}{gray}{0.85}
\definecolor{RED}{HTML}{D62728}
\definecolor{GREEN}{HTML}{009900}
\definecolor{ORANGE}{HTML}{FF8000}
\definecolor{LIGHTBLUE}{HTML}{66B2FF}

\usepackage{url}

\usepackage{tabu}

\usepackage{hhline}

\usepackage{mathtools}

\usepackage{float}

\begin{document}
\title{Joint \textcolor{black}{NN-Supported} Multichannel Reduction of Acoustic Echo, Reverberation and Noise}

\author{Guillaume~Carbajal,~\IEEEmembership{Student Member,~IEEE,}
        Romain~Serizel,~\IEEEmembership{Member,~IEEE,}
        Emmanuel~Vincent, ~\IEEEmembership{Senior Member,~IEEE,}
        and~Eric~Humbert,~\IEEEmembership{Member,~IEEE}% <-this % stops a space
        \thanks{M. Carbajal and E. Humbert are with Invoxia SAS, 8 Esplanade de la Manufacture, 92130 Issy-les-Moulineaux, France (email: guillaume.carbajal@invoxia.com, eric.humbert@invoxia.com). R. Serizel and E. Vincent are with Université de Lorraine, CNRS, Inria, LORIA, F-54000 Nancy, France (email: romain.serizel@loria.fr, emmanuel.vincent@inria.fr).}% <-this % stops a space
}


\markboth{IEEE/ACM TRANSACTIONS ON AUDIO, SPEECH, AND LANGUAGE PROCESSING, July~2020}%
{Shell \MakeLowercase{\textit{et al.}}: Bare Demo of IEEEtran.cls for IEEE Journals}



\maketitle

%The reduction of each of these three distortion sources involves one or multiple filters. 
%the residual echo, the residual reverberation, and the noise signal 
\begin{abstract}
We consider the problem of simultaneous reduction of acoustic echo, reverberation and noise. In real scenarios, these distortion sources may occur simultaneously and reducing them implies combining the corresponding distortion-specific filters. As these filters interact with each other, they must be jointly optimized. We propose to model the target and residual signals after linear echo cancellation and dereverberation using a multichannel Gaussian modeling framework and to jointly represent their spectra by means of a neural network. We develop an iterative block-coordinate ascent algorithm to update all the filters. We evaluate our system on real recordings of acoustic echo, reverberation and noise acquired with a smart speaker  in various situations. The proposed approach outperforms in terms of overall distortion a cascade of the individual approaches and a joint reduction approach which does not rely on a spectral model of the target and residual signals.
\end{abstract}


% \fi
\IEEEpeerreviewmaketitle

\section{Introduction}

\IEEEPARstart{I}{n} hands-free telecommunications, a speaker from a near-end point interacts with another speaker at a far-end point. The near-end speaker can be a few meters away from the microphones and the interactions can be subject to several distortion sources such as background noise, acoustic echo and near-end reverberation. Each of these distortion sources degrades speech quality, intelligibility and listening comfort, and must be reduced.
%The filters are obtained by optimizing their coefficients in the minimum mean square error (MMSE) or in the maximum likelihood (ML) sense. 
%Nonlinear filters introduce speech artifacts and there is a trade-off to be found between these artifacts and the reduction of distortion source. 
%\cite{yang_statistical_2017, valin_adjusting_2007} 
\par
Single- and multichannel filters have been used to reduce each of these distortion sources independently. They can be categorized into short nonlinear filters that vary quickly over time and long linear filters that are time-invariant (or slowly time-varying). Short nonlinear filters are generally used for noise reduction \cite{vincent_audio_2018}. They are robust to the fluctuations and nonlinearities inherent to real signals. Long linear filters can be required for dereverberation \cite{naylor_speech_2010} and echo reduction \cite{hansler_acoustic_2004}. They are able to reduce most of the distortion sources in time-invariant conditions without introducing any artifact, or musical noise, in the near-end signal. %However, they assume a linear perturbation model, which does not fit real conditions. In such conditions, nonlinear filters are needed.
\par
When several distortion sources occur simultaneously, reducing them requires cascading the distortion-specific filters. However, as these filters interact with each other, tuning them independently can be suboptimal and even lead to additional distortions. Several joint approaches that handle two distortion sources have been proposed, namely for joint dereverberation and source separation/noise reduction \cite{erkelens_correlation-based_2010, kodrasi_joint_2016, schwartz, dietzen_joint_2018, yoshioka_blind_2011, kagami_joint_2018}, for joint echo and noise reduction \cite{jeannes_combined_2001, gustafsson_psychoacoustic_2002, herbordtt_joint_2005, reuven_joint_2007, togami_frequency_2014, nathwani_joint_2018}, and for joint echo reduction and dereverberation \cite{takeda_ica-based_2009, togami_reverb_echo}.
\par
A joint approach for single-channel echo reduction, dereverberation and noise reduction was proposed by Habets et al. \cite{habets}. However, the linear echo cancellation filter was ignored in the optimization process. To the best of our knowledge, only Togami et al. proposed a solution optimizing two linear filters and a nonlinear postfilter for reducing echo, reverberation and noise \cite{togami_simultaneous_2014}. They represented the filter interactions by modeling the target and residual signals after echo cancellation and dereverberation within a multichannel Gaussian framework. However, no model was proposed for the short term spectra of these signals. This results in misestimation of the linear filters and the nonlinear postfilter. %, especially in time-varying conditions. 
\par
%echo reduction \cite{schwarz_spectral_2013, lee_dnn-based_2015, madrid_portillo, carbajal_multiple-input_2018}, dereverberation, \cite{wu_reverberation-time-aware_2017_spatial, santos_speech_2017, wake_semi-blind_2017, ernst_speech_2018, mack_single-channel_2018, kinoshita_neural_2017}, noise reduction \cite{lu_speech_2013, erdogan_phase-sensitive_2015, park_fully_2016, zhao_convolutional-recurrent_2018, wang_rank-1_2017, heymann_beamnet_2017} and source separation \cite{hershey_deep_2015, chen_deep_2017, wang_end--end_2018, nugraha, erdogan_improved_2016, heymann_neural_2016, drude_dual_2018, leglaive_semi-supervised_2018}. A few DNN-based approaches have been proposed 
Recently, neural networks (NN) have shown promising results to estimate the short term spectra of speech and distortion sources for joint dereverberation and source separation/noise reduction \cite{williamson_speech_2017, zhao_two-stage_2017}, and for joint echo and noise reduction \cite{seo_integrated_2018, zhang_deep_2018}. However, these approaches have only focused on reducing two distortion sources.
\par
%These recordings include the nonlinearities of the distortion sources. 
In this article, we propose an NN-supported approach for joint multichannel reduction of echo, reverberation and noise. We simultaneously model the spatial and spectral parameters of the target and residual signals within a multichannel Gaussian framework and we derive an iterative a block-coordinate ascent (BCA) algorithm to update the echo cancellation, dereverberation and noise/residual reduction filters. We evaluate our system on real recordings of acoustic echo, near-end reverberation and background noise acquired with a smart speaker in various situations. We experimentally show the effectiveness of our proposed approach compared with a cascade of individual approaches and Togami et al.'s joint reduction approach \cite{togami_simultaneous_2014}.
\par
The rest of this article is organized as follows. In Section II, we describe existing enhancement methods designed for the separate reduction of echo, reverberation or noise, and Togami et al.'s approach. We explain our joint approach using an NN spectral model within a BCA algorithm in Section III. In Section IV we detail our NN-based joint spectral model. Section V describes the experimental settings for the training and evaluation of our approach. Section VI shows the results of our approach compared to the cascade of individual approaches and Togami et al.'s approach. Finally Section VII concludes the article and provides future directions.

\section{Background \label{sec:background}}

%, and the corresponding signal model and notations will also be reused. 
In this section, we first describe multichannel approaches for the separate reduction of echo, reverberation or noise. These approaches will be used as building blocks for our solution and a basis for comparison in our experiments. We then describe Togami et al.’s joint approach. We adopt the following notations through the article: scalars are represented by plain letters, vectors by bold lowercase letters, and matrices by bold uppercase letters. The symbol $(\cdot)^*$ refers to complex conjugation, $(\cdot)^T$ to matrix transposition, $(\cdot)^H$ to Hermitian transposition, $\text{tr}(\cdot)$ to the trace of a matrix, $\rVert\cdot \rVert$ to the Euclidean norm and $ \otimes$ to the Kronecker product. The identity matrix is denoted as $\mathbf{I}$. Its dimension is either implied by the context or explicitly specified by a subscript.
% The underscore notation  $\underline{(\cdot)}$ refers to a concatenation of consecutive time frames of a signal or filter. 
%The operation $\text{vec}(\cdot)$ refers to a conversion of a matrix to a vector and $\text{mat}(\cdot)$ to the conversion of vector to a matrix, both defined in the supporting document \cite{carbajal_supporting_2019}. 

\subsection{Echo reduction \label{sec:echo_reduc}}

The echo reduction problem is defined as follows. Denoting by $M$ the number of channels (microphones), the mixture $\mathbf{d}^\text{echo}(t) \in \mathbb{R}^{M \times 1}$ observed at the microphones at time $t$ is the sum of the near-end signal $\mathbf{s}(t) \in \mathbb{R}^{M \times 1}$ and the acoustic echo $\mathbf{y}(t) \in \mathbb{R}^{M \times 1}$:
\begin{equation}
\mathbf{d}^\text{echo}(t) = \mathbf{s}(t) +\mathbf{y}(t).
\end{equation}
%Here, the conditions are assumed to be time-invariant. 
The acoustic echo $\mathbf{y}(t)$ is a nonlinearly distorted version of the observed far-end signal  $x(t) \in \mathbb{R}$ played by the loudspeaker, which is assumed to be single-channel. The echo signal can be expressed as
%The conditions are often time-varying, e.g. the near-end speaker is moving which modifies the echo path, hence the RIR depends on $t$. 
\begin{equation}
\mathbf{y}(t) = \sum_{\tau=0}^{\infty} \mathbf{a}_\text{y}(\tau) x(t - \tau) \textcolor{black}{+ \mathbf{y}_\text{nl}(t)}.
\end{equation}
The linear part corresponds to the linear convolution of $x(t)$ and the $M$-dimensional room impulse response (RIR) $ \mathbf{a}_\text{y}(\tau)  \in \mathbb{R}^{M \times 1} $, or echo path, modeling the acoustic path from the loudspeaker (including the loudspeaker response) to the microphones. \textcolor{black}{The nonlinear part is denoted by $\mathbf{y}_\text{nl}(t) \in \mathbb{R}^{M \times 1}$.} The signals are transformed into the time-frequency domain by the short-time Fourier transform (STFT)
\begin{equation}
\mathbf{d}^\text{echo}(n,f) =\mathbf{s}(n,f) + \mathbf{y}(n,f),
\end{equation}
at time frame index $n \in [0,N-1]$ and frequency bin index $f \in [0, F-1]$, where $F$ is the number of frequency bins and $N$ the number of time frames of the utterance. As the far-end signal $x(n,f) \in \mathbb{C}$ is known, the goal is to recover the $M$-dimensional near-end speech $\mathbf{s}(n,f) \in \mathbb{C}^{M \times 1}$ from the mixture $\mathbf{d}^\text{echo}(n,f) \in \mathbb{C}^{M \times 1}$  by identifying the echo path $\{ \mathbf{a}_\text{y}(n,f) \}_{n, f}$. The underlying idea is to estimate $\mathbf{y}(n,f) \in \mathbb{C}^{M \times 1}$ with a long, multiframe linear echo cancellation filter $\mathbf{\underline{H}}(f) = [\mathbf{h}(0,f) \ldots \mathbf{h}(K-1,f) ] \in \mathbb{C}^{M \times K}$ applied on the $K$ previous frames of the far-end signal $x(n,f)$, and to subtract the resulting signal $\mathbf{\widehat{y}}(n,f)$ from $\mathbf{d}^\text{echo}(n,f)$:
\begin{equation}
\mathbf{e}^\text{echo}(n,f) 	= \mathbf{d}^\text{echo}(n,f) - \underbrace{\sum_{k=0}^{K-1} \mathbf{{h}}(k,f) x(n-k,f)}_{= \mathbf{\widehat{y}}(n,f)}. \label{eq:echo_reduc_1}
\end{equation}
%The far-end signal is assumed to be $x(n,f) = 0$ for $n<0$. 
where $\mathbf{h}(k,f) \in \mathbb{C}^{M \times 1}$ is the $M$-dimensional vector corresponding to the $k$-th tap of $\mathbf{\underline{H}}(f)$. Note that the tap $k$ is measured in frames and the underscore notation in $\mathbf{\underline{H}}(f)$ denotes the concatenation of the $K$ taps of $\mathbf{h}(k,f)$. Since the far-end signal $x(n,f)$ is known, the filter $\mathbf{\underline{H}}(f)$ is usually estimated adaptively in the minimum mean square error (MMSE) sense \cite{hansler_acoustic_2004}. Adaptive MMSE optimization typically relies on adaptive algorithms such as least mean squares (LMS) which adjust the filter $\mathbf{\underline{H}}(f)$ in an online manner by stochastic gradient descent with a time-varying step size \cite{hansler_acoustic_2004}. These algorithms have low complexity and fast convergence, which makes them particularly suitable for time-varying conditions. Yang et al. provide a comprehensive review of optimal step size selection for echo reduction \cite{yang_statistical_2017}.
\par
In practice, the output signal $\mathbf{e}^\text{echo}(n,f)$ is not equal to the near-end speech $\mathbf{s}(n,f)$, not only because of the estimation error, but also because of the smaller length of $\mathbf{\underline{H}}(f)$ compared to the true echo path and of nonlinearities $\mathbf{y}_\text{nl}(n,f)$ that cannot be modeled by $\mathbf{\underline{H}}(f)$ \cite{habets}. As a result, a residual echo $\mathbf{z}(n,f)$ remains that can be expressed as \cite{hansler_acoustic_2004} 
\begin{equation}
\mathbf{e}^\text{echo}(n,f) - \mathbf{s}(n,f) = \underbrace{\mathbf{y}(n,f) - \mathbf{\widehat{y}}(n,f)}_{= \mathbf{z}(n,f)}.
\end{equation}
To overcome this limitation, a (nonlinear) residual echo suppression postfilter $\mathbf{W}^\text{echo}(n,f) \in \mathbb{C}^{M \times M}$ is typically applied:
\begin{equation}
\mathbf{\widehat{s}}(n,f) = \mathbf{W}^\text{echo}(n,f) \mathbf{e}^\text{echo}(n,f).
\end{equation}
%the underlying idea is to first estimate $\mathbf{z}(n,f)$ with a model, and then use this estimate to adapt simultaneously $\mathbf{\underline{H}}(f)$ and $\mathbf{W}_\text{echo}(n,f)$.  
% Valin proposed a similar approach that estimates $\mathbf{z}(n,f)$ as a function of $\mathbf{\widehat{y}}(n,f)$ \cite{valin_adjusting_2007}, which makes possible to derive simultaneously $\mathbf{H}(n,f)$ and $\mathbf{W}^\text{echo}(n,f)$ \cite{carbajal_multiple-input_2018}. 
There exist multiple approaches to derive $\mathbf{W}^\text{echo}(n,f)$ \cite{hansler_acoustic_2004}. Recently, direct estimation of $\mathbf{W}^\text{echo}(n,f)$ using an NN has shown good performance in the single-channel case \cite{lee_dnn-based_2015, carbajal_multiple-input_2018}. However, when $\mathbf{\underline{H}}(f)$ changes, $\mathbf{z}(n,f)$ also changes and the postfilter $\mathbf{W}^\text{echo}(n,f)$ must be adapted consequently. Estimating $\mathbf{\underline{H}}(f)$ and $\mathbf{W}^\text{echo}(n,f)$ separately is thus suboptimal. Joint optimization of $\mathbf{\underline{H}}(f)$ and $\mathbf{W}^\text{echo}(n,f)$ was investigated in the MMSE and the maximum likelihood (ML) sense \cite{enzner_frequency-domain_2006, togami_echo_2011}.
\iffalse
The joint optimization of $\mathbf{\underline{H}}(f)$ and $\mathbf{W}^\text{echo}(n,f)$ was investigated in the MMSE and ML sense \cite{enzner_frequency-domain_2006, togami_echo_2011}. In the MMSE sense, Enzner and Vary proposed a linear model for deriving $\mathbf{z}(n,f)$ as a function of $x(n,f)$ to jointly update $\mathbf{\underline{H}}(f)$ via the step size of an adaptive algorithm, and the postfilter $\mathbf{W}^\text{echo}(n,f)$ via a Wiener filtering rule  \cite{enzner_frequency-domain_2006}. In the ML sense, Togami and Hori proposed to model the echo path $\mathbf{a}_y$ as a nonzero-mean Gaussian variable \cite{togami_echo_2011}, that is an extension of the multichannel local Gaussian model where the sources are modeled as zero-mean Gaussian variables \cite{duong_under_2010}. This approach enables to derive both $\mathbf{\underline{H}}(f)$ and $\mathbf{W}^\text{echo}(n,f)$. They modeled the spectral parameters of $\mathbf{a}_y$ as a function of $x(n,f)$ and jointly updated $\mathbf{\underline{H}}(f)$ and $\mathbf{W}^\text{echo}(n,f)$ via an exact EM algorithm.
\fi
\par
In Section V, we will use adaptive MMSE optimization for estimating the echo cancellation filter  $\mathbf{\underline{H}}(f)$ as a part of the cascade of the individual approaches to which we compare our joint approach.

%In our experiments, we will use Valin's approach for estimating linear echo cancellation filter  $\mathbf{\underline{H}}(f)$ as a part of the cascade of the individual approaches to which we compare our approach for the joint reduction of echo, reverberation and noise \cite{valin_adjusting_2007}.

\subsection{Near-end dereverberation \label{sec:dereverb}} 
%Here, the conditions are also assumed to be time-invariant. 
The near-end dereverberation problem is defined as follows.  The signal $\mathbf{d}^\text{rev}(t)$  observed at the microphones at time $t$ is just the reverberant near-end signal $\mathbf{s}(t)$, which is obtained by linear convolution of the anechoic near-end signal $u(t) \in \mathbb{R}$ and the $M$-dimensional RIR $\mathbf{a}_\text{s}(\tau) \in \mathbb{R}^{M \times 1}$:
\begin{equation}
\mathbf{d}^\text{rev}(t) = \mathbf{s}(t)  = \sum_{\tau=0}^{\infty} \mathbf{a}_\text{s}(\tau) u(t - \tau). \label{eq:reverb}
\end{equation}
%where $\mathbf{s}(t)$ is obtained by linear convolution of the anechoic near-end signal $u(t)$ and the $M$-dimensional RIR $\mathbf{a}_s$.
This signal can be decomposed as
\begin{equation}
\mathbf{s}(t) = \underbrace{\sum_{0 \leq \tau \leq t_\text{e}}\mathbf{a}_\text{s}(\tau) u(t - \tau)}_{= \mathbf{s}_\text{e}(t)} + \underbrace{\sum_{\tau > t_\text{e}} \mathbf{a}_s(\tau) u(t - \tau)}_{= \mathbf{s}_\text{l}(t)}, \label{eq:se_sl}
\end{equation}
where $\mathbf{s}_\text{e}(t)$ denotes the early near-end signal component, $\mathbf{s}_\text{l}(t)$ the late reverberation component, and $t_\text{e}$ is the mixing time. The component $\mathbf{s}_\text{e}(t)$ comprises the main peak of the RIR (the direct path) and the early reflections within a delay $t_e$ which contribute to speech quality and intelligibility. The component $\mathbf{s}_\text{l}(t)$ comprises all the later reflections which degrade intelligibility. In the time-frequency domain, the reverberant near-end speech can thus be expressed as
\begin{equation}
\mathbf{s}(n,f) = \mathbf{s}_\text{e}(n,f) + \mathbf{s}_\text{l}(n,f).
\end{equation}
% Motivation of WPE: One drawback as regards methods using instantaneous mixture models is that they are weak against long reverberation.
 The goal is to recover the early near-end component $\mathbf{s}_\text{e}(n,f)$ from the reverberant near-end signal $\mathbf{s}(n,f)$. Naylor et al. provided a comprehensive review of dereverberation approaches \cite{naylor_speech_2010}. Among them, the weighted prediction error (WPE) method \cite{nakatani_speech_2010} estimates $\mathbf{s}_\text{l}(n,f)$ by inverse filtering with a long, multiframe linear filter $\mathbf{\underline{G}}(f) = [\mathbf{G}(\Delta,f) \ldots \mathbf{G}(\Delta+L-1,f)] \in \mathbb{C}^{M \times M L}$ applied on the $L$ previous frames of the mixture signal $\mathbf{s}(n-\Delta,f)$ defined in (\ref{eq:reverb}). The delay $\Delta$ is introduced to avoid over-whitening of the near-end speech. The resulting signal $\mathbf{\widehat{s}}_\text{l}(n,f)$ is then subtracted from the mixture signal $\mathbf{s}(n,f)$ defined in (\ref{eq:reverb}):
%											&= \mathbf{s}(n,f) - , \label{eq:dereverb_2}
\begin{equation}
\mathbf{r}^\text{rev}(n,f) = \mathbf{s}(n,f) - \underbrace{\sum_{l=\Delta}^{\Delta+L-1} \mathbf{G}(l,f) \mathbf{s}(n-l,f)}_{\mathbf{\widehat{s}}_\text{l}(n,f)}. \label{eq:dereverb_1}
\end{equation}
 %The reverberant near-end signal is assumed to be $\mathbf{s}(n,f)=\mathbf{0}$ for $n<0$. 
where $\mathbf{G}(l,f) = [\mathbf{g}_1(l,f) \ldots \mathbf{g}_M(l,f)]  \in \mathbb{C}^{M \times M}$ is the $M \times M$-dimensional matrix corresponding to the $l$-th tap of $\mathbf{\underline{G}}(f)$, and $\mathbf{g}_m(l,f) \in \mathbb{C}^{M \times 1}$ is the $m$-th channel vector of  $\mathbf{G}(l,f)$. As the component $\mathbf{s}_\text{e}(n,f)$ is not an observed signal, Nakatani et al. estimated the filter $\mathbf{\underline{G}}(f)$ in the ML sense by modeling $\mathbf{s}_\text{e}(n,f)$ as a directional source \cite{nakatani_speech_2010}.
However, they did not impose any constraint on its short-term spectrum which results in limited dereverberation \cite{nakatani_speech_2010, yoshioka_generalization_2012}. Other authors have assumed a model of the short-term spectrum. Yoshioka et al. used an all-pole model \cite{yoshioka_blind_2011}, Kagami et al. used nonnegative matrix factorization (NMF) \cite{kagami_joint_2018}, Juki\'c et al. used sparse priors \cite{jukic_multi-channel_2015}, and Kinoshita et al. used an NN \cite{kinoshita_neural_2017}.
\par
For several reasons, including the smaller length of the filter compared to true near-end RIR and potentially time-varying conditions, a residual late reverberation component $\mathbf{s}_\text{r}(n,f)$ remains \cite{furuya_robust_2007, togami_optimized_2013, cohen_combined_2017} that can be expressed as
\begin{equation}
\mathbf{r}^\text{rev}(n,f) - \mathbf{s}_\text{e}(n,f) =  \underbrace{\mathbf{s}_\text{l}(n,f) -  \mathbf{\widehat{s}}_\text{l}(n,f)}_{= \mathbf{s}_\text{r}(n,f)}.
\end{equation}
To overcome this limitation, a (nonlinear) residual reverberation suppression postfilter $\mathbf{W}^\text{rev}(n,f) \in \mathbb{C}^{M \times M}$ is applied on the signal $\mathbf{r}^\text{rev}(n,f)$:
\begin{equation}
\mathbf{\widehat{s}}_\text{e}(n,f) = \mathbf{W}^\text{rev}(n,f) \mathbf{r}^\text{rev}(n,f).
\end{equation}
There are multiple approaches to derive $\mathbf{W}^\text{rev}(n,f)$ \cite{furuya_robust_2007, cohen_combined_2017}. However, when $\mathbf{\underline{G}}(f)$ changes, $\mathbf{s}_\text{r}(n,f)$ also changes and the postfilter $\mathbf{W}^\text{rev}(n,f)$ must be adapted consequently. Estimating $\mathbf{\underline{G}}(f)$ and $\mathbf{W}^\text{rev}(n,f)$ separately is thus suboptimal. Joint optimization of $\mathbf{\underline{G}}(f)$ and $\mathbf{W}^\text{rev}(n,f)$ was investigated in the ML sense  \cite{togami_optimized_2013}. 
\iffalse
However, estimating $\mathbf{\underline{G}}(f)$ and $\mathbf{W}^\text{rev}(n,f)$ separately is suboptimal. Indeed when $\mathbf{\underline{G}}(f)$ changes, $\mathbf{s}_\text{r}(n,f)$ also changes and the postfilter $\mathbf{W}^\text{rev}(n,f)$ must be adapted consequently. The joint optimization of $\mathbf{\underline{G}}(f)$ and $\mathbf{W}^\text{rev}(n,f)$ was investigated in the ML sense  \cite{togami_optimized_2013}. Similarly to echo reduction, Togami et al. proposed to model the inverse of the RIR $\mathbf{a}_s$ as a nonzero-mean Gaussian variable \cite{togami_echo_2011}. They modeled the spectral parameters of the inverse of the RIR $\mathbf{a}_s$ as a function of $s(n,f)$ and they jointly derived $\mathbf{\underline{G}}(f)$ and $\mathbf{W}^\text{rev}(n,f)$ via an exact EM algorithm.
\fi
\par
In section V, we will use WPE for estimating the dereverberation filter  $\mathbf{\underline{G}}(f)$ as a part of the cascade of the individual approaches to which we compare our joint approach.

\subsection{Noise reduction \label{sec:noise}}
%Here, the conditions are also assumed to be time-invariant. 
The noise reduction problem is defined as follows. In the time-frequency domain, the $M$-channel mixture $\mathbf{d}^\text{noise}(n,f)$ observed at the microphones is the sum of the near-end signal $\mathbf{s}(n,f)$ and a noise signal $\mathbf{b}(n,f) \in \mathbb{C}^{M \times 1}$: 
\begin{equation}
\mathbf{d}^\text{noise}(n,f) = \mathbf{s}(n,f) + \mathbf{b}(n,f).
\end{equation}
%\begin{figure}[h]
%\centering
%\includegraphics[width=.8\textwidth]{diagrams/multichannel_denoising.pdf}
%\caption{\label{fig:noise_reduc} General setting for noise reduction.}
%\end{figure}
%We use the notation $\mathbf{c}$ for a source. In this subsection, $\mathbf{c}$ can be either the near-end signal $\mathbf{s}$ or the noise signal $\mathbf{b}$
Note that the noise signal $\mathbf{b}(n,f)$  can be either spatially diffuse or localized. The goal is to recover the near-end speech $\mathbf{s}(n,f)$ from the mixture $\mathbf{d}^\text{noise}(n,f)$. This is typically achieved by applying a short nonlinear filter $\mathbf{W}^\text{noise}_s(n,f) \in \mathbb{C}^{M \times M}$ on $\mathbf{d}^\text{noise}(n,f)$:
\begin{equation}
\widehat{\mathbf{s}}(n,f) = \mathbf{W}^\text{noise}(n,f) \mathbf{d}^\text{noise}(n,f).
\end{equation}
The filter can be estimated in the MMSE or ML sense. Gannot el al. provide a comprehensive review of spatial filtering solutions \cite{gannot_consolidated_2017}. One family of solutions relies on multichannel time-varying Wiener filtering, where the filter is derived from a local Gaussian model of the target and noise sources \cite{duong_under_2010}.
\iffalse
The near-end signal $\mathbf{s}(n,f)$ is modeled as
\begin{equation}
\mathbf{s}(n,f) \sim \mathcal{N}\Big(\mathbf{0}, v_s(n,f) \mathbf{R}_s(f) \Big),
\end{equation}
where $v_\text{s}(n,f) \in \mathbb{R}_{+}$ and $\mathbf{R}_{s}(f) \in \mathbb{C}^{M \times M}$ denote the PSD and SCM of $\mathbf{s}(n,f)$, respectively. Similarly, the noise signal $\mathbf{b}(n,f)$ is modeled as
\begin{equation}
\mathbf{b}(n,f) \sim \mathcal{N}\Big(\mathbf{0}, v_b(n,f) \mathbf{R}_b(f) \Big).
\end{equation}
The mixture signal $\mathbf{d}(n,f)$ is consequently distributed as
\begin{equation}
\mathbf{d}(n,f) \sim \mathcal{N}_{\mathbb{C}} \Big(\mathbf{d}(n,f); \mathbf{0}, v_{s}(n,f) \mathbf{R}_{s}(f) + v_{b}(n,f) \mathbf{R}_{b}(f)  \Big),
\end{equation}
and the multichannel Wiener filter $\mathbf{W}^\text{noise}_s(n,f)$ is formulated as
\begin{equation}
\mathbf{{W}}_{s}^\text{noise}(n,f) = v_{s}(n,f) \mathbf{R}_{s}(f) \Big(v_{s}(n,f) \mathbf{R}_{s}(f) + v_{b}(n,f) \mathbf{R}_{b}(f) \Big)^{-1}. \label{eq:wiener_noise}
\end{equation}
%\mathbf{\widehat{W}}_{s}^\text{noise}(n,f) = \max_{ \substack{ \\ v_{s_d}(n,f) \\ \mathbf{R}_{s_d}(f)}} 
%\begin{equation}
%\mathcal{L} \left(\mathbf{d}^\text{noise}(n,f)  \Big|  \Big\{ v_{c}(n,f), \mathbf{R}_{c}(f) \Big\}_{c \in \{s, b\}}  \right) = \sum_{n=0}^{N-1} \log  \mathcal{N}_{\mathbb{C}} \bigg(\mathbf{d}^\text{noise}(n,f);  \mathbf{0}, \sum_{c \in \{s, b\}} v_{c}(n,f) \mathbf{R}_{c}(f) \bigg) \label{eq:likelihood_noise}.
%\end{equation}
\fi
The spectral parameters (short term power spectra) and the spatial parameters (spatial covariance matrices) of this model are estimated in the ML sense. Since there is no closed form solution, the ML parameters are estimated using an EM algorithm.
\iffalse
In the E-step, the second-order posterior moment of the near-end signal $\mathbf{s}$ is computed as
\begin{equation}
\mathbf{\widehat{R}}_s(n,f) = \mathbf{\widehat{s}}(n,f) \mathbf{\widehat{s}}(n,f)^H + \Big( \mathbf{I} - \mathbf{W}_s^\text{noise}(n,f) \Big) v_s(n,f) \mathbf{R}_s(f). \label{eq:posterior_wiener_noise}
\end{equation}
In the M-step, the PSD and SCM are updated as
\begin{align}
v_s(n,f) &= \frac{1}{M} \text{tr} \left(\mathbf{{R}}_{s}(f)^{-1} \mathbf{\widehat{R}}_s(n,f)\right), \label{eq:vc_wiener_noise} \\
\mathbf{{R}}_{s}(f) &= \frac{1}{N} \sum_{n=1}^N \frac{1}{v_{s}(n,f)} \mathbf{\widehat{R}}_{s}(n,f). \label{eq:Rc_wiener_noise}
\end{align}
(\ref{eq:posterior_wiener_noise})--(\ref{eq:Rc_wiener_noise}) can be derived similarly for the noise signal $\mathbf{b}(n,f)$.
\fi
%, seki_generalized_2018
When no constraint is imposed on the spectral or spatial parameters, the EM algorithm operates in each frequency bin $f$ independently which results in a permutation ambiguity in the separated components at each frequency bin $f$ and requires additional permutation alignment. Alternatively, the spectral parameters can be estimated with a model. Ozerov et al. used NMF \cite{ozerov_multichannel_2010}, Nugraha et al. used an NN \cite{nugraha}, and recently variational autoencoders were used \cite{leglaive_semi-supervised_2018}.
\par
In Section V, we will use multichannel time-varying Wiener filtering as a part of the cascade of the individual approaches to which we compare our joint approach.

\subsection{Joint reduction of echo, reverberation and noise \label{sec:togami}}
% TODO : conditions are not time varying (cf end of Togami)
%When fluctu- ation of the acoustic transfer function is non-stationary, speech enhancement performance will degrade.
%Here, the conditions are also assumed to be time-invariant. 
\begin{figure}[b]
\centering
\includegraphics[width=.3\textwidth]{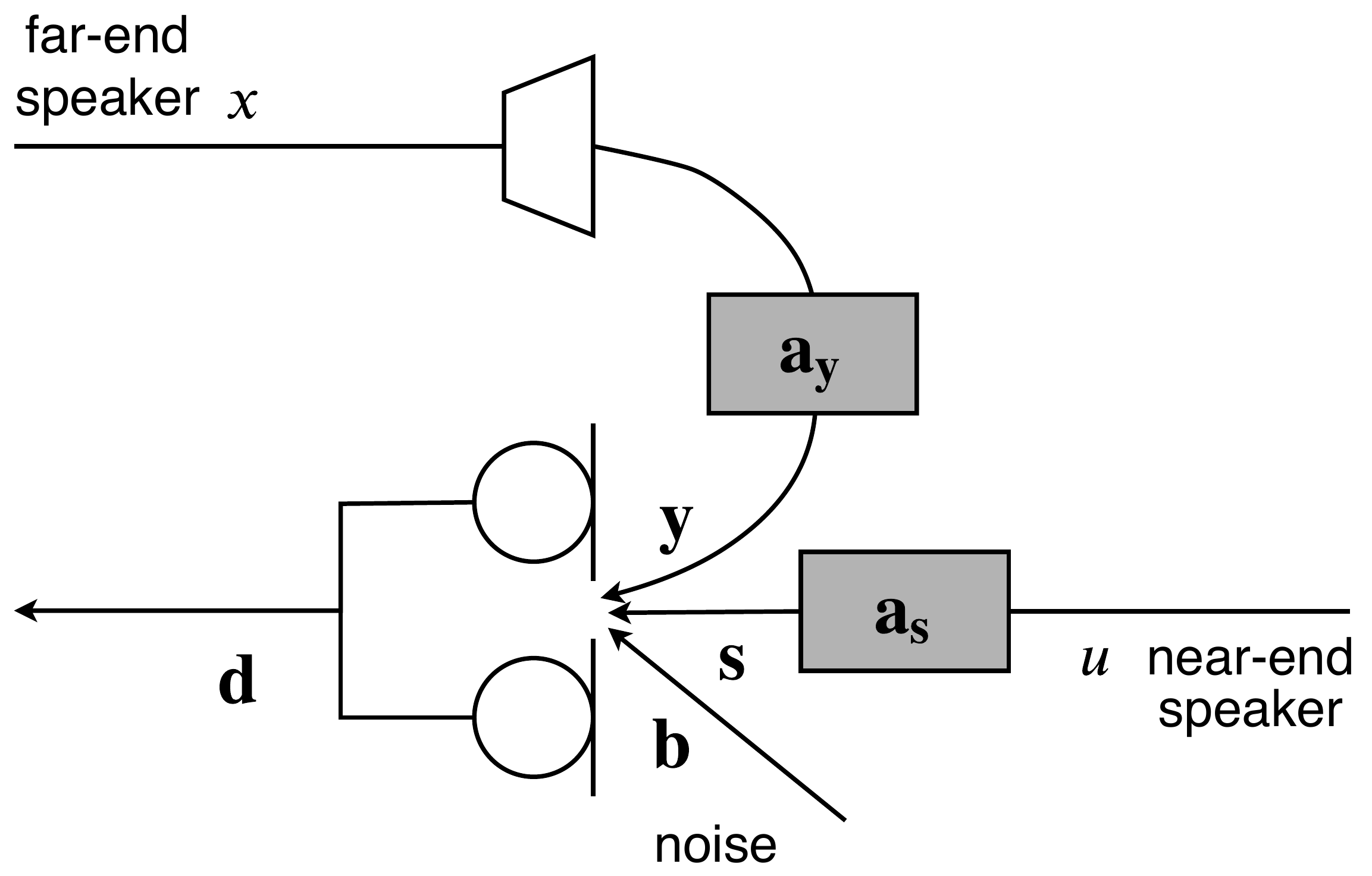}
\caption{\label{fig:echo_reverb_noise} Acoustic echo, reverberation and noise problem.}
%\vspace{-.5cm}
\end{figure}
In real scenarios, all the distortions introduced above can be simultaneously present as illustrated in Fig.\ \ref{fig:echo_reverb_noise}. The mixture $\mathbf{d}(n,f)$  observed at the microphones is thus the sum of the acoustic echo $\mathbf{y}(n,f)$, the reverberant near-end signal $\mathbf{s}(n,f)$  and the noise $\mathbf{b}(n,f)$
\begin{align}
\mathbf{d}(n,f) &= \mathbf{s}(n,f) + \mathbf{y}(n,f) + \mathbf{b}(n,f) \label{eq:mixture} \\
						&= \mathbf{s}_\text{e}(n,f) + \mathbf{s}_\text{l}(n,f) + \mathbf{y}(n,f) + \mathbf{b}(n,f).
\end{align}
The goal is to recover the  early near-end component $\mathbf{s}_\text{e}(n,f)$ from the mixture $\mathbf{d}(n,f)$.
\par
\begin{figure}[t]
\centering
\includegraphics[width=.45\textwidth]{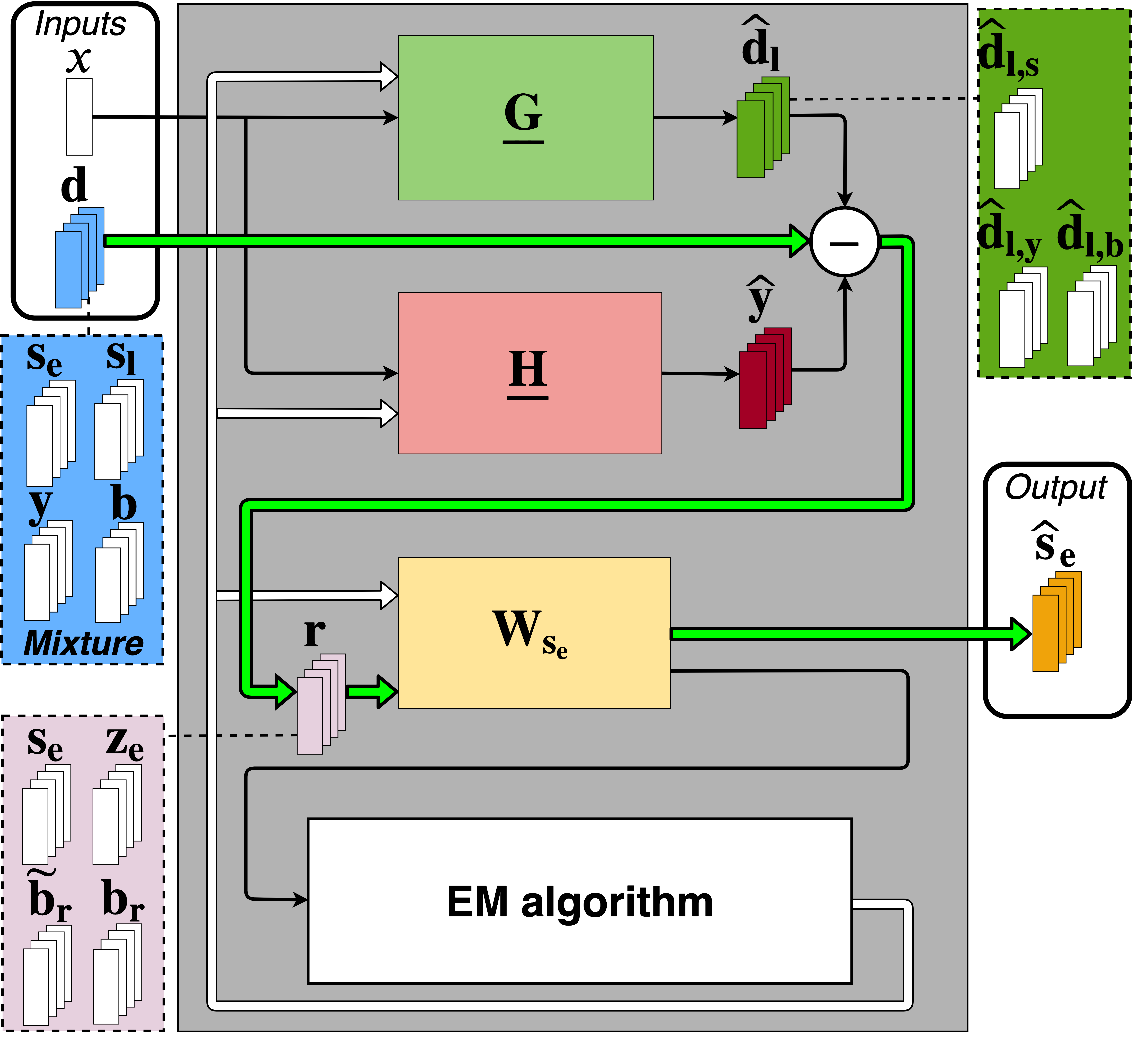}
\caption{Togami et al.'s approach for joint reduction of echo, reverberation and noise  \cite{togami_simultaneous_2014}. The bold green arrows denote the filtering steps. The dashed lines denote the latent signal components. The thin black arrows denote the signals used for the filtering steps and for the joint update. The bold white arrows denote the filter updates. \label{fig:togami}}
\end{figure}

Togami et al. proposed a joint approach combining an echo cancellation filter $\mathbf{\underline{H}}(f)$ (see Section \ref{sec:echo_reduc}), a dereverberation filter $\mathbf{\underline{G}}(f)$ (see Section \ref{sec:dereverb}), and a nonlinear multichannel Wiener postfilter $\mathbf{W}_{s_\text{e}}(n,f)$ (see Section \ref{sec:noise})\cite{togami_simultaneous_2014}. The approach is illustrated in Fig.\ \ref{fig:togami}. In the first step, they apply the echo cancellation filter $\mathbf{\underline{H}}(f)$ as in (\ref{eq:echo_reduc_1}) and subtract the resulting echo estimate $\mathbf{\widehat{y}}(n,f)$ from the mixture signal $\mathbf{d}(n,f)$. In parallel, the authors apply the dereverberation filter $\mathbf{\underline{G}}(f)$ on the mixture signal $\mathbf{d}(n,f)$ as in (\ref{eq:dereverb_1}) and subtract the resulting late reverberation estimate $\mathbf{\widehat{d}}_\text{l}(n,f)$ from $\mathbf{d}(n,f)$. The resulting signal $\mathbf{r}(n,f)$ after echo cancellation and dereverberation is then
%\mathbf{\widehat{y}}(n, f)
\iffalse
\begin{align}
\begin{split}
\mathbf{r}(n, f) = \mathbf{d}(n, f) &- \underbrace{ \sum_{k=0}^{K-1} \mathbf{h}(k,f) x(n-k,f)}_{= \mathbf{\widehat{y}}(n, f)} \\
 &- \underbrace{\sum_{l=\Delta}^{\Delta+L-1} \mathbf{G}(l,f) \mathbf{d}(n-l,f)}_{= \mathbf{\widehat{d}}_\text{l}(n,f)}.
\end{split} \label{eq:togami_dereverberation}
\end{align}
\fi
\begin{equation}
\mathbf{r}(n, f) = \mathbf{d}(n, f) - \mathbf{\widehat{y}}(n, f) - \underbrace{\sum_{\mathclap{l=\Delta}}^{\mathclap{\Delta+L-1}} \mathbf{G}(l,f) \mathbf{d}(n-l,f)}_{= \mathbf{\widehat{d}}_\text{l}(n,f)}. \label{eq:togami_dereverberation}
\end{equation}
%Note that the filters $\mathbf{\underline{H}}(f)$ and $\mathbf{\underline{G}}(f)$ operate independently on the mixture signal $\mathbf{d}(n,f)$, thus they do not interact with each other.
Due to the reasons mentioned in Sections \ref{sec:echo_reduc} and \ref{sec:dereverb}, and to the presence of the noise signal $\mathbf{b}(n,f)$, undesired residual signals remain and can be expressed as %(see Fig.\ \ref{fig:togami})
\begin{equation}
\mathbf{r}(n,f) - \mathbf{s}_\text{e}(n,f) = \mathbf{z}_\text{e}(n,f) + \mathbf{\widetilde{b}}_\text{r}(n,f)  + \mathbf{b}_\text{r}(n,f). \label{eq:residual_mixture}
\end{equation}
%but does not mean that the echo $\mathbf{y}(n,f)$ or the noise $\mathbf{b}(n,f)$ are actually dereverberated, as the dereverberation filter $\mathbf{G}(\Delta, f)$ aims at reducing the near-end reverberation
The signals $\mathbf{z}_\text{e}(n,f)$, $\mathbf{\widetilde{b}}_\text{r}(n,f)$  and $\mathbf{b}_\text{r}(n,f)$ are defined as
\begin{align}
\mathbf{z}_\text{e}(n,f) &= \mathbf{y}_\text{e}(n,f) - \mathbf{\widehat{y}}(n,f), \label{eq:togami_ze} \\
\mathbf{\widetilde{b}}_\text{r}(n,f) &= \mathbf{s}_\text{l}(n,f) - \mathbf{\widehat{d}}_{\text{l}, s}(n,f) + \mathbf{y}_\text{l}(n,f) - \mathbf{\widehat{d}}_{\text{l}, y}(n,f), \label{eq:togami_br_tilde} \\
\mathbf{b}_\text{r}(n,f) &= \mathbf{b}(n,f) - \mathbf{\widehat{d}}_{\text{l}, b}(n,f),
\end{align}
%Because of (\ref{eq:mixture}), the signal $\mathbf{\widehat{d}}_\text{l}(n,f)$ has thus $3$ components:
%the application of the dereverberation filter $\mathbf{G}(\Delta, f)$ on the mixture signal $\mathbf{d}(n,f)$
%&= \mathbf{\widehat{d}}_{\text{l}, s}(n,f) + \mathbf{\widehat{d}}_{\text{l}, y}(n,f) + \mathbf{\widehat{d}}_{\text{l}, b}(n,f).
%\begin{align}
%\mathbf{\widehat{d}}_\text{l}(n,f) &= \sum_{l=\Delta}^{\Delta+L-1} \mathbf{G}(l,f) \Big(\mathbf{s}(n-l,f) + \mathbf{y}(n-l,f) + \mathbf{b}(n-l,f) \Big).				
%\end{align}
%Note that the filters $\mathbf{\underline{H}}(f)$ and $\mathbf{G}(\Delta, f)$ operate independently on the mixture signal $\mathbf{d}(n,f)$, thus they do not interact with each other and no model is required for these specific interactions. 
%Note that the filters $\mathbf{\underline{H}}(f)$ and $\mathbf{G}(\Delta, f)$ operate independently on the mixture signal $\mathbf{d}(n,f)$, thus they do not interact with each other and no model is required for these specific interactions. 
%,but does not mean that the echo $\mathbf{y}(n,f)$ or the noise $\mathbf{b}(n,f)$ are actually dereverberated, as the dereverberation filter $\mathbf{G}(\Delta, f)$ aims at reducing the near-end reverberation.
%Note that the term \textit{dereverberated} means "after applying the dereveberation filter", but does not mean that the echo $\mathbf{y}(n,f)$ or the noise $\mathbf{b}(n,f)$ are actually dereverberated, as the dereverberation filter $\mathbf{G}(\Delta, f)$ aims at reducing the near-end reverberation.
where the signals $\mathbf{y}_\text{e}(n,f)$ and $\mathbf{y}_\text{l}(n,f)$ denote the early component and the late reverberation of the echo $\mathbf{y}(n,f)$, respectively, $\mathbf{\widehat{d}}_{\text{l}, s}(n,f) =  \sum_{l=\Delta}^{\Delta+L-1} \mathbf{G}(l,f) \mathbf{s}(n-l,f)$,  $\mathbf{\widehat{d}}_{\text{l}, y}(n,f) = \sum_{l=\Delta}^{\Delta+L-1} \mathbf{G}(l,f) \mathbf{y}(n-l,f)$ and $ \mathbf{\widehat{d}}_{\text{l}, b}(n,f) = \sum_{l=\Delta}^{\Delta+L-1} \mathbf{G}(l,f) \mathbf{b}(n-l,f)$  are the latent components of $ \mathbf{\widehat{d}}_\text{l}(n,f)$ resulting from (\ref{eq:togami_dereverberation}), and $\mathbf{b}_\text{r}(n,f)$ is the \textit{dereverberated} noise signal. The term \textit{dereverberated} means "after applying the dereverberation filter".
\par
To recover the early near-end signal component $\mathbf{s}_\text{e}(n,f)$ from the signal $\mathbf{r}(n,f)$, the authors applied a multichannel Wiener postfilter $\mathbf{W}_{s_\text{e}}(n,f) \in \mathbb{C}^{M \times M}$ on the signal $\mathbf{r}(n,f)$:
\begin{equation}
\mathbf{\widehat{s}}_\text{e}(n,f) = \mathbf{W}_{s_\text{e}}(n,f) \mathbf{r}(n,f).
\end{equation}
The authors estimated $\mathbf{\underline{H}}(f)$, $\mathbf{\underline{G}}(f)$ and $\mathbf{{W}}_{s_\text{e}}(n, f)$ by modeling $\mathbf{s}_\text{e}(n,f)$ and $\mathbf{b}_r(n,f)$ as zero-mean multichannel Gaussian variables, and $\mathbf{z}_\text{e}(n,f)$ and $\mathbf{\widetilde{b}}_r(n,f)$  as nonzero-mean multichannel Gaussian variables \cite{togami_simultaneous_2014}. They used an EM algorithm to jointly optimize the spectral and spatial parameters of this model in the ML sense.
\par
However, their approach suffers from several limitations. First, they did not impose any constraint on the spectral parameters of the target $\mathbf{s}_\text{e}(n,f)$ and the \textit{dereverberated} noise signal $\mathbf{b}_\text{r}(n,f)$. Secondly, the signal components $\mathbf{s}_\text{l}(n,f)$ and $\mathbf{y}_\text{l}(n,f)$ in $\mathbf{\widetilde{b}}_\text{r}(n,f)$ are not separately modeled, i.e. these components share the same spatial parameters, which is not the case in practice. These two limitations result in misestimation of the filters $\mathbf{\underline{H}}(f)$, $\mathbf{\underline{G}}(f)$ and the postfilter $\mathbf{W}_{s_\text{e}}(n,f)$. Thirdly, because the filters $\mathbf{\underline{H}}(f)$ and $\mathbf{\underline{G}}(f)$ operate independently on the mixture signal $\mathbf{d}(n,f)$, their respective components $\mathbf{\widehat{y}}(n,f)$ and $\mathbf{\widehat{d}}_{\text{l}, y}(n,f)$ subtracted from the echo $\mathbf{y}(n,f)$ in (\ref{eq:togami_ze}) and (\ref{eq:togami_br_tilde}) might interfere with each other. Finally, since the echo $\mathbf{y}(n,f)$ is often much louder than the near-end speech $\mathbf{s}(n,f)$ and the noise signal $\mathbf{b}(n,f)$ in $\mathbf{d}(n,f)$, the dereverberation filter $\mathbf{\underline{G}}(f)$ here mainly reduces the late reverberation of the echo $\mathbf{y}_\text{l}(n,f)$ instead of the reverberation  of the near-end speech $\mathbf{s}_\text{l}(n,f)$. 
%On the one hand, the dereverberation filter $\mathbf{H}^{togami}(f)$ cannot be applied after the echo reduction filter $\mathbf{G}^{togami}(\Delta, f)$ as they were not designed to handle interactions between each other. As the echo $\mathbf{y}(n,f)$ is much louder than the other signals of the mixture $\mathbf{d}(n,f)$, the signal component $\mathbf{\widehat{s}}_l(n,f)$ in $\mathbf{\widehat{d}}_l(n,f)$ is suboptimal compared to the scenario where the dereverberation filter is applied after the echo reduction filter. On the other hand, the approach is suboptimal for handling the interactions between the filters and the postfilter, and thus is not robust to time-varying scenarios.
%\vspace{-0.3cm}
% TODO : change denomination of algorithm
% TODO :l'absence de modélisation séparée des 3 résiduels doit être pointée comme une limitation ci-dessous (en particulier cela suppose que les 3 résiduels ont les mêmes propriétés spatiales, ce qui n'est évidemment pas vrai dans la pratique).
% TODO

\section{NN-supported BCA algorithm for joint reduction of echo, reverberation and noise \label{sec:dnn_em}}
In this section, we propose a joint NN-supported model to estimate the spectral parameters of the target and the residual signals. We derive an NN-supported BCA algorithm for joint reduction of echo, reverberation and noise that exploits these estimated spectral parameters for an accurate derivation of the echo cancellation and dereverberation filters and the nonlinear postfilter.

\subsection{Model}
\begin{figure}[t]
%\vspace{-.3cm}
\centering
\includegraphics[width=.45\textwidth]{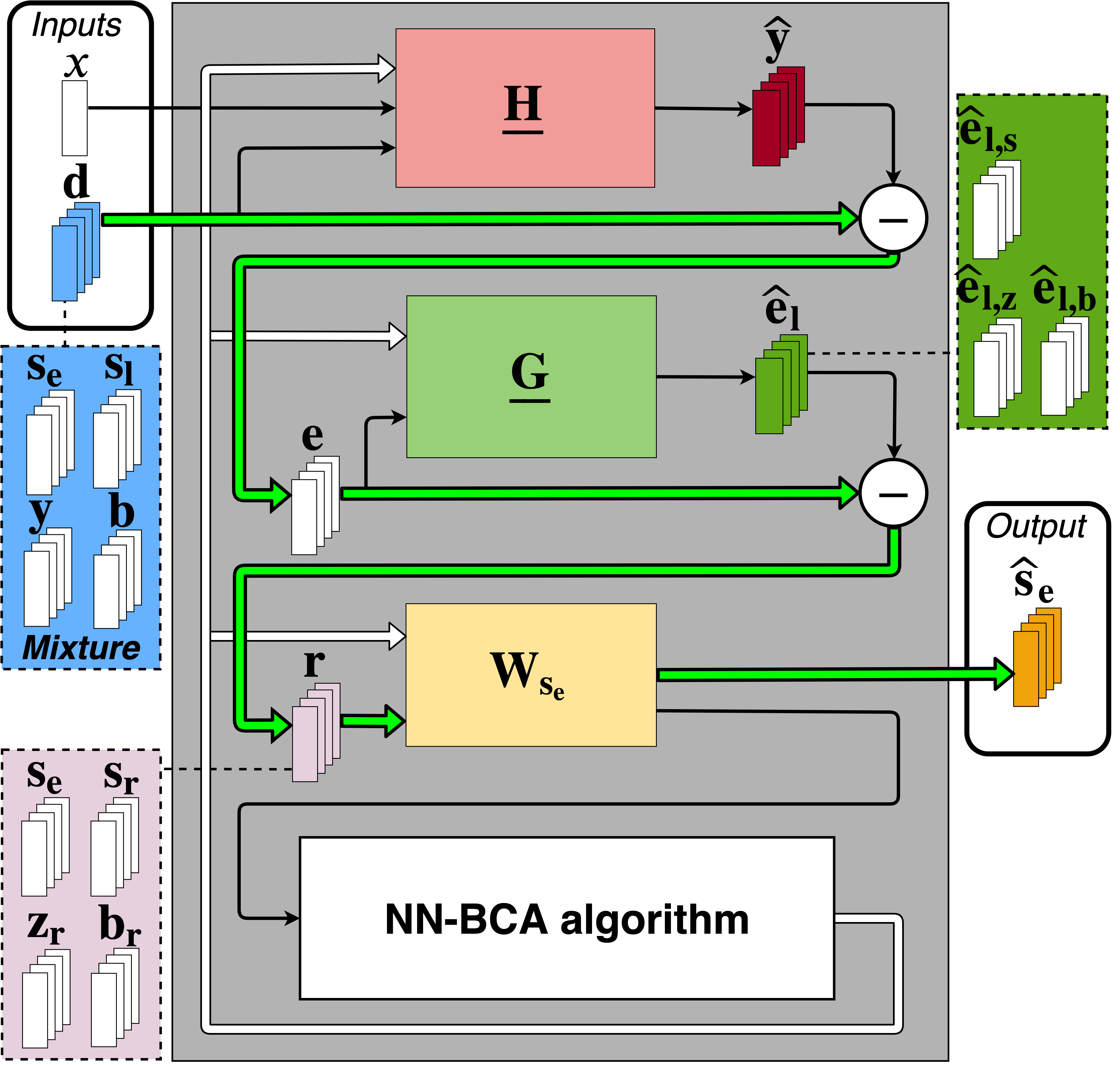}
\caption{Proposed approach. Arrows and lines have the same meaning as in Fig. \ref{fig:togami}.\label{fig:joint}}
%\vspace{-.37cm}
\vspace{-.17cm}
\end{figure}

The approach is illustrated in Fig.\ \ref{fig:joint}. In the first step, we apply the echo cancellation filter $\mathbf{\underline{H}}(f)$ as in (\ref{eq:echo_reduc_1}) and subtract the resulting echo estimate $\mathbf{\widehat{y}}(n,f)$ from $\mathbf{d}(n,f)$:
\begin{equation}
\mathbf{e}(n,f) 	= \mathbf{d}(n,f) - \underbrace{\sum_{k=0}^{K-1} \mathbf{{h}}(k,f) x(n-k,f)}_{= \mathbf{\widehat{y}}(n,f)}. \label{eq:mix_echo_reduc}
\end{equation}
%\underbrace{\sum_{k=0}^{K-1} \mathbf{{h}}(k,f) x(n-k,f)}_{=\mathbf{\widehat{y}}(n,f)} \label{eq:mix_echo_reduc} \\
%\underbrace{\mathbf{y}(n,f) - \mathbf{\widehat{y}}(n,f)}_{= \mathbf{z}(n,f)}.\label{eq:mix_post_echo_reduc}
%			&= \mathbf{s}(n,f) + \mathbf{b}(n,f) + \mathbf{z}(n,f).\label{eq:mix_post_echo_reduc}
% - \mathbf{\widehat{y}}(n,f)
The resulting signal $\mathbf{e}(n,f)$ contains the near-end signal $\mathbf{s}(n,f)$, the residual echo $\mathbf{z}(n,f)$ and the noise signal $\mathbf{b}(n,f)$. Unlike Togami et al. \cite{togami_simultaneous_2014}, we do not apply the dereverberation filter $\mathbf{\underline{G}}(f)$ on the mixture signal $\mathbf{d}(n,f)$, but on the signal $\mathbf{e}(n,f)$ and subtract the resulting late reverberation estimate $\mathbf{\widehat{e}}_\text{l}(n,f)$ from $\mathbf{e}(n,f)$. To the best of our knowledge, this is the first work where the dereverberation filter $\mathbf{\underline{G}}(f)$ is applied after the echo cancellation filter $\mathbf{\underline{H}}(f)$ in the context of joint echo reduction of echo, reverberation and noise. The resulting signal $\mathbf{r}(n,f)$ is thus expressed as \linebreak
\begin{equation}
\mathbf{r}(n, f) = \mathbf{e}(n, f) - \underbrace{\sum_{l=\Delta}^{\Delta+L-1} \mathbf{G}(l,f) \mathbf{e}(n-l,f)}_{=\mathbf{\widehat{e}}_\text{l}(n,f)}. \label{eq:my_dereverberation}					
\end{equation}
%The dereverberation filter should mainly aim at reducing the near-end reverberation $\mathbf{s}_\text{l}(n,f)$. 
 % For this reason, the ones mentioned in Sections \ref{sec:echo_reduc} and \ref{sec:dereverb}, and the presence of the noise signal $\mathbf{b}(n,f)$, undesired signals remain \cite{togami_simultaneous_2014} and can be expressed as
%Note that the filters $\mathbf{\underline{H}}(f)$ and $\mathbf{\underline{G}}(f)$ operate successively, thus they interact with each other.
Since the linear filters $\mathbf{\underline{H}}(f)$ and $\mathbf{\underline{G}}(f)$ are causal, we make the assumption that the observed signals $\mathbf{d}(n,f)$ and $x(n,f)$ are equal to zero for $n < 0$. Since the residual echo $\mathbf{z}(n,f)$ in $\mathbf{e}(n, f)$ is a reduced version of the echo $\mathbf{y}(n, f)$ in $\mathbf{d}(n, f)$, the dereverberation filter $\mathbf{\underline{G}}(f)$ should achieve a greater reduction of the near-end late reverberation $\mathbf{s}_\text{l}(n,f)$ than in Togami et al.'s approach \cite{togami_simultaneous_2014}. Due to the reasons mentioned in Sections \ref{sec:echo_reduc} and \ref{sec:dereverb}, and to the presence of the noise signal $\mathbf{b}(n,f)$, undesired residual signals remain and can be expressed as 
\begin{equation}
\mathbf{r}(n,f) - \mathbf{s}_\text{e}(n,f) = \mathbf{s}_\text{r}(n,f) + \mathbf{z}_\text{r}(n,f)  + \mathbf{b}_\text{r}(n,f), \label{eq:residual_mixture}
\end{equation}
%We denote them as in (\ref{eq:residual_mixture}) but we define them differently as
\textcolor{black}{where $\mathbf{s}_\text{r}(n,f)$ is the residual late reverberation near-end component (see Section \ref{sec:dereverb}), $\mathbf{z}_\text{r}(n,f)$ is the \textit{dereverberated} residual echo which represents the residual echo remaining after linear dereverberation reduced its linear component (see Section \ref{sec:echo_reduc}), and $\mathbf{b}_\text{r}(n,f)$ is the \textit{dereverberated} noise which represents the residual noise remaining after linear dereverberation reduced its stationary component.} The signals $\mathbf{s}_\text{r}(n,f)$, $\mathbf{z}_\text{r}(n,f)$ and  $\mathbf{b}_\text{r}(n,f)$ are defined as
\begin{align}
\mathbf{s}_\text{r}(n,f) &= \mathbf{s}_\text{l}(n,f) - \mathbf{\widehat{e}}_{\text{l}, s}(n,f) \label{eq:true_sr},\\
\mathbf{z}_\text{r}(n,f) &= \mathbf{{z}}(n,f) - \mathbf{\widehat{e}}_{\text{l}, z}(n,f) \label{eq:true_zr},\\
\mathbf{b}_\text{r}(n,f) &= \mathbf{b}(n,f) - \mathbf{\widehat{e}}_{\text{l}, b}(n,f), \label{eq:true_br}
\end{align}
%\begin{align}
%\mathbf{\widehat{e}}_\text{l}(n,f) &= \sum_{l=\Delta}^{\Delta+L-1} \mathbf{G}(l,f) \Big(\mathbf{s}(n-l,f) + \mathbf{z}(n-l,f) + \mathbf{b}(n-l,f) \Big).				
%\end{align}
%Note also that since the filters $\mathbf{\underline{H}}(f)$ and $\mathbf{G}(\Delta, f)$ operate successively, the \textit{dereverberated} residual echo $\mathbf{z}_r(n,f)$ defined in (\ref{eq:true_zr}) is lower than in the case of Togami et al. (\ref{eq:togami_zr}) \cite{togami_simultaneous_2014}.
%&= \mathbf{\widehat{e}}_{\text{l}, s}(n,f) + \mathbf{\widehat{e}}_{\text{l}, z}(n,f)+ \mathbf{\widehat{e}}_{\text{l}, b}(n,f)
%\begin{align}
%\begin{split}
%\mathbf{\widehat{e}}_\text{l}(n,f) = \sum_{l=\Delta}^{\Delta+L-1} \mathbf{G}(l,f) \Big(&\mathbf{s}(n-l,f) +  \mathbf{z}(n-l,f) \\
%&+ \mathbf{b}(n-l,f) \Big).
%\end{split}				
%\end{align}
where the signals $\mathbf{\widehat{e}}_{\text{l}, s}(n,f) = \sum_{l=\Delta}^{\Delta+L-1} \mathbf{G}(l,f) \mathbf{s}(n-l,f)$,  $\mathbf{\widehat{e}}_{\text{l}, z}(n,f) = \sum_{l=\Delta}^{\Delta+L-1} \mathbf{G}(l,f) \mathbf{z}(n-l,f) $ and $ \mathbf{\widehat{e}}_{\text{l}, b}(n,f) = \sum_{l=\Delta}^{\Delta+L-1} \mathbf{G}(l,f) \mathbf{b}(n-l,f) $  are the latent components of $ \mathbf{\widehat{e}}_\text{l}(n,f)$ resulting from (\ref{eq:my_dereverberation}). To recover the signal $\mathbf{s}_\text{e}(n,f)$ from the signal $\mathbf{r}(n,f)$, we apply a multichannel Wiener postfilter $\mathbf{W}_{s_\text{e}}(n,f) \in \mathbb{C}^{M \times M}$ on the signal $\mathbf{r}(n,f)$ as 
\begin{equation}
\mathbf{\widehat{s}}_\text{e}(n,f) = \mathbf{W}_{s_\text{e}}(n,f) \mathbf{r}(n,f). \label{eq:se_est}
\end{equation}
\indent Inspired by WPE for dereverberation \cite{nakatani_speech_2010}, we estimate $\mathbf{\underline{H}}(f)$, $\mathbf{\underline{G}}(f)$ and $ \mathbf{W}_{s_\text{e}}(n,f)$ by modeling the target $\mathbf{s}_\text{e}(n,f)$ and the three residual signals $\mathbf{s}_\text{r}(n,f)$, $\mathbf{z}_\text{r}(n,f)$ and $\mathbf{b}_\text{r}(n,f)$ with a multichannel local Gaussian framework. In the following we use the general notation $\mathbf{c}(n,f)$ to denote each one of these four signals, and consider them as \emph{sources} to be separated. Each of these four sources is modeled as
\begin{equation}
\mathbf{c}(n,f) \sim \mathcal{N}_{\mathbb{C}}\left(\mathbf{0}, v_c(n,f) \mathbf{R}_c(f) \right), \label{eq:lgm}
\end{equation}
where $v_{c}(n,f) \in \mathbb{R}_+$ and $\mathbf{R}_{c}(f) \in \mathbb{C}^{M \times M}$ denote the power spectral density (PSD) and the spatial covariance matrix (SCM) of the source, respectively \cite{duong_under_2010}. The multichannel Wiener filter for the source $\mathbf{c}(n,f)$ is formulated as
\begin{equation}
\medmuskip=2mu   % reduce spacing around binary operators
\thickmuskip=3mu % reduce spacing around relational operators
\renewcommand\arraystretch{1.5}
\mathbf{W}_{c}(n,f) = v_c(n,f) \mathbf{R}_c(f) \Bigg(\sum_{\mathclap{c' \in \mathcal{C}}} v_{c'}(n,f) \mathbf{R}_{c'}(f)  \Bigg)^{-1}, \label{eq:wiener}
\end{equation}
\linebreak where $\mathcal{C} = \{\mathbf{s}_\text{e}, \mathbf{s}_\text{r}, \mathbf{z}_\text{r}, \mathbf{b}_\text{r}\}$ denotes all four sources in signal $\mathbf{r}(n,f)$. The postfilter $\mathbf{W}_{s_\text{e}}(n,f)$ is a specific case of (\ref{eq:wiener}) where $\mathbf{c}(n,f) = \mathbf{s}_\text{e}(n,f)$.

\subsection{Likelihood}

In order to estimate the parameters of this model, we must first express its likelihood. Following (\ref{eq:mix_echo_reduc}), (\ref{eq:my_dereverberation}), (\ref{eq:residual_mixture}) and (\ref{eq:lgm}), the log-likelihood of the observed sequence $\mathcal{O} = \{\mathbf{d}(n,f), x(n,f) \}_{n, f}$ is given by
\begin{align}
\begin{split}
\mathcal{L} &\Big(\mathcal{O};\Theta_{H}, \Theta_{G}, \Theta_c \Big) \\
 &=  \sum_{f=0}^{F-1} \sum_{n=0}^{N-1} \log p\Big(\mathbf{d}(n,f) \Big| \mathbf{d}(n-1,f), \ldots, \mathbf{d}(0,f), \\
  & \qquad \qquad \qquad \qquad \qquad \quad x(n,f), \ldots, x(0,f) \Big),
\end{split}
\\
 &= \sum_{f=0}^{F-1} \sum_{n=0}^{N-1} \log  \mathcal{N}_{\mathbb{C}} \Big(\mathbf{d}(n,f); \bm{\mu}_{\mathbf{d}}(n, f), \mathbf{R}_{\mathbf{d}\mathbf{d}}(n,f)\Big), \label{eq:likelihood_1}
\end{align}
where
\begin{equation}
\bm{\mu}_{\mathbf{d}}(n, f) = \sum_{\mathclap{k=0}}^{\mathclap{K-1}} \mathbf{{h}}(k,f) x(n-k,f) + \sum_{\mathclap{l=\Delta}}^{\mathclap{\Delta+L-1}} \mathbf{G}(l,f) \mathbf{e}(n-l,f),
\end{equation}
\begin{flalign}
\mathbf{R}_{\mathbf{d}\mathbf{d}}(n,f) = \sum_{c' \in \mathcal{C}} v_{c'}(n,f) \mathbf{R}_{c'} (f),&&\label{eq:mix_cov}
\end{flalign}
\iffalse
\begin{flalign}
\begin{split}
\bm{\mu}_{\mathbf{d}}(n, f) &= \sum_{k=0}^{K-1} \mathbf{{h}}(k,f) x(n-k,f) \\
						&\qquad + \sum_{l=\Delta}^{\Delta+L-1} \mathbf{G}(l,f) \mathbf{e}(n-l,f),
\end{split}
\\
\mathbf{R}_{\mathbf{d}\mathbf{d}}(n,f) &= \sum_{c' \in \mathcal{C}} v_{c'}(n,f) \mathbf{R}_{c'} (f), \label{eq:mix_cov} &&
\end{flalign}
\fi
%The likelihood gives us a common criterion whereby the energy of the mixture signal $\mathbf{d}(n,f)$ is weighted by the energy of the signals $\mathbf{s}_\text{e}(n,f)$, $\mathbf{s}_\text{r}(n,f)$, $\mathbf{z}_\text{r}(n,f)$ and $\mathbf{b}_\text{r}(n,f)$.
and $\Theta_{H} = \{\mathbf{\underline{H}} (f) \}_f$, $\Theta_{G} = \{\mathbf{\underline{G}}(f) \}_f$ and $\Theta_c = \{ v_{c}(n,f), \mathbf{R}_{c}(f) \}_{c, n, f}$ are the parameters to be estimated. The resulting ML optimization problem has no closed form solution, hence we need to estimate the parameters via an iterative procedure.
%\begin{split}
%\mathcal{L} \left(\mathcal{D};\Theta_{H, G}, \Theta_c \right)   & = \sum_{f=0}^{F-1} \sum_{n=0}^{N-1} \log  \mathcal{N}%_{\mathbb{C}} \Big(\mathbf{d}(n,f); \bm{\mu}_d(n, f), \mathbf{R}_d(n,f)\Big) \label{eq:likelihood_1}
%\end{split}
%\begin{split}
%  & = \sum_{f=0}^{F-1} \sum_{n=0}^{N-1} \log |\mathbf{R}_d(n,f)|^{-1} \\
%              &\qquad - \Big(\mathbf{d}(n,f) - \bm{\mu}_d(n, f)\Big)^H \mathbf{R}_d^{-1}(n,f) \Big(\mathbf{d}(n,f) - \bm{\mu}_d(n, f)\Big),
%\end{split} 
%\\

% TODO : parler la procedure iterative closed-form etc. etc.
%Avant de parler de NN-EM, il faut souligner qu'en plus le modèle de signal ne contraint pas les variances et dire que nous proposons d'utiliser un NN pour modéliser conjointement les variances vc des 4 composantes, c'est-à-dire vse, vsr, vzr, vbr. Ce NN va contraindre leurs valeurs au moment de l'estimation. 

%We propose to extend Nugraha et al.'s NN-EM algorithm \cite{nugraha} to optimize jointly the linear filter parameters $\Theta_{H}$, $\Theta_{G}$ and the source parameters $\Theta_c$ for simultaneous reduction of echo, reverberation and noise.

% TODO : change name of subsection
\subsection{Iterative optimization algorithm}
\begin{figure}[t]
%\vspace{-.3cm}
\centering
\includegraphics[width=.5\textwidth]{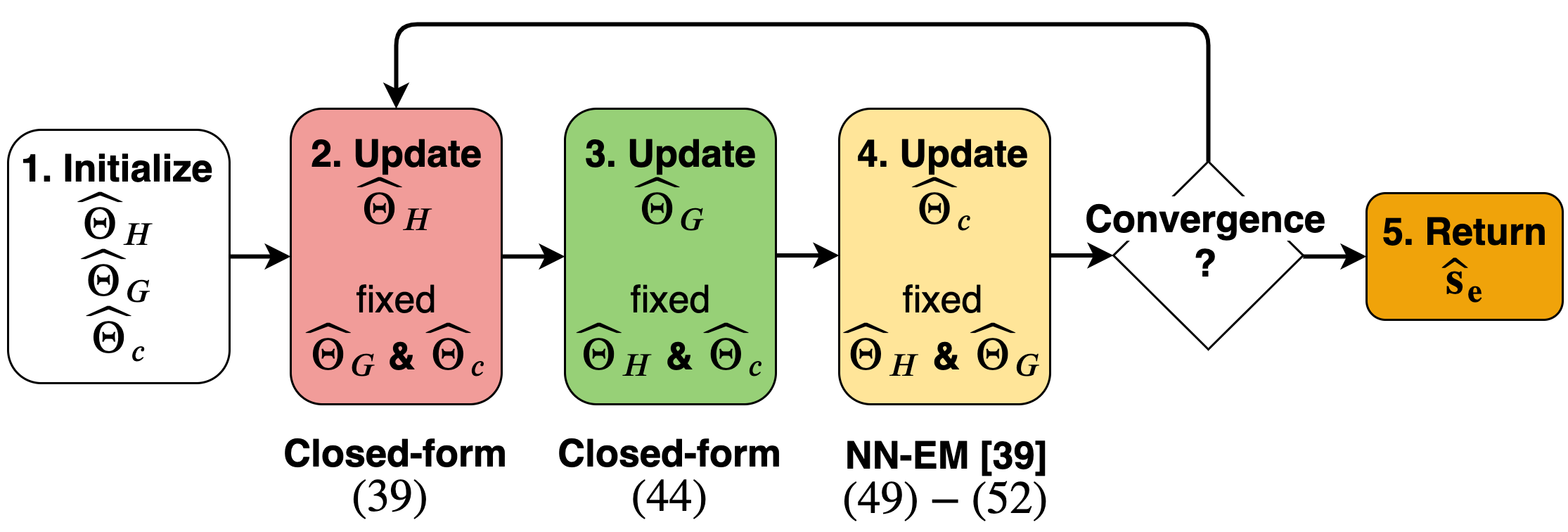}
\caption{\label{fig:flowchart} Flowchart of the proposed BCA algorithm.}
\vspace{-.15cm}
\end{figure}
%The algorithm iteratively increases the log-likelihood (\ref{eq:likelihood_1}) by using a BCA method. 
We propose a BCA algorithm for likelihood optimization. Each iteration $i$ comprises the following three maximization steps:
\begin{align}
\widehat{\Theta}_{H} &\gets \argmax_{\Theta_H} \mathcal{L} \left(\mathcal{O};\Theta_{H}, \widehat{\Theta}_{G},  \widehat{\Theta}_c \right), \label{eq:maximize_H}\\
\widehat{\Theta}_{G} &\gets \argmax_{\Theta_G} \mathcal{L} \left(\mathcal{O};\widehat{\Theta}_{H}, \Theta_{G}, \widehat{\Theta}_c \right), \label{eq:maximize_G} \\
\widehat{\Theta}_{c} &\gets \argmax_{\Theta_c} \mathcal{L} \left(\mathcal{O}; \widehat{\Theta}_{H}, \widehat{\Theta}_{G}, \Theta_c \right).  \label{eq:maximize_c}
\end{align}
The solutions of (\ref{eq:maximize_H}) and (\ref{eq:maximize_G}) are closed-form. As there is no closed-form solution for  (\ref{eq:maximize_c}), we propose to use a modified version of Nugraha et al.'s NN-EM algorithm \cite{nugraha}. The overall flowchart of the proposed algorithm is shown in Fig.\ \ref{fig:flowchart}. Note that it is also possible to optimize the parameters ${\Theta}_{H}$, ${\Theta}_{G}$ and ${\Theta}_{c}$ with the EM algorithm by adding a nuisance term to (\ref{eq:residual_mixture}) \cite{ozerov_multichannel_2010}. However, this approach would be less efficient to derive the filter parameters ${\Theta}_{H}$ and ${\Theta}_{G}$. In the next subsections, we provide the initialization and the update rules for steps (\ref{eq:maximize_H})--(\ref{eq:maximize_c}) of our proposed algorithm at iteration $i$. The derivation of these update rules is detailed in our companion technical report \cite[Sec. 3]{carbajal_supporting_2019}. \textcolor{black}{ At each iteration $i$, we use the dereverberation filter parameters ${\Theta}_{G}$ and the source parameters ${\Theta}_{c}$ of the previous iteration $i-1$.}

\subsubsection{Initialization \label{sec:initialization}}

We initialize the linear filters $\mathbf{\underline{H}}(f)$ and $\mathbf{\underline{G}}(f)$ to $\mathbf{\underline{H}}_0(f)$ and $\mathbf{\underline{G}}_0(f)$, respectively. The PSDs $v_c(n,f)$ of the four sources are jointly initialized using a pretrained NN denoted as $\text{NN}_0$ and the SCMs $\mathbf{R}_c(f)$ as the identity matrix $\mathbf{I}_M$. The inputs, the targets and the architecture of $\text{NN}_0$ are described in Section IV below.

\subsubsection{Echo cancellation filter parameters $\Theta_H$}
% TODO : define h as one vector
The echo cancellation filter $\mathbf{\underline{H}}(f)$ is updated as
\begin{equation}
\underline{\mathbf{h}}(f)  = \mathbf{P}(f)^{-1} \mathbf{p}(f), \label{eq:update_h}
\end{equation}
where
\begin{align}
\mathbf{P}(f) &= \sum_{n=0}^{N-1}  \mathbf{\underline{X}}_\text{r}(n,f)^H \mathbf{R}_{\mathbf{d} \mathbf{d}} (n,f)^{-1} \mathbf{\underline{X}}_\text{r}(n, f), \label{eq:update_P} \\
\mathbf{p}(f) &=\sum_{n=0}^{N-1} \mathbf{\underline{X}}_\text{r}(n,f)^H \mathbf{R}_{\mathbf{d} \mathbf{d}} (n,f)^{-1} \mathbf{r}_d(n,f). \label{eq:update_p}
\end{align}
%where $\mathbf{\underline{h}}(f) = \text{vec}\big(\mathbf{\underline{H}}(f)\big) \in \mathbb{C}^{M K \times 1}$ is a vectorized version of the filter $\mathbf{\underline{H}}(f)$ \cite{carbajal_supporting_2019}
 $\mathbf{\underline{h}}(f) = [\mathbf{h}(0,f)^T \ldots \mathbf{h}(K-1,f)^T]^T \in \mathbb{C}^{M K \times 1}$ is a vectorized version of $\mathbf{\underline{H}}(f)$, $\mathbf{\underline{X}}_\text{r}(n,f) = \left[\mathbf{{X}}_\text{r}(n,f) \ldots \mathbf{{X}}_\text{r}(n-K+1,f) \right] \in \mathbb{C}^{M \times M K}$ results from the $K$ taps $\mathbf{{X}}_\text{r}(n-k,f) \in \mathbb{C}^{M \times M}$. \textcolor{black}{The $K$ taps $\mathbf{{X}}_\text{r}(n-k,f)$ are \emph{dereverberated}  versions of $x(n-k,f)$ obtained by applying the dereverberation filter $\mathbf{\underline{G}}(f)$ on $x(n-k,f)$:}
 %$L$ previous frames of $x(n-k,f)$ as
 \begin{equation}
 \mathbf{{X}}_\text{r}(n-k,f) = x(n-k,f)\mathbf{I}_M  - \sum_{l=\Delta}^{\Delta+L-1} x(n-k-l,f) \mathbf{G}(l, f).
 \end{equation}
\textcolor{black}{$\mathbf{r}_d(n,f)$ is a \emph{dereverberated} version of $\mathbf{d}(n,f)$ obtained by applying the dereverberation filter $\mathbf{\underline{G}}(f)$ on $\mathbf{d}(n,f)$ without prior echo cancellation:}
\begin{equation}
\mathbf{r}_d(n,f) = \mathbf{d}(n,f) - \sum_{k=\Delta}^{\Delta+L-1} \mathbf{G}(k,f) \mathbf{d}(n-k,f)
\end{equation}
Note that the update of the echo cancellation filter $\mathbf{\underline{H}}(f)$ is influenced by the dereverberation filter $\mathbf{\underline{G}}(f)$ through the terms \mbox{$\mathbf{\underline{X}}_\text{r}(n,f)$} and $\mathbf{r}_d(n,f)$.  \textcolor{black}{This update prevents the echo cancellation filter $\mathbf{\underline{H}}(f)$ from reducing the component of the echo $\mathbf{y}(n,f)$ already reduced by the dereverberation filter $\mathbf{\underline{G}}(f)$.}
\par
\textcolor{black}{The update of the echo cancellation filter $\mathbf{\underline{H}}(f)$ also depends on the PSDs $v_c(n,f)$ and the SCMs $\mathbf{R}_c(f)$ through the term $\mathbf{R}_{\mathbf{d}\mathbf{d}}(n,f)$ defined in \eqref{eq:mix_cov}. As the post-filter $\mathbf{W}_c(n,f)$ is used for the updates of both of the PSDs $v_c(n,f)$ (see Section \ref{sec:inputs} below) and the SCMs $\mathbf{R}_c(f)$ (see Section \ref{sec:variance} below), the update of the echo cancellation filter $\mathbf{\underline{H}}(f)$ is also influenced by the post-filter $\mathbf{W}_c(n,f)$.}

\subsubsection{Dereverberation filter parameters $\Theta_G$}
 Similarly to WPE for dereverberation \cite{yoshioka_generalization_2012}, the dereverberation filter $\mathbf{\underline{G}}(f)$ is updated as
 \begin{equation}
\underline{\mathbf{g}}(f) = \mathbf{Q}(f)^{-1} \mathbf{q}(f), \label{eq:update_g} 
 \end{equation}
 where
\begin{align}
\mathbf{Q}(f) &= \sum_{n=0}^{N-1} \mathbf{\underline{E}}(n,f)^H \mathbf{R}_{\mathbf{d} \mathbf{d}}  (n,f)^{-1} \mathbf{\underline{E}}(n,f), \label{eq:update_Q} \\
\mathbf{q}(f) &=\sum_{n=0}^{N-1} \mathbf{\underline{E}}(n,f)^H \mathbf{R}_{\mathbf{d} \mathbf{d}}  (n,f)^{-1}  \mathbf{e}(n,f). \label{eq:update_q}
\end{align}
%where $\mathbf{\underline{g}}(\Delta, f) = \text{vec}\big(\mathbf{\underline{G}}(f)\big) \in \mathbb{C}^{M^2 L \times 1}$ is a vectorized version of the filter $\mathbf{\underline{G}}(f)$ \cite{carbajal_supporting_2019}, 
%and $\mathbf{\underline{E}}(n-\Delta, f) = \text{mat}\big(\mathbf{\underline{e}}(n-\Delta,f)\big) \in \mathbb{C}^{M \times M^2 L}$ is a multichannel version of $\mathbf{\underline{e}}(n-\Delta,f) = [\mathbf{e}(n-\Delta,f)^T \ldots \mathbf{e}(n-\Delta-L+1,f)^T]^T \in \mathbb{C}^{M^2 L \times 1}$\cite{carbajal_supporting_2019}.
$\mathbf{\underline{g}}(f) = [\mathbf{g}_1(\Delta,f)^T \ldots \mathbf{g}_M(\Delta,f)^T \ldots \ldots \mathbf{g}_1(\Delta+L-1,f)^T \ldots \mathbf{g}_M(\Delta+L-1,f)^T]^T \in \mathbb{C}^{M^2 L \times 1}$  is a vectorized version of $\mathbf{\underline{G}}(f)$, and $\mathbf{\underline{E}}(n, f) = \left[\mathbf{{E}}(n-\Delta,f) \ldots \mathbf{{E}}(n-\Delta-L+1,f) \right] \in \mathbb{C}^{M \times M^2 L}$ results from the $L$ taps $\mathbf{E}(n-l,f) \in \mathbb{C}^{M \times M^2}$ obtained as
\begin{equation}
\mathbf{E}(n-l,f) = \mathbf{I}_M \otimes \mathbf{e}(n-l,f)^T. 
\end{equation}
The update of the dereverberation filter $\mathbf{\underline{G}}(f)$ is influenced by the echo cancellation filter $\mathbf{\underline{H}}(f)$ through the terms $\mathbf{e}(n,f)$. \textcolor{black}{Similarly to the echo cancellation filter $\mathbf{\underline{H}}(f)$, the update of the dereverberation filter $\mathbf{\underline{G}}(f)$ is also influenced by the post-filter $\mathbf{W}_c(n,f)$ through the PSDs $v_c(n,f)$ and the SCMs $\mathbf{R}_c(f)$ used in the term $\mathbf{R}_{\mathbf{d}\mathbf{d}}(n,f)$ defined in \eqref{eq:mix_cov}.}

\subsubsection{Variance and spatial covariance parameters $\Theta_c$ \label{sec:variance}}
As there is no closed-form solution for the log-likelihood optimization with respect to $\Theta_c$, we estimate the variance and spatial covariance parameters using an EM algorithm.  Given the past sequence of the mixture signal $\mathbf{d}(n,f)$, the far-end signal $x(n,f)$ and its past sequence, and the linear filters $\underline{\mathbf{H}}(f)$ and $\underline{\mathbf{G}}(f)$, the residual mixture signal $\mathbf{r}(n, f)$ is conditionally distributed as
\begin{equation}
\begin{split}
\mathbf{r}(n, f) \Big| \mathbf{d}(n-1&, f), \ldots, \mathbf{d}(0, f), x(n,f), \ldots, x(0,f), \\
 &\underline{\mathbf{H}}(f), \underline{\mathbf{G}}(f) \sim \mathcal{N}_{\mathbb{C}} \Big(\mathbf{0}, \mathbf{R}_{\mathbf{d}\mathbf{d}} (n,f)  \Big).
\end{split}
\end{equation}
% TODO : Avant de parler de NN-EM, il faut souligner qu'en plus le modèle de signal ne contraint pas les variances et dire que nous proposons d'utiliser un NN pour modéliser conjointement les variances vc des 4 composantes, c'est-à-dire vse, vsr, vzr, vbr. Ce NN va contraindre leurs valeurs au moment de l'estimation. 
The signal model is conditionally identical to the local Gaussian modeling framework for source separation \cite{duong_under_2010}. However, this framework does not constrain the PSDs or the SCMs which results in a permutation ambiguity (see Section \ref{sec:noise}). Instead, after each update of the linear filters $\mathbf{\underline{H}}(f)$ and $\mathbf{\underline{G}}(f)$, we propose to use one iteration of Nugraha et al.'s NN-EM algorithm to update the PSDs and the SCMs of the target and residual signals $\mathbf{s}_\text{e}(n,f)$,  $\mathbf{s}_\text{r}(n,f)$,  $\mathbf{z}_\text{r}(n,f)$ and $\mathbf{b}_\text{r}(n,f)$  \cite{nugraha}. In the E-step, each of these four sources $\mathbf{c}(n,f)$ is estimated as
\begin{equation}
\mathbf{\widehat{c}}(n,f) = \mathbf{W}_c(n,f) \mathbf{r}(n,f) \label{eq:source_est},
\end{equation}
 and its second-order posterior moment $\mathbf{\widehat{R}}_c(n,f)$ as 
 \begin{equation}
\mathbf{\widehat{R}}_c(n,f) = \mathbf{\widehat{c}}(n,f) \mathbf{\widehat{c}}(n,f)^H + \Big( \mathbf{I} - \mathbf{W}_c(n,f) \Big) v_c(n,f) \mathbf{R}_c(f).
\label{eq:posterior_wiener}
\end{equation}
 \iffalse
\begin{equation}
\begin{split}
\mathbf{\widehat{R}}_c(n,f) = &\mathbf{\widehat{c}}(n,f) \mathbf{\widehat{c}}(n,f)^H \\
&+ \Big( \mathbf{I} - \mathbf{W}_c(n,f) \Big) v_c(n,f) \mathbf{R}_c(f).
\end{split}
\label{eq:posterior_wiener}
\end{equation}
\fi
In the M-step, we consider a weighted form of update for the SCMs $\mathbf{R}_c(f)$ \cite{nugraha_multichannel_2016}
\begin{equation}
\mathbf{{R}}_{c}(f) = \left(\sum_{n=0}^{N-1} w_c(n, f) \right)^{-1} \sum_{n=0}^{N-1} \frac{w_c(n, f)}{v_{c}(n,f)} \mathbf{\widehat{R}}_{c}(n,f), \label{eq:weighted_Rc_wiener}
\end{equation}
where $w_c(n, f)$ denotes the weight of the source $\mathbf{c}(n,f)$. When $w_c(n, f) = 1$ , (\ref{eq:weighted_Rc_wiener}) reduces to the exact EM algorithm \cite{duong_under_2010}. Here, we use $w_c(n,f) = v_c(n,f)$ \cite{liutkus_scalable_2015, nugraha_multichannel_2016}. Experience shows that this weighting trick mitigates inaccurate estimates in certain time-frequency bins and increases the importance of the bins for which $v_c(n,f)$ is large. As the PSDs are constrained, we also need to constrain $\mathbf{{R}}_{c}(f)$ so as to encode only the spatial information of the sources. We modify (\ref{eq:weighted_Rc_wiener}) by normalizing $\mathbf{{R}}_{c}(f)$ after each update \cite{nugraha_multichannel_2016}:
%\mathbf{\widetilde{R}}_{c}(f) &= \left(\sum_{n=0}^{N-1} v_c(n, f) \right)^{-1} \sum_{n=1}^N \mathbf{\widehat{R}}_{c}(n,f), \label{eq:weighted_Rc_wiener} \\
\begin{equation}
\mathbf{{R}}_c(f) \gets \frac{M}{\text{tr}\left(\mathbf{R}_{c}(f) \right)} \mathbf{R}_{c}(f). \label{eq:weighted_norm_Rc_wiener}
\end{equation}
The PSDs $v_c(n,f)$ of the four sources are jointly updated using a pretrained NN denoted as $\text{NN}_i$, with $i\geq 1$ the iteration index. The inputs, the targets and the architecture of $\text{NN}_i$ are described in Section IV below. 
%For each iteration $i$ of the BCA algorithm, we perform $1$ iteration of this modified version of NN-EM.
\subsubsection{Estimation of the final early near-end component $\mathbf{s}_\text{e}(n,f)$}
% TODO : Ajoute une partie 3.4 pour expliquer comment, une fois que l'algo a convergé, le signal rehaussé est finalement estimé.
%When we know \widehat{Theta}_c, we compute h.... then apply ... then compute g ... then apply ... then compute W... then apply
% TODO : We initialize the PSDs using a pretrained NN denoted by $\text{NN}_0$ and the SCMs as the identity matrix $\mathbf{I}_M$. 
% TODO : After parameter optimization, the output signal is set to               , which is the MMSE estimate of the   th source signal in the E step. 
\textcolor{black}{Once the proposed iterative optimization algorithm has converged after $I$ iterations, we have estimates of the PSDs $v_c(n,f)$, the SCMs $\mathbf{R}_c(f)$ and the dereverberation filter $\mathbf{\underline{G}}(f)$. We can perform one more iteration of the NN-supported BCA algorithm to derive the final filters $\mathbf{\underline{H}}(f)$, $\mathbf{\underline{G}}(f)$ and $\mathbf{W}_{s_\text{e}}(n,f)$. Ultimately, we obtain the target estimate $\mathbf{\widehat{s}}_\text{e}(n,f)$ using (\ref{eq:mix_echo_reduc}), (\ref{eq:my_dereverberation}) and (\ref{eq:source_est}).} For the detailed pseudo-code of the algorithm, please refer to the supporting document \cite[Sec.3.5]{carbajal_supporting_2019}.

\section{NN spectral model}
In this section, we define the inputs, the targets and the architecture of the NN used to initialize and update the target and residual PSDs.
\begin{figure}[t]
\centering
\subfloat[][Early near-end speech $\mathbf{s}_\text{e}$]{
\includegraphics[width=.42\textwidth, valign=t]{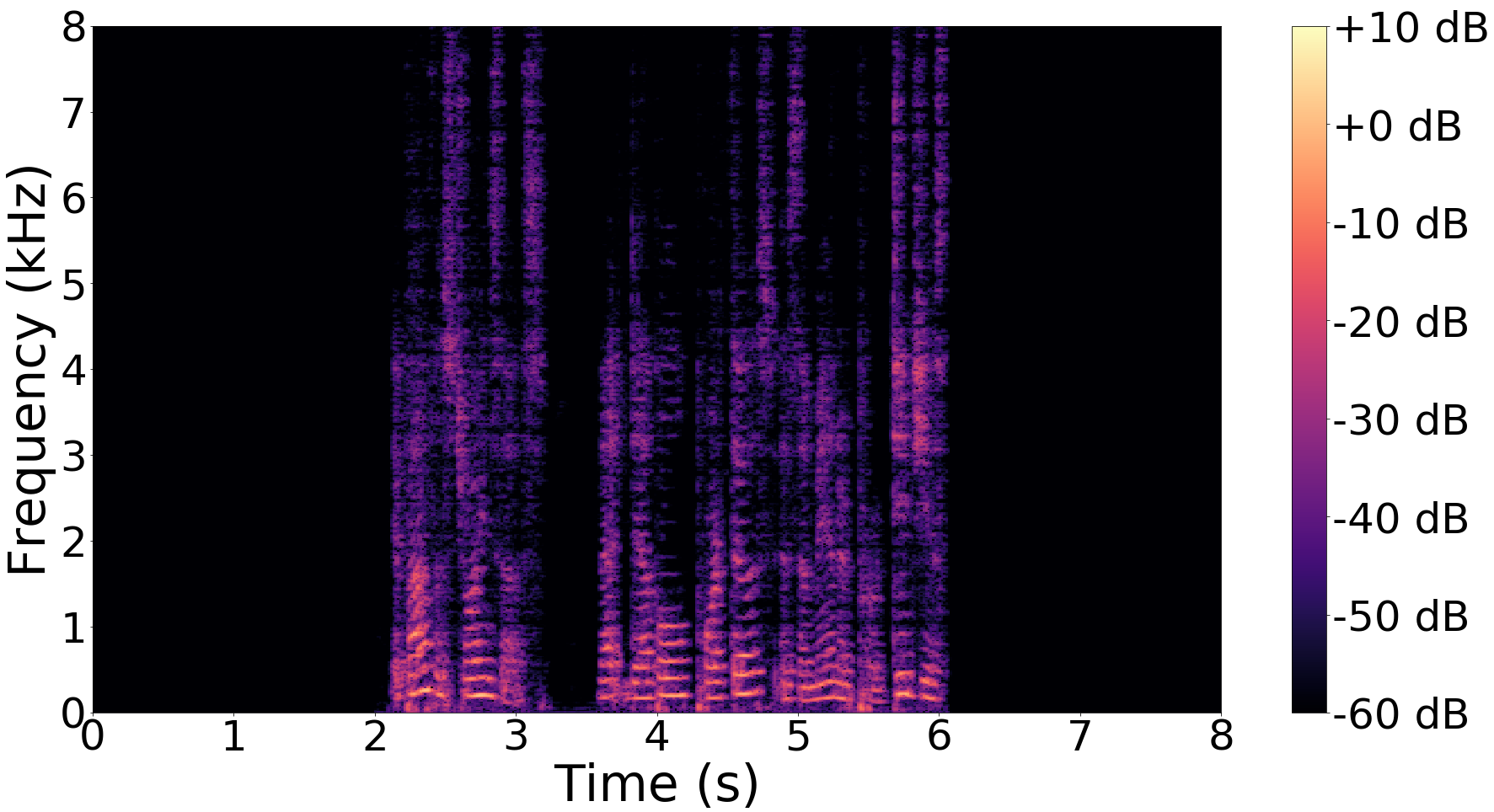}
}%
\\
\subfloat[][Near-end residual late reverberation $\mathbf{s}_\text{r}$]{
\includegraphics[width=.42\textwidth, valign=t]{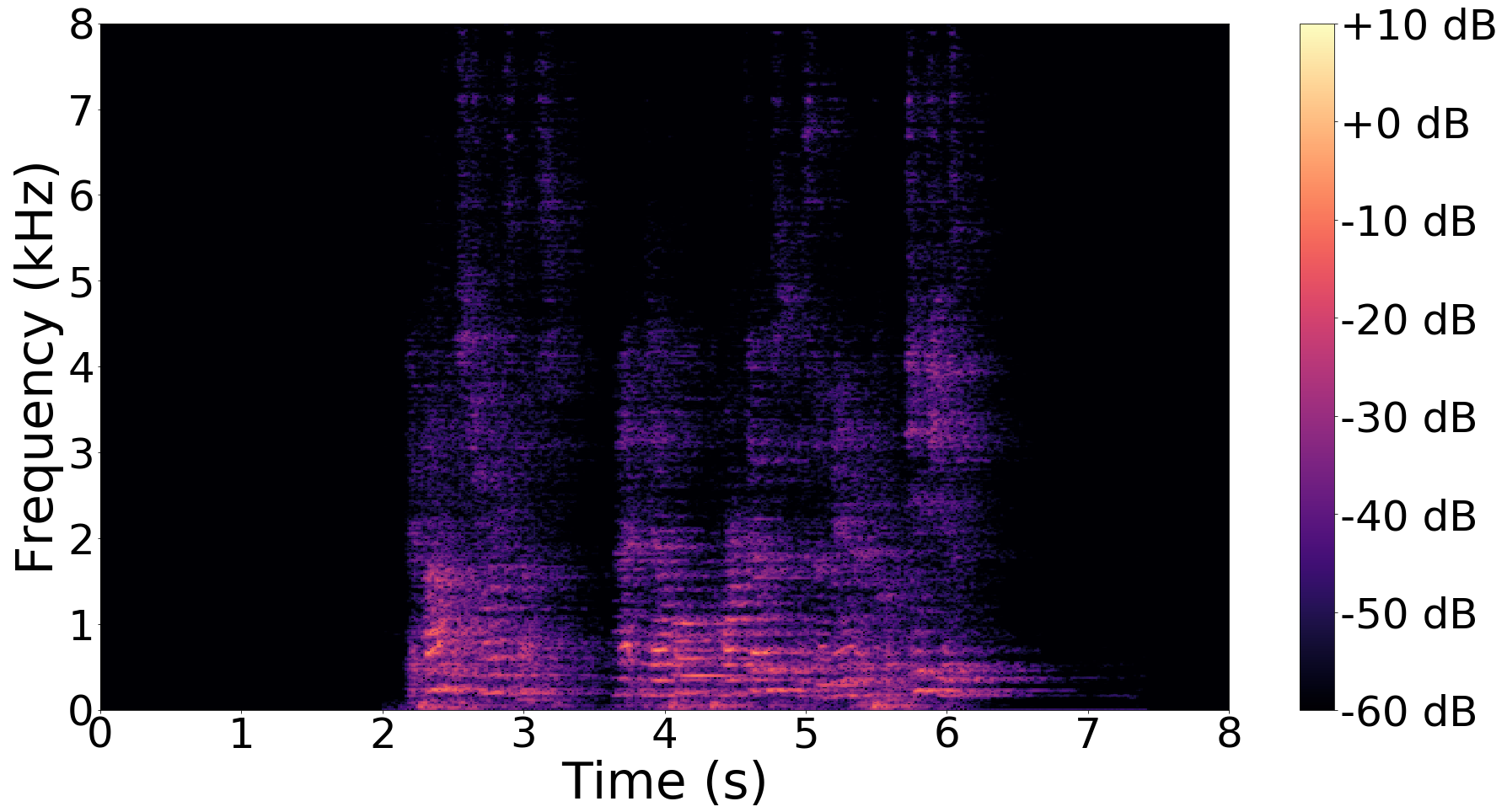}
}%
\\
\subfloat[][\textit{Dereverberated} residual echo $\mathbf{z}_\text{r}$]{
\includegraphics[width=.42\textwidth, valign=t]{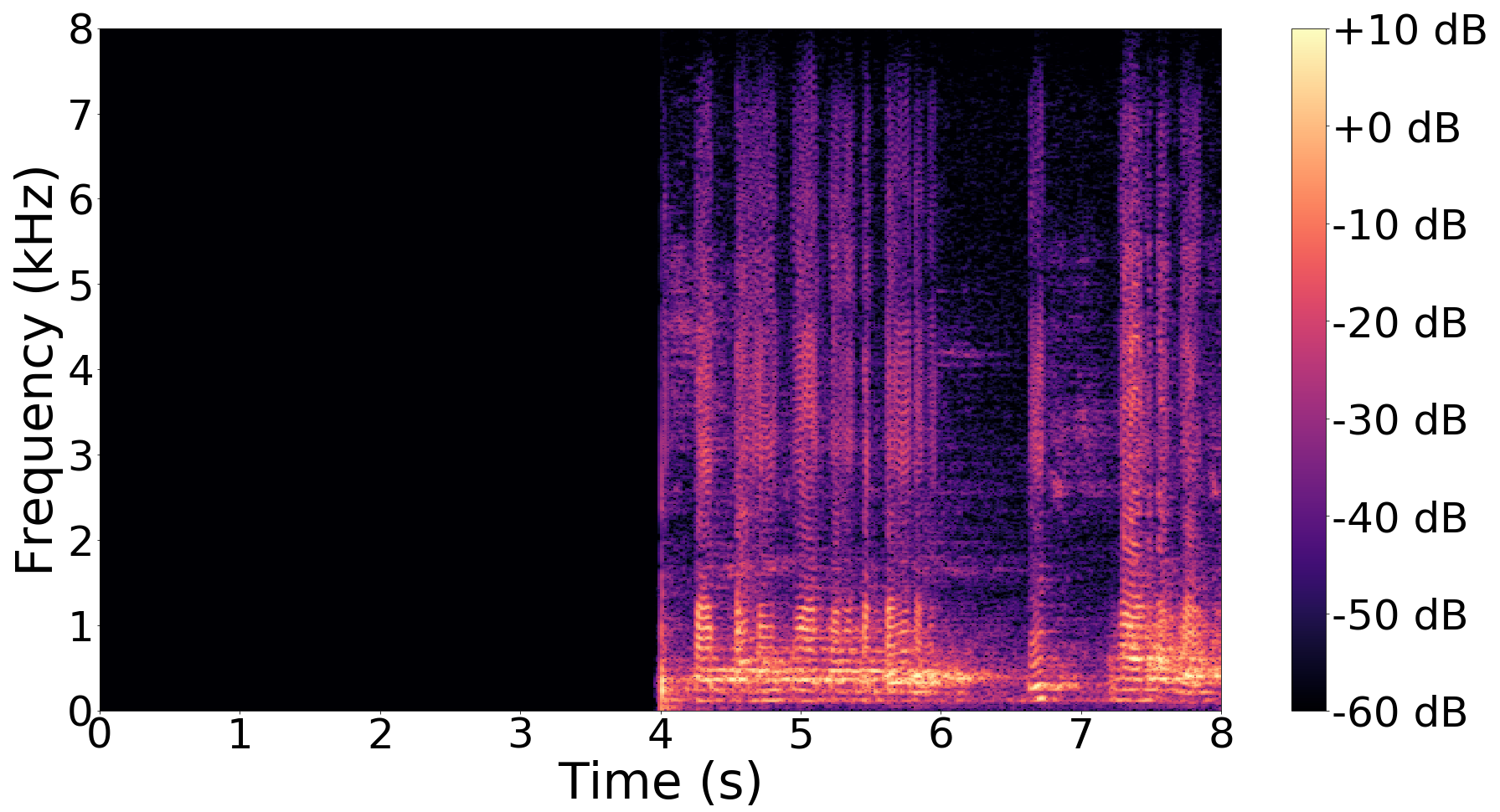}
}%
\\
\subfloat[][\textit{Dereverberated} noise $\mathbf{b}_\text{r}$]{
\includegraphics[width=.42\textwidth, valign=t]{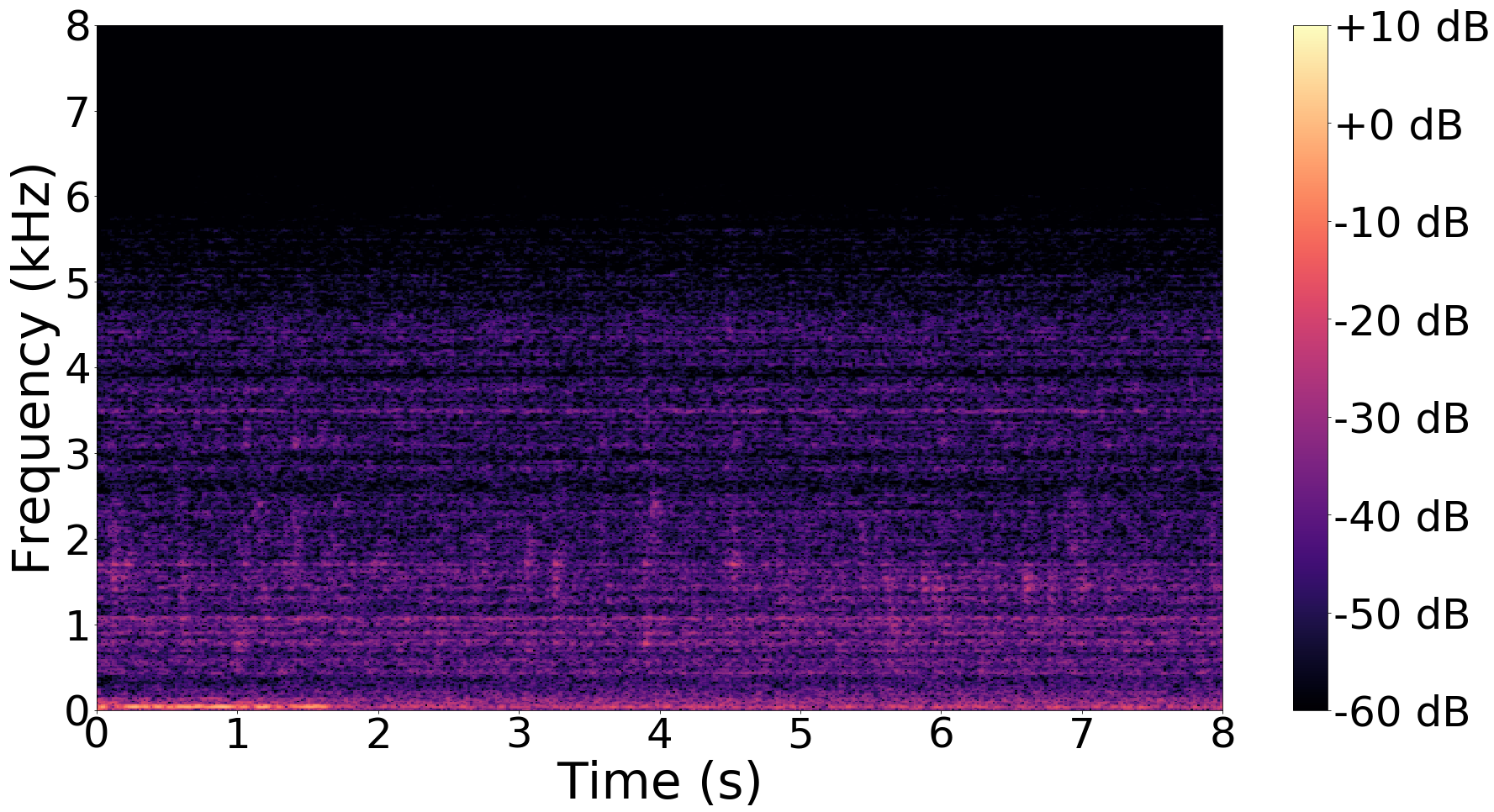}
}%
\caption{\label{fig:example_targets} Example ground truth target and residual signal PSDs in the training set.}
%\vspace{-.5cm}
\end{figure}

\subsection{Targets \label{sec:targets}}

% TODO : we predict the 4 targets jointly
%\label{eq:psd_nugraha}
% TODO : replace NN estimation by
%At initialization, we set the linear filters $\mathbf{\underline{H}}(f)$ and $\mathbf{\underline{G}}(f)$ to zero, and we compute the PSDs as $v_c(n,f) = \frac{1}{M} \rVert \mathbf{c}(n,f) \rVert^2$ using $\mathbf{s}_\text{e}(n,f)$,  $\mathbf{s}_\text{l}(n,f)$, $\mathbf{y}(n,f)$ and $\mathbf{b}(n,f)$
Estimating $\sqrt{v_c(n,f)}$ has been shown to provide better results than estimating the power spectra $v_c(n,f)$, as the square root compresses the signal dynamics \cite{nugraha}. Therefore we define $\Big[\sqrt{v_{{s}_\text{e}}(n,f)} \sqrt{v_{s_\text{r}}(n,f)} \sqrt{v_{z_\text{r}}(n,f)} \sqrt{v_{b_\text{r}}(n,f)} \Big]$ as the targets for the NN. Nugraha et al. defined the ground truth PSDs as $v_c(n,f) = \frac{1}{M} \rVert \mathbf{c}(n,f) \rVert^2$  \cite{nugraha}. We thus need to know the ground truth source signals $\mathbf{c}(n,f)$.
\par
The ground truth latent signals $\mathbf{s}_\text{r}(n,f)$, $\mathbf{z}_\text{r}(n,f)$ and $\mathbf{b}_\text{r}(n,f)$ are unkown. However, in the training and validations sets, we can know the ground truth early near-end signal $\mathbf{s}_\text{e}(n,f)$ and the signals $\mathbf{s}_\text{l}(n,f)$, $\mathbf{y}(n,f)$ and $\mathbf{b}(n,f)$ (see Section \ref{sec:dataset}). These last three signals correspond to the values of $\mathbf{s}_\text{r}(n,f)$, $\mathbf{z}_\text{r}(n,f)$ and $\mathbf{b}_\text{r}(n,f)$, respectively, when the linear filters $\mathbf{\underline{H}}(f)$ and $\mathbf{\underline{G}}(f)$ are equal to zero. To derive the ground truth latent signals $\mathbf{s}_\text{r}(n,f)$, $\mathbf{z}_\text{r}(n,f)$ and $\mathbf{b}_\text{r}(n,f)$, we propose to use an iterative procedure similar to the NN-supported BCA algorithm \textcolor{black}{(see Fig.\ \ref{fig:flowchart})}, where the linear filters $\mathbf{\underline{H}}(f)$ and $\mathbf{\underline{G}}(f)$ are initialized to zero.
\par
At each iteration, we derive the linear filters $\mathbf{\underline{H}}(f)$ and $\mathbf{\underline{G}}(f)$ \textcolor{black}{ as in steps 2 and 3 of Fig.\ \ref{fig:flowchart}, respectively}. We update $\mathbf{s}_\text{r}(n,f)$, $\mathbf{z}_\text{r}(n,f)$ and $\mathbf{b}_\text{r}(n,f)$ by applying the linear filters $\mathbf{\underline{H}}(f)$ and $\mathbf{\underline{G}}(f)$ to each of the signals $\mathbf{s}_\text{l}(n,f)$, $\mathbf{y}(n,f)$ and $\mathbf{b}(n,f)$ as in (\ref{eq:true_sr}), (\ref{eq:true_zr}) and (\ref{eq:true_br}). \textcolor{black}{To obtain the ground truth PSDs ${v_{c}(n,f)}$, we replace NN-EM at step 4 of Fig.\ \ref{fig:flowchart} by an \emph{oracle} estimation using Duong et al.'s EM algorithm \cite{duong_under_2010}.} For the detailed pseudo-code of the iterative procedure, please refer to the supporting document \cite[Sec. 4.1]{carbajal_supporting_2019}. After a few iterations, we observed the convergence of the latent variables $\mathbf{s}_\text{r}(n,f)$, $\mathbf{z}_\text{r}(n,f)$ and $\mathbf{b}_\text{r}(n,f)$. \textcolor{black}{In particular, we found that the \emph{dereverberated} residual echo $\mathbf{z}_\text{r}(n,f)$ decreases over the iterations.} Fig.\ \ref{fig:example_targets} shows an example of the PSD spectrograms after convergence.
\subsection{Inputs \label{sec:inputs}}
		%$\Big[\norm{\mathbf{\widehat{y}}(n,f)}, \norm{\mathbf{e}(n,f)}, \norm{\mathbf{\widehat{e}}_\text{l}(n,f)}, \norm{\mathbf{r}(n,f)}\Big]  \gets \text{avg}\Big(|\mathbf{d}(n,f)|\Big)$
%		
% TODO : at each iteration we consider all signals. at iteration 0, h and g are initialized to some value.
We use magnitude spectra as inputs for $\text{NN}_\text{0}$ and $\text{NN}_i$ rather than power spectra, since they have been shown to provide better results when the targets are the magnitude spectra $\sqrt{v_c(n,f)}$ \cite{nugraha}. We concatenate these spectra to obtain the inputs. The different inputs are summarized in Fig.\ \ref{fig:lstm}. We consider first the far-end signal magnitude $|x(n,f)|$ and a single-channel signal magnitude $|\widetilde{d}(n,f)|$ obtained from the corresponding multichannel mixture signal $\mathbf{d}(n,f)$ as \cite{nugraha_multichannel_2016}
\begin{equation}
|\widetilde{d}(n,f)| = \sqrt{\frac{1}{M} \rVert \mathbf{d}(n,f) \rVert^2}. \label{eq:input_type_I}
\end{equation}
Additionally we use the magnitude spectra $|\widetilde{y}(n,f)|$, $|\widetilde{e}(n,f)|$, $|\widetilde{e}_\text{l}(n,f)|$ and $|\widetilde{r}(n,f)|$ obtained from the corresponding multichannel signals after each linear filtering step $\mathbf{\widehat{y}}(n,f)$, $\mathbf{e}(n,f)$, $\mathbf{\widehat{e}}_\text{l}(n,f)$, $\mathbf{r}(n,f)$. Indeed in our previous work on single-channel echo reduction, using the estimated echo magnitude as an additional input was shown to improve the estimation \cite{carbajal_multiple-input_2018}. We refer to the above inputs as type-I inputs. We consider additional inputs to improve the estimation. In particular, we use the magnitude spectra $\sqrt{v_c^\text{unc}(n,f))}$ of the source unconstrained PSDs obtained as
\begin{equation}
v_c^\text{unc}(n,f) = \frac{1}{M} \text{tr} \left(\mathbf{{R}}_{c}(f)^{-1} \mathbf{\widehat{R}}_c(n,f)\right). \label{eq:input_type_II}
\end{equation}
Indeed these inputs partially contain the spatial information of the sources and have been shown to improve results in source separation \cite{nugraha}. We refer to the inputs obtained from (\ref{eq:input_type_II}) as type-II inputs. For $\text{NN}_0$, we only use type-I inputs, as type-II inputs are not available at initialization. For $\text{NN}_i$ with $i\geq 1$, we use both type-I and type-II inputs.

%\begin{table}[h]
%\centering
%\begin{TAB}(r,3cm,0.7cm)[3pt]{|l|c|c|}{|c|cc|}% (rows,min,max)[tabcolsep]{columns}{rows}
% Inputs &$\text{NN}_0$ & $\text{NN}_i$\\
%type I &$|\widetilde{d}|$, $|x|$ & $|\widetilde{d}|$, $|x|$, $|\widetilde{y}(n,f)|$, $|\widetilde{e}(n,f)|$, $|\widetilde{e}_\text{l}(n,f)|$, $|\widetilde{r}(n,f)|$   \\
%type II &	 & $\sqrt{v_c^\text{unc}(n,f)}$
%\end{TAB}
%\caption{NN inputs.\label{tab:NN_inputs}}
%\end{table}

\subsection{Cost function}
% TODO : redire qu'on predit les 4 a la fois
Let $|\widetilde{c}(n,f)|$ denote the NN output for source $\mathbf{c}(n,f)$. As mentioned above, we use $\text{NN}_0$ and $\text{NN}_i$ to jointly predict the $4$ spectral parameters $\Big[|\widetilde{s}_\text{e}(n,f)| |\widetilde{s}_\text{r}(n,f)| |\widetilde{z}_\text{r}(n,f)|  |\widetilde{b}_\text{r}(n,f)|\Big]$ (see Fig.\ \ref{fig:lstm}). We use the Kullback-Leibler divergence as the training loss, which has shown to provide the best results for NN training among several other losses \cite{nugraha}:
\begin{equation}
\begin{split}
\mathcal{D}_{KL} = \frac{1}{4 F N} \sum_{c,n,f} \Big(&\sqrt{v_c(n,f)} \log \frac{\sqrt{v_c(n,f)}}{|\widetilde{c}(n,f)|} \\
&- \sqrt{v_c(n,f)} + |\widetilde{c}(n,f)| \Big).
\end{split}
\label{eq:kullback}
\end{equation}

\subsection{Architecture}
% TODO : diagram of NN_i

\begin{figure}[t]
\centering
\includegraphics[width=.35\textwidth, valign=t]{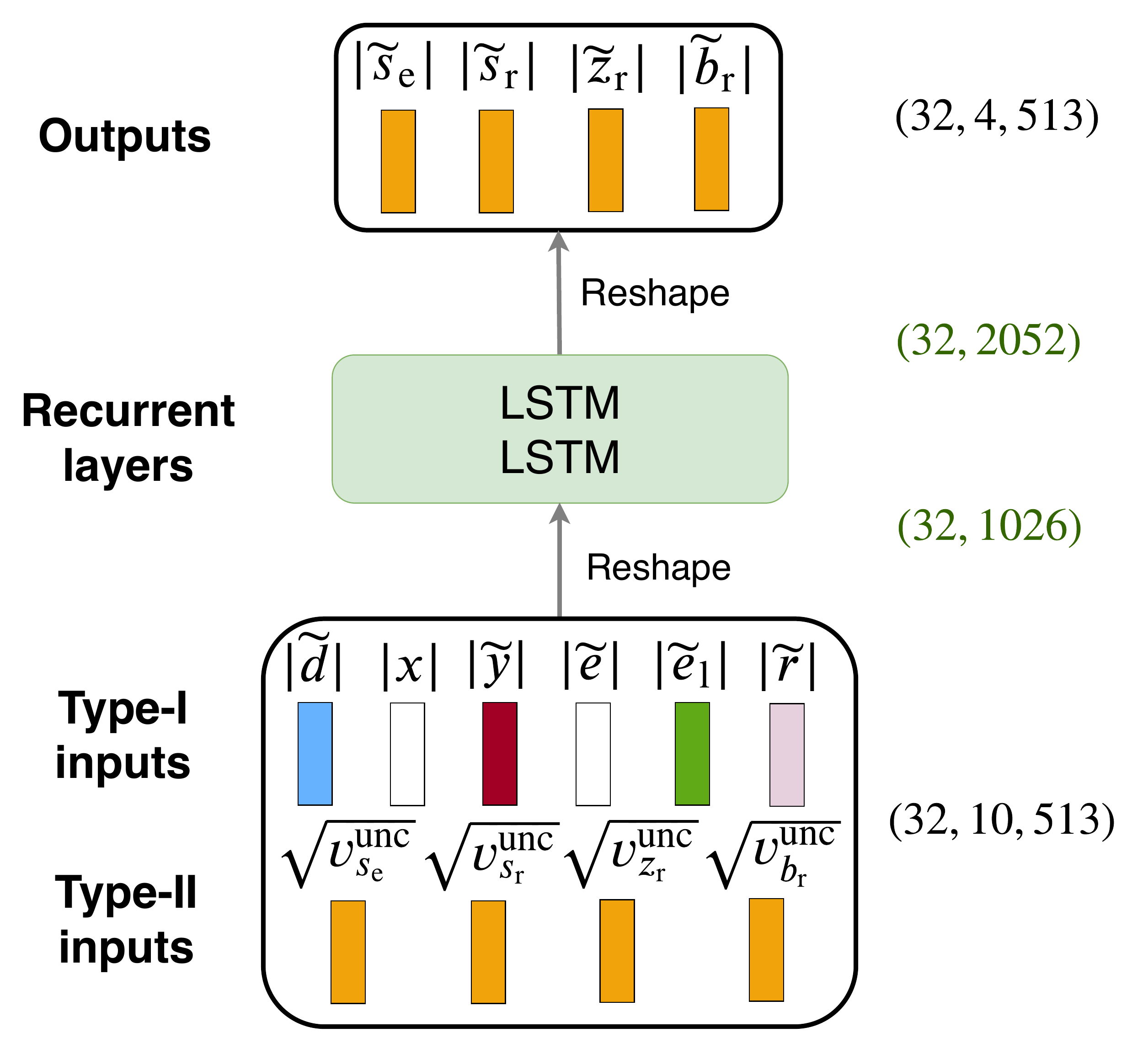}
\caption{\label{fig:lstm} Architecture of $\text{NN}_i$ with a sequence length of $32$ timesteps and $F = 513$ frequency bins.}
%\vspace{-.25cm}
\end{figure}

The neural network follows a long-short-term-memory (LSTM) network architecture. \textcolor{black}{We consider $2$ LSTM layers (see Fig.\ \ref{fig:lstm}).} The number of inputs is $6F$ for $\text{NN}_0$ and $10F$ for $\text{NN}_i$. \textcolor{black}{The number of outputs is $4F$.} Other network architectures are not considered here as the performance comparison between different architectures is beyond the scope of this article.

\section{Experimental protocol}

In this section we describe the datasets, the metrics, the baselines and the hyperparameter settings used to evaluate the proposed algorithm.
%the echo reduction, dereverberation and noise reduction setting parameters of the proposed joint approach are investigated using objective and perceptual metrics.
%In Section V-A the considered acoustic systems, algorithmic settings, and metrics are introduced.
%In Section V-B the influence of the linear filters on the performance of the proposed approach is investigated in far-field environment with different noise, echo and reverberation levels in stationary and nonstationary conditions.
%In Section V-C the influence of the NN inputs on the performance of the proposed techniques is also examined in those conditions.
%\vspace{-.2cm}
\subsection{Scenario}
\begin{figure}[t]
\centering
\includegraphics[width=.5\textwidth]{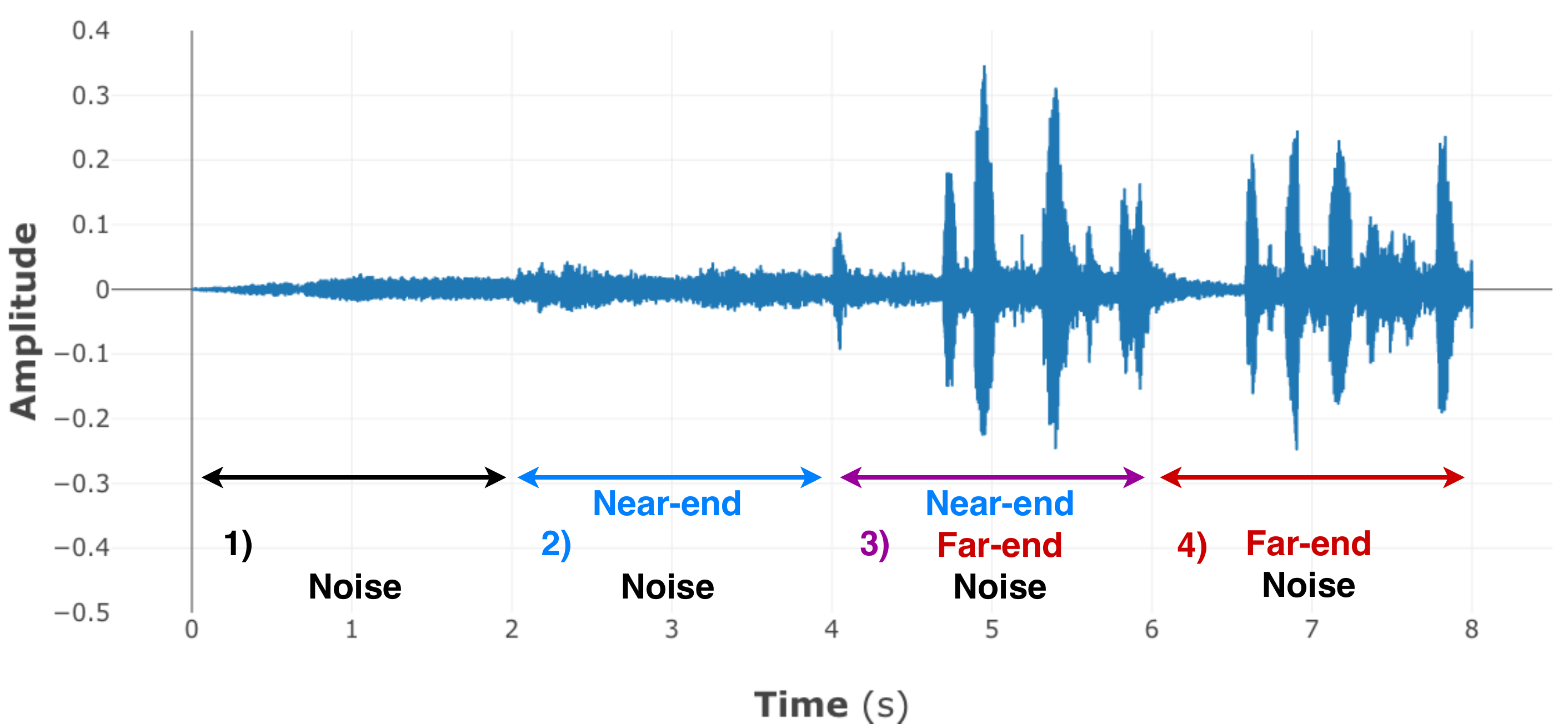}
\caption{\label{fig:utterance_train} Example utterance (only one channel shown).}
%\vspace{-.6cm}
\end{figure}

% TODO : scenario considered
% TODO : modify --> same scenario for test set
We consider the scenario where a near-end speaker interacts with a far-end speaker using a hands-free communication system at a distance of $1.5$~m in a noisy environment. Each utterance has $8$-s duration and contains $4$~s of near-end speech and $4$~s of far-end speech overlapping for $2$~s. Background noise is present during the whole utterance. Each utterance is hence composed of $4$ periods of $2$~s as shown in Fig.\ \ref{fig:utterance_train}: 1) noise only, 2) noise and near-end speech, 3) noise, near-end and far-end speech, 4) noise and far-end speech.
\subsection{Datasets \label{sec:dataset}}
\subsubsection{Overall description}
%The NNs for the source separation evaluation were trained on both the real and simulated training sets (tr05_real and tr05_simu) with the real and simulated development sets (dt05_real and dt05_simu) as validation data. Conversely, we trained the NNs for the speech recognition evaluation on the real training set only (tr05_real) and validated them on the real development set only (dt05_real). The same NNs were also used for the performance comparison to the general iterative EM algorithm. See [29] for the perfomance comparison between these two different settings.
%The training set is used to train the NN and the validation set is used as cross validation for the NN and to tune hyperparameters. The test set is used to evaluate our proposed approach.  The characteristics of these subsets are summarized in Table \ref{tab:train_val_test}.
%in time-invariant conditions
We created three disjoint datasets for training, validation and test, whose characteristics are summarized in Table \ref{tab:train_val_test}. We considered $M=3$ microphones. For each dataset, we separately recorded or simulated the acoustic echo $\mathbf{y}(t)$, the near-end speech $\mathbf{s}(t)$ and the noise $\mathbf{b}(t)$ using clean speech and noise signals as base material and we computed the mixture signal $\mathbf{d}(t)$ as in (\ref{eq:mixture}). This protocol is required to obtain the ground truth target and residual signals for training and evaluation, which is not possible with real-world recordings for which these ground truth signals are unknown. The training and validation sets correspond to time-invariant acoustic conditions, while the test set includes both a time-invariant and a time-varying subset. The recording and simulation parameters \textcolor{black}{(e.g. simulated room charateristics, position of the sources)} are detailed in our companion technical report\cite[Sec. 7.1]{carbajal_supporting_2019}. 

\begin{table}[t]
\center
\begin{tabular}{|lc|c|c|c|}
\hline
\multicolumn{2}{|l|}{\textbf{Dataset}} &\textbf{Training} &\textbf{Validation} & \textbf{Test} \\
\hline
\multirow{3}{*}{Signals} & $\mathbf{y}$ & \multicolumn{2}{c|}{recorded} & \multirow{3}{*}{recorded} \\
									&	$\mathbf{s}$ &  \multicolumn{2}{c|}{simulated $\mathbf{a}_s$} & \\
									&	$\mathbf{b}$ & \multicolumn{2}{c|}{simulated} & \\
\hline
\multicolumn{2}{|l|}{Rooms} & $1$-$2$-$3$ & $1$-$2$ & $4$ \\
\hline
\multicolumn{2}{|l|}{\# speaker pairs} & $79$ & $27$ & $25$ \\
\hline
\multicolumn{2}{|l|}{\# utterances} & \num{13572} & \num{4536} & \num{4500}  \\
\hline
\multicolumn{2}{|l|}{\# noise samples} & $36$ & $36$ & $6$ \\
\hline
\multicolumn{2}{|l|}{SER range (dB)} & \multicolumn{2}{c|}{$[-45, +6]$}  & $[-45, -7]$ \\
\hline
\multicolumn{2}{|l|}{SNR range (dB)} &\multicolumn{2}{c|}{$[-21, +24]$}  & $[-20, +13]$ \\
\hline
\end{tabular}
\caption{Dataset characteristics. \label{tab:train_val_test}}
%\vspace{-.27cm}
\end{table}

% TODO : creation of the signals
% TODO : creation of y
% TODO : The separate recordings were used to obtain the ground truths of the targets $v_c(n,f)$. 
% TODO : phrase Romain pas lisible

\paragraph{Clean speech and noise signals}
Clean speech signals were taken from the train-clean-360 subset of the Librispeech corpus \cite{panayotov_librispeech:_2015}, which consists of $921$ speakers reading books for $25$~min each on average. We selected $262$ speakers and grouped them into $131$ disjoint pairs for training, validation, and test. We alternately considered each speaker as near-end or far-end and picked several non-overlapping \mbox{$4$-s} speech samples for each pair. Each $4$-s sample was used only once in the whole dataset. Regarding the noise signals, we considered $6$ types of domestic noise: babble, dishwasher, fridge, microwave, vacuum cleaner and washing machine. We randomly selected $78$ non-overlapping $8$-s noise samples from $1.7$~h of YouTube videos and grouped them into disjoint subsets for training, validation, and test.
\begin{figure}[t]
\centering
\includegraphics[width=.4\textwidth]{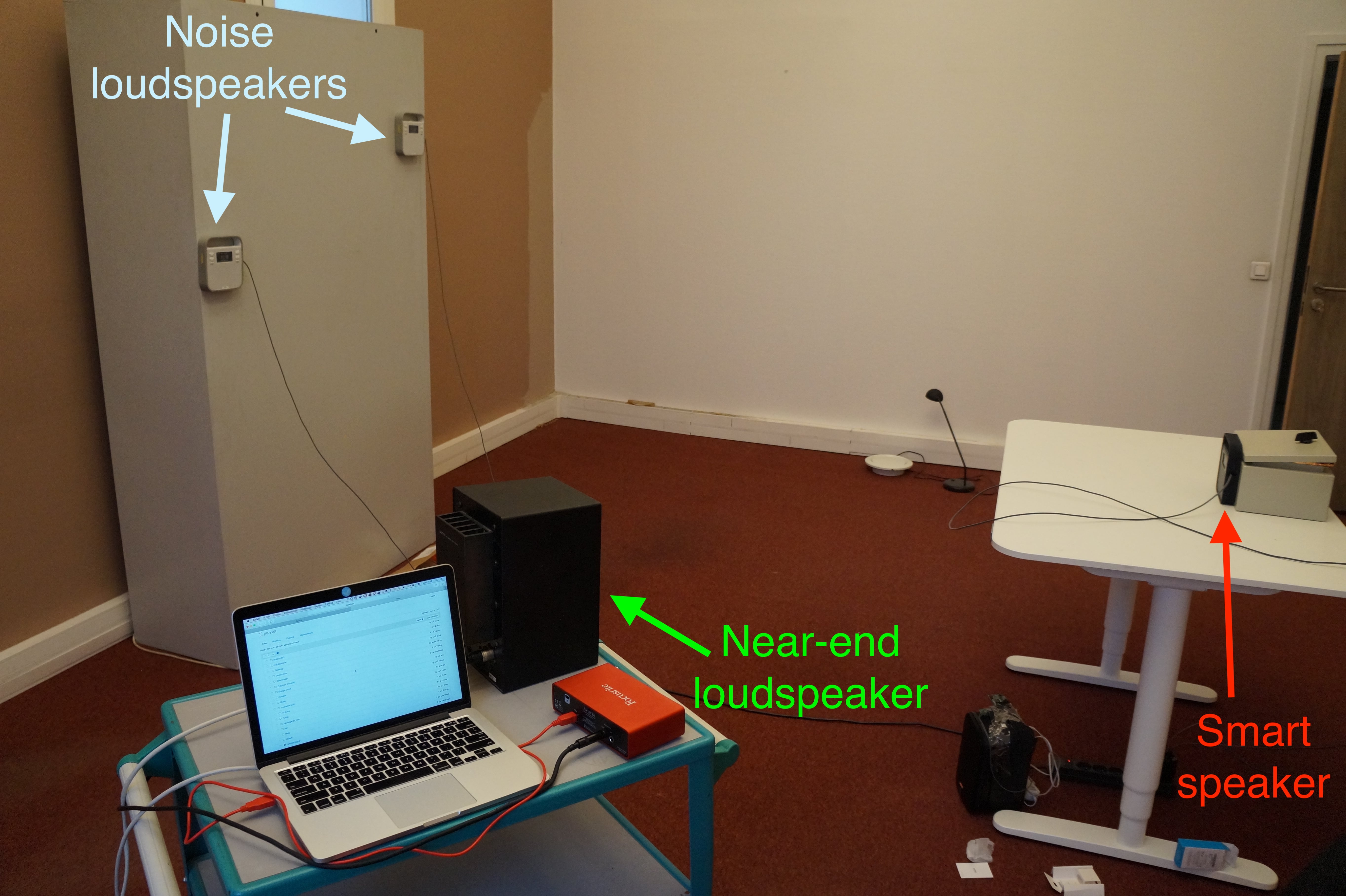}
\caption{Recording setup for the test set. \label{fig:test_setup}}
%\vspace{-.25cm}
\end{figure}
\paragraph{Real echo recordings}
% The distance between the loudspeaker (playing the far-end signal) and the microphones was $11$~cm, and the distance between the microphones was $3$~cm. One of the microphones exhibited some recording problem due to occlusion and was discarded. We only used the signals of $M=3$ microphones. 
%The smart speaker was placed on a table in the center of each room at $2$ differents positions to increase the diversity of the echo paths. 
%The far-end speech was played by the loudspeaker at $3$ different loudness levels to increase the diversity of nonlinearities: the louder the far-end speech, the larger the nonlinearities.
To create the acoustic echo $\mathbf{y}(t)$, Togami et al. convolved far-end speech signals $x(t)$ with simulated echo paths $\mathbf{a}_y(\tau)$ which do not include any nonlinearity \cite{togami_simultaneous_2014}. In real hands-free systems, the acoustic echo contains nonlinearities caused by the nonlinear response of the loudspeaker, enclosure vibrations and hard clipping effects due to amplification (see Section \ref{sec:echo_reduc}). In order to achieve more realistic test conditions, we created the acoustic echo by recording the acoustic feedback from the loudspeaker to the microphones of a real hands-free system. The far-end speech was played and recorded at a rate of $16$~kHz with a Triby, a smart speaker device developed by Invoxia. A configuration of the echo recording setup is given in Fig. \ref{fig:test_setup}. The recordings were done with the same Triby in 4 rooms with different size and reverberation time ($\text{RT}_{60}$) listed in Table \ref{tab:room}. 
%For further details regarding the echo recording setup, please refer to the supporting document \cite{carbajal_supporting_2019}. 

% TODO : rooms 1,2,3,4
% TODO : volume (challenging conditions)

% TODO : finish completing
\begin{table}[t]
\center
\begin{tabular}{c c c}
\toprule
\textbf{Room} &\textbf{Size} (m) & $\textbf{RT}_{60}$ (s) \\
\midrule
$1$ & $4.4 \times 4.2 \times 4$  & $1.0$ \\
$2$ & $3.8 \times 2.5 \times 3.5$ & $0.5$ \\
$3$ & $3.4 \times 2.1 \times 3.3$ & $0.8$ \\
$4$ & $5.9 \times 4.6 \times 4$ &  $1.3$ \\
\bottomrule
\end{tabular}
\caption{Room characteristics. \label{tab:room}}
%\vspace{-.17cm}
\end{table}

\paragraph{Reverberant near-end speech and noise}
The creation procedures for $\mathbf{s}(t)$ and $\mathbf{b}(t)$ differ for each dataset and are described in the following subsections.

\subsubsection{Training set}
% TODO : creation of s with RT60
% TODO : RIR noise
% The simulated rooms had the same reverberation characteristics ($\text{RT}_{60}$ per octave) as the real rooms where the echo was recorded. These characteristics are provided in the supporting document \cite{carbajal_supporting_2019}. However the simulated room size was chosen randomly within $\SI{\pm 20}{\percent}$ of the real room size. The dimensions of the simulated microphone array were identical to those of the real Triby, and its simulated position and orientation were similar. The position of the near-end speaker was chosen randomly on a semicircle of $1.5$~m radius centered in the middle of the microphone array and at least $10$~cm from the walls (see Fig.\ \ref{fig:nearend_simu}). 
%For each RIR, a random room and a random near-end position were generated to increase the diversity of RIRs. 
For the training set, the echo recordings were done in rooms $1$, $2$ and $3$ (see Table \ref{tab:room}). To create the reverberant near-end speech $\mathbf{s}(t)$, we convolved anechoic near-end speech $u(t)$ with near-end RIRs $\mathbf{a}_s(\tau)$ simulated to match the echo recording properties using the Roomsimove toolbox \cite{roomsimove} \cite[Sec. 7.1]{carbajal_supporting_2019}. Among the $79$ pairs of speakers used for training, $54$ were used in rooms $1$ and $2$. We played and recorded \num{4536} far-end signals and we simulated \num{4536} near-end RIRs in each of these $2$ rooms. The remaining $25$ pairs were used in room $3$. We played and recorded \num{4500} far-end signals and we simulated \num{4500} near-end RIRs in this room.
\par
To create the noise signal $\mathbf{b}(t)$, we convolved a randomly chosen noise sample among the $36$ noise samples ($6$ per noise type) used for training \textcolor{black}{with the average of the late tail of two distinct RIRs randomly picked among $42$ measured RIRs. This procedure approximates a spatially diffuse noise signal. To obtain the $42$ measured RIRs, we measured $14$ RIRs in each of the rooms $1$, $2$ and $3$.}
\par
The levels of the recorded far-end, the near-end speech and the noise signal were chosen randomly such that the signal-to-echo ratio (SER) varied from $-45$~dB to $+6$~dB and the signal-to-noise ratio (SNR) varied from $-21$~dB to $+24$~dB. These conditions are very challenging, especially as reverberation dominates in the reverberant near-end speech $\mathbf{s}(t)$. In total, we obtained \num{13572} utterances which amount to roughly $32$~h of audio.

%For further details regarding the simulation setup, please refer to the supporting document \cite{carbajal_supporting_2019}. 
%We measure using the swept-sine technique at a sampling frequency $f_s = 16$~kHz  with Triby.
% TODO : volume and challenging conditions
% TODO : compute residual levels of train / validation set
% TODO : rooms -->1,2,3
% ranged between $8$~dB and $34$~dB.
%levels were chosen randomly between $-11$~dB and $+15$~dB.
%between $-9$~dB and $+10$dB. As a result, 
% TODO : RT60 per octave bands
% TODO : corpus
% TODO : noise + RIR

\iffalse
\subfloat[Near-end simulation settings. \label{fig:nearend_simu}]{
\includegraphics[width=.3\textwidth]{diagrams/nearend_simu.pdf}
}%
\\
\fi

\subsubsection{Validation set}
% TODO : volume --> same as training set
 % TODO : only time-invariant procedure
 % TODO : volume
The validation set was generated in a similar way as the training set, using $27$ speaker pairs and $36$ noise samples that are not in the training set. The echo recordings were done in rooms $1$ and $2$, and the near-end RIRs were simulated similarly to the training set procedure. We played and recorded \num{4536} far-end signals and we simulated \num{4536} near-end RIRs in each room. To create the diffuse noise, we used the same $42$ measured RIRs as in the training set. The levels of the recorded far-end speech, the near-end speech and the noise signal were chosen in the same range as the training set, resulting in the same challenging SER and SNR conditions. In total, we obtained \num{4536} utterances which amount to roughly $10$~h of audio. 

\subsubsection{Time-invariant test set}
% TODO : scenario time-invariant + time-varying
%using (\ref{eq:nearend_path})--(\ref{eq:late_path}).
%We thus did not consider periods with only noise signal in the utterances. Each utterance has a $6$~s duration and is divided into $3$ periods of $2$~s: 1) noise signal and near-end speech, 2) noise signal, near-end speech and echo signal, 3) noise and echo signals.
% TODO : creation of signals
% TODO : expliciter la presence des 3 loudspeakers en meme temps (important ou pas?)
%The acoustic echo was played and recorded with the same smart speaker as in the training set. 
%For the time-invariant scenario, we recorded the mixture $\mathbf{d}(n,f)$, and separately the acoustic echo $\mathbf{y}(n,f)$, the near-end speech $\mathbf{s}(n,f)$ and the noise signal $\mathbf{b}(n,f)$. 
% For further details regarding the test set recording setup, please refer to the supporting document \cite{carbajal_supporting_2019}. 
The time-invariant test set was built from real recordings only, using $25$ speaker pairs and $6$ noise samples that are neither in the training nor in the validation sets. The echo, the near-end speech and the noise were all recorded in room $4$ (see Table \ref{tab:room}) using the setup shown in Fig.\ \ref{fig:test_setup}. The reverberant near-end speech $\mathbf{s}(t)$ was obtained by playing anechoic speech with a Yamaha MSP5 Studio loudspeaker at a single loudness level. The noise signal $\mathbf{b}(t)$ was obtained by picking a random original noise signal and playing it through $4$ Triby loudspeakers simutaneously. The noise signals resulting from this procedure are less diffuse than in the training and validation sets. The recorded levels were such that the resulting SER varied from $-45$~dB to $-7$~dB and the SNR varied from $-20$~dB to $+13$~dB. These challenging conditions are comprised within those of the training and validation sets. We played and recorded \num{4500} far-end speech, near-end speech, and noise signals, hence we obtained a total of \num{4500} $8$-s utterances amounting to $10$~h of audio.

% TODO : give SER and SNR input
%ranged between $+9$~dB and $+34$~dB. 
% ranged recorded between $-11$~dB and $+15$~dB. 
%ranged between $-9$~dB and $+10$dB. As a result,

\subsubsection{Time-varying test set}
In order to evaluate our approach in time-varying acoustic conditions, we also considered the scenario when the near-end speaker speaks for $4$~s, moves to a different position, and speaks for $4$~s again. To do so, we concatenated pairs of $8$-s near-end and echo recordings from the time-invariant test set corresponding to the same near-end and far-end speakers and microphone array positions, but to two different positions of the loudspeaker playing the near-end speech. The two recordings summed with an $16$~s recorded noise signal. This resulted in \num{2250} $16$-s utterances or roughly $10$~h of audio.
%This realistic protocol makes it possible to estimate accurately the ground truth target and residual signals $\mathbf{s}_\text{e}(n,f)$, $\mathbf{s}_\text{r}(n,f)$,  $\mathbf{z}_\text{r}(n,f)$ and $\mathbf{b}_\text{r}(n,f)$ at each position.
%We only evaluate the performance of our approach when the near-end and/or the far-end speakers are active. For the time-invariant conditions, we evaluate the system at convergence. Therefore we consider the same type of scenario as in the training and validation sets, since the first $2$~s noise-only period allows the system to converge with respect to the noise.
% TODO : pour scenario time-invariant
% TODO : explain the near-end position 2 were used with same speaker but different positions
%\vspace{-.25cm}
\subsection{Evaluation metrics}
% TODO : citer emmanuel 2007 first stereo... (average by channel)
% TODO : parler du ERR (citer naylor_speech_2010) (modifier DRR en ERR) + citer yoshioka blind 2011
% TODO : citer Le Roux
% TODO : cite segmental SNR for all metrics --> donner la formule ?

\subsubsection{\color{black}{ Early near-end components}}

\begin{table}[t]
\centering
\color{black}
{\tabulinesep=1.3mm
\setlength{\tabcolsep}{4pt}
\begin{tabu}{|c|c|l|}
  \hline
\textbf{Overall}  & \multirow{2}{*}{\textbf{SI-SDR}} & \multirow{2}{*}{$10 \log_{10} \displaystyle \frac{\rVert s_\text{e}^\text{post} \rVert^2}{\rVert {s}_\text{l}^\text{post} + {y}^\text{post} + {b}^\text{post} + s_\text{e}^\text{art} \rVert^2}$} \\
  \textbf{distortion} & & \\
  \hhline{|=|=|=|}
\multirow{2}{*}{\textcolor{RED}{Echo}}  &\textcolor{RED}{ERLE} & $10 \log_{10} \displaystyle \frac{\rVert y \rVert^2}{\rVert y^\text{post} \rVert^2}$ \\
  \cline{2-3}
  &  \textcolor{RED}{SER} & $10 \log_{10} \displaystyle \frac{\rVert s_\text{e}^\text{post} \rVert^2}{\rVert y^\text{post} \rVert^2}$ \\
    \hline
  \textcolor{GREEN}{Rever-}& \multirow{2}{*}{\textcolor{GREEN}{ELR}} & \multirow{2}{*}{$10 \log_{10} \displaystyle \frac{\rVert s_\text{e}^\text{post} \rVert^2}{\rVert s_\text{l}^\text{post} \rVert^2}$}  \\
  \textcolor{GREEN}{beration}  & & \\
    \hline
  \textcolor{ORANGE}{Noise}  & \textcolor{ORANGE}{SNR} & $10 \log_{10} \displaystyle \frac{\rVert s_\text{e}^\text{post} \rVert^2}{\rVert b^\text{post} \rVert^2}$ \\
    \hline
Artifacts & SI-SAR & $10 \log_{10} \displaystyle \frac{\rVert s_\text{e}^\text{post} \rVert^2}{\rVert  s_\text{e}^\text{art} \rVert^2}$  \\

  \hline
\end{tabu}}
\caption{\label{tab:metrics}Evaluation metrics. The formulas are given in the single-channel case ($M=1$) and the channel index $m$ is omitted for conciseness.}
\end{table}
\textcolor{black}{ 
The estimated early near-end signal $\mathbf{\widehat{s}}_\text{e}(t)$ has $5$ components
\begin{equation}
\mathbf{\widehat{s}}_\text{e}(t) = \mathbf{{s}}_\text{e}^\text{post}(t) + \mathbf{{s}}_\text{l}^\text{post}(t) + \mathbf{{y}}^\text{post}(t) + \mathbf{{b}}^\text{post}(t) + \mathbf{{s}}_\text{e}^\text{art}(t),  \label{eq:post_res}
\end{equation}
where $\mathbf{s}_\text{e}^\text{post}(t)$ is the potentially attenuated early near-end signal,  $\mathbf{s}_\text{l}^\text{post}(t)$, $\mathbf{{y}}^\text{post}(t)$ and $\mathbf{{b}}^\text{post}(t)$ are the post-residual distortion sources that are ideally equal to zero vectors, and $\mathbf{{s}}_\text{e}^\text{art}(t)$ denotes the artifacts introduced in the early near-end signal $\mathbf{s}_\text{e}(t)$.  This definition of the $5$ components of the estimated target $\mathbf{\widehat{s}}_\text{e}(t)$ is an extension of Le Roux et al.'s component definition in noise reduction to multiple distorsion sources \cite{roux_sdr_2019}. For a detailed derivation of the components, please refer to the supporting document \cite[Sec. 7.2]{carbajal_supporting_2019}}

\subsubsection{\color{black}Definition of the metrics}

%From the $5$ components in \eqref{eq:post_res}, we can define $6$ scale-invariant metrics to assess the overall reduction and the reduction of each distortion source. 
The objective metrics are summarized in Table \ref{tab:metrics} in the single channel case ($M=1$). In the multichannel case ($M>1$), we compute each metric on each channel $m$ separately and we average the results over the $M$ channels.
\par
%In the following, for the sake of conciseness, the formula for each metric is given in the single-channel case and the channel index is omitted.
% TODO modifier le calcul des composantes
%The metrics depend on the 4 components of $\mathbf{\widehat{s}}_\text{e}(n,f)$ in (\ref{eq:post_res}). The signals $\mathbf{b}^\text{post}_\text{r}(n,f)$ and $\mathbf{z}^\text{post}_\text{r}(n,f)$ are computed by applying the filters $\mathbf{\underline{H}}(f)$, $\mathbf{\underline{G}}(f)$ and $\mathbf{W}_{s_\text{e}}(n,f)$ to $\mathbf{b}(n,f)$ and $\mathbf{y}(n,f)$, respectively. The components $\mathbf{s}^\text{post}_\text{e}(n,f)$ and $\mathbf{s}^\text{post}_\text{r}(n,f)$ are comprised in $\mathbf{\widehat{s}}_{\text{e}, s}(n,f)$, which is the resulting signal of the filters applied to $\mathbf{s}(n,f)$. The $2$ components are computed using Yoshioka et al.'s method \cite{yoshioka_blind_2011, yoshioka_generalization_2012}. It consists in performing the same procedure as (\ref{eq:nearend_path})--(\ref{eq:late_path}) to signal $\mathbf{\widehat{s}}_{\text{e}, s}$ instead of $\mathbf{s}(n,f)$.
\textcolor{black}{We evaluate the proposed joint approach in terms of the overall distortion, which is measured with the scale-invariant signal-to-distortion ratio (SI-SDR) \cite{roux_sdr_2019}. The overall distortion takes the three distortion sources and the artifacts into account. To analyze the distribution of the overall distortion over the distortion sources and the artifacts, we use $5$ additional metrics.} For echo reduction, we use the SER and the echo return loss enhancement (ERLE) \cite{hansler_acoustic_2004}. %After echo cancellation, it is defined as
%\begin{equation}
%\text{ERLE} = 10 \log_{10} \frac{\norm{{y}}^2}{\norm{{{z}}}^2}.
%\end{equation}
Dereverberation is assessed by the early-to-late reverberation ratio (ELR) \cite{naylor_speech_2010}. We use this metric instead of the direct-to-reverberant ratio (DRR) \cite{naylor_speech_2010} since early reflections are part of the target signal to be estimated. % (see Section \ref{sec:dereverb}). %It is defined as the ratio between the early near-end component and the near-end late reverberation:
%\begin{equation}
%\text{ELR} = 10 \log_{10} \frac{\norm{{s}_\text{e}}^2}{\norm{{{s}}_\text{l}}^2}.
%\end{equation}
%supprimé cette formule qui diffère de celle dans le tableau
For noise reduction, we use the SNR. The artifacts are measured with the scale-invariant signal-to-artifacts ratio (SI-SAR) \cite{roux_sdr_2019}. 
\subsubsection{\color{black}Evaluation period}
%Note that the input ERR is defined as
%\begin{equation}
%\text{ERR}_\text{i} = 10 \log_{10} \frac{\norm{s_\text{e}}^2}{\norm{s_\text{l}}^2}.
%\end{equation}
% and applied it to each channel of $\mathbf{s}_\text{e}$ as 
%\begin{equation}
%\widetilde{s}_\text{e} (m, t) = \left.\frac{\langle \widehat{s}_\text{e}(m), s_\text{e}(m) \rangle}{\norm{s_\text{e}(m)}} \right\vert_{\textit{time period}} s_\text{e}(m,t)
%\end{equation}
%In case filters provide attenuation of signals, we use processing of Vincent el al. \cite{vincent_performance_2006}. We compute metrics on each channel and average the results over the channels. In the training and validation set, we have access to all information accurately as we produced all signals separately and we have access to the true near-end RIR.
\textcolor{black}{The SI-SDR, ELR, SNR and SI-SAR are evaluated during both \textit{single-talk} (near-end speech only) and \textit{double-talk} (simultaneous near-end and far-end speech).} The SER is only evaluated during \textit{double-talk}, while the ERLE is evaluated during both \textit{double-talk} and \textit{far-end talk} (far-end speech only).
\par
Since the performance may vary depending on the presence of acoustic echo which is the loudest signal, we compute the metrics separately for \textit{near-end talk}, \textit{double-talk} and \textit{far-end talk}. \textcolor{black}{In particular, each metric depends on a scaling factor $\gamma_c$ \cite{carbajal_supporting_2019}. We assume that $\gamma_c$ is constant during each \textit{near-end talk}, \textit{double-talk} or \textit{far-end talk} period. However, $\gamma_c$ may vary from one period to another.} Finally, we average each metric over all periods in the fashion of the segmental SNR \cite{loizou_speech_2007}. 

\subsubsection{\color{black}Ground truth signals}

\textcolor{black}{All the above metrics are based on the ground truth signals $\mathbf{s}_\text{e}(t)$, $\mathbf{s}_\text{l}(t)$, $\mathbf{y}(t)$ et $\mathbf{b}(t)$ (see Table \ref{tab:metrics}).} The dataset generation procedure readily provides ground truth signals for the echo $\mathbf{y}(t)$ and the noise $\mathbf{b}(t)$. To define the ground truth signals of the target $\mathbf{s}_\text{e}(t)$ and late reverberation $\mathbf{s}_\text{l}(t)$, we set the mixing time as $t_\text{e} = 64$~ms. We computed these two components using (\ref{eq:se_sl}) \textcolor{black}{which requires the ground truth near-end RIR $\mathbf{a}_s(\tau)$. In the test set, since the ground truth near-end RIR $\mathbf{a}_s(\tau)$ is unknown, we derive it using the evaluation procedure proposed by Yoshioka et al. for ELR\footnote{The authors denoted this metric by DRR instead of ELR.} when $\mathbf{a}_s(\tau)$ is unknown \cite[Sec. VII.A]{yoshioka_blind_2011} \cite[Sec. VI.A]{yoshioka_generalization_2012}. This evaluation procedure determines the ground truth $\mathbf{a}_s(\tau)$ by performing MMSE optimization between the reverberant near-end speech $\mathbf{s}(t)$ (output signal) and the anechoic near-end speech $u(t)$ (input signal).}
\subsection{Baselines}
%In contrast to the proposed algorithm, the EM algorithm can di- rectly make use of spatial information contained in the multi-channel observations. A further difference is that the algorithm does not re- quire a training phase since all necessary statistics are captured dur- ing runtime. Its main drawback is obviously the fact, that, due to the independent treatment of each frequency, it does not make use of the typical spectral structure of speech. (CITE HEYMANN ??)

%As a second comparison, we consider a variant of our proposed NN-supported BCA algorithm, where the echo cancellation and dereverberation filters are both applied on the mixture signal $\mathbf{d}(n,f)$ similarly to Togami et al.\cite{togami_simultaneous_2014}. 
% TODO : cite Table
% TODO : Therefore, we will use SpeexDSP\footnote{https://github.com/xiph/speexdsp}, an implementation of this approach, 
% TODO : To the best of our knowledge, the assumption $\mathbf{R}_{s_\text{e}}(f) \neq \mathbf{I}_M$ was not tested in the literature. Therefore, we will use our implementation of this approach 
\textcolor{black}{Hereafter we denote our joint NN-supported approach as \textit{NN-joint}. We compare it with four baselines:
\begin{enumerate}
\item \textit{Togami}: our implementation of Togami et al.'s approach \cite{togami_simultaneous_2014},
\item \textit{Cascade}: a cascade approach where the echo cancellation filter $\mathbf{\underline{H}}(f)$, the dereverberation filter $\mathbf{\underline{G}}(f)$ and the Wiener postfilter $\mathbf{W}_{s_\text{e}}(n,f)$ are estimated and applied one after another. Echo cancellation relies on SpeexDSP\footnote{https://github.com/xiph/speexdsp}, which implements Valin's adaptive approach and is particularly suitable for time-varying conditions \cite{valin_adjusting_2007} (see Section \ref{sec:echo_reduc}). Dereverberation relies on our implementation of WPE \cite{nakatani_speech_2010, yoshioka_generalization_2012} (see Section \ref{sec:dereverb}). The multichannel Wiener postfilter is computed using our implementation of Nugraha et al.'s NN-EM approach \cite{nugraha} (see Section \ref{sec:noise}).
\item \textit{NN-parallel}: a variant of \emph{NN-joint} where the echo cancellation filter $\mathbf{\underline{H}}(f)$ and the dereverberation filter $\mathbf{\underline{G}}(f)$ are applied in parallel as Togami et al.'s approach (see Fig.\ \ref{fig:togami}),
\item \textit{NN-cascade}: a variant of \textit{Cascade} where the echo cancellation filter $\mathbf{\underline{H}}(f)$ is estimated using an NN-supported approach similar to \emph{NN-joint} instead of Valin's adaptive approach. As WPE dereverberates similarly to its NN-supported counterpart in the multichannel case \cite{kinoshita_neural_2017}, \emph{NN-cascade} corresponds to a cascade variant of \emph{NN-joint} which estimates each filter separately using NN-supported optimization algorithms.
% TODO : Variant with only NN-supported echo cancellation is described in supporting document
% TODO : corresponds basically to a variant of our NN-supported approach where the filters are estimated separetly. However no NN to estimate reverberation filter...
% TODO : explain why dereverberation not with NN.
\end{enumerate}
For the detailed description of the model and optimization algorithm of \textit{NN-parallel} and \textit{NN-cascade}, please refer to the supporting document \cite[Sec. 5 and 6]{carbajal_supporting_2019}.}

% TODO : parler de Valin et SpeexDSP
%In our experiments, we found that Valin's approach for estimating $\mathbf{\underline{H}}(f)$ \cite{valin_adjusting_2007} greatly reduces the acoustic echo $\mathbf{y}(n,f)$ in various situations. Therefore, we will use SpeexDSP\footnote{https://github.com/xiph/speexdsp}, an implementation of this approach, for the linear echo reduction filter as a part of the combination of individual approaches to which we compare our approach for the joint reduction of echo, reverberation and noise.

\subsection{Hyperparameter settings \label{sec:hyperparameters}}
The hyperparameters of the three approaches are set as follows.
% TODO : matching errors
%la définition de la target ne fait pas partie de l'algo mais des datasets
\subsubsection{Initialization of the linear filters}
% TODO : we apply SpeexDSP on each channel 
For echo cancellation, we compute $\mathbf{\underline{H}}_0(f)$ by applying SpeexDSP on each channel of $\mathbf{d}(n,f)$. Since SpeexDSP relies on half-overlapping rectangular STFT windows, we use a window of length \num{512} and hopsize \num{256}. We set the filter length to $0.208$~s in the time domain, that is $K=13$ frames. As SpeexDSP is an online algorithm, we apply it twice to each utterance to ensure convergence. For dereverberation, we compute $\mathbf{\underline{G}}_0(f)$ by performing $3$ iterations of WPE on the signal $\mathbf{e}(t)$ output by SpeexDSP. We use the STFT with a Hanning window of length \num{1024} and hopsize \num{256}. We set the filter length to $0.208$~s in the time domain, that is $L=10$ frames, and the delay to $\Delta=3$ frames.

\subsubsection{\color{black}Hyperparameters of the NNs}
We consider $1026$ units for the hidden layer of the LSTM structure. \textcolor{black}{Regarding the activation functions, we use rectified linear units (ReLU) for the cell state of the layers and sigmoids for the gates.} NN training is done by backpropagation with a minibatch size of $16$ sequences, a fixed sequence length of $32$ frames and the Adam parameter update algorithm with default settings \cite{kingma_adam:_2014}. To avoid gradient explosion with long sequences, we use gradient clipping with a threshold of $1.0$. Training is stopped when the loss on the validation set stops decreasing for $5$ epochs. 

%\subsubsection{Hyperparameters of the NN-supported BCA algorithm}
\subsubsection{Hyperparameters of NN-joint}
The STFT coefficients are computed with a Hanning window of length \num{1024} and hopsize \num{256} resulting in $F = 513$ frequency bins. The length of the echo cancellation filter $\mathbf{\underline{H}}(f)$ ($0.208$~s in the time domain) now corresponds to $K=10$ frames. The hyperparameters of the dereverberation filter $\mathbf{\underline{G}}(f)$ are identical to those of WPE. At training time, we perform $3$ iterations of the iterative procedure to derive the ground truth PSDs (see Section \ref{sec:targets}) \cite{carbajal_supporting_2019}. At test time, we perfom $I=3$ iterations of the proposed NN-supported BCA algorithm with $1$ spatial and $1$ spectral update for each iteration $i$ (see Fig.\ \ref{fig:flowchart}).
%\subsubsection{Hyperparamètres de l'algorithme NN-BCA \label{sec:hyperparam_dnn-bca_ch4}}
% TODO : dire qu'on entraine 3 NNs (nous faisons 2 itérations sur le test, ce qui correspond à 3 NNs entrainés durant l'apprentissage)
%Pour l'algorithme NN-BCA, la TFCT est calculée avec une fenêtre de Hanning de longueur $T_\text{TFCT} =$ \num{1024} échantillons et un pas d'avancement $P=256$ échantillons, pour obtenir $F=513$ bandes de fréquence. La longueur du filtre d'annulation d'écho $\mathcal{H}^{(i)}(f)$ ($0,208$~s dans le domaine temporel) correspond ici à $K=10$ trames. Les hyperparamètres du filtre de déréverbération $\mathcal{G}^{(i)}(f)$ sont identiques à ceux de $\mathcal{G}^{(0)}(f)$. Pour déterminer les cibles des NNs durant l'apprentissage, nous réalisons $3$ itérations de la procédure de détermination des vérités terrain des DSPs (voir la partie \ref{sec:targets}).

%\subsubsection{Hyperparameters of Togami et al.'s joint approach \cite{togami_simultaneous_2014} \label{sec:togami_hpp}}
\subsubsection{Hyperparameters of Togami \label{sec:togami_hpp}}
%For the variant of our approach, we take the same parameters of the whole procedure. 
% TODO : configuration of baseline --> cascade approach with different STFT and hop but same filter length.
% TODO : number of iterations of Togami
% TODO : inputs NN nugraha
% TODO : finish cascade approach --> if WPE not the same with Delta, might remove the early component ?
% TODO : initialisation of Togami
% TODO : at iteration 0, everything is initialized with Speex and WPE.
% TODO : Togami --> we initialize signal s and b with NN with inputs = ....
%For each algorithm of the baseline, we consider the same filter lengths as our approach. 
%Togami et al.'s approach requires initial estimates of the PSDs of  
%while the postfilter $\mathbf{W}_{s_\text{e}}(n,f)$ is initialized with initial 
\emph{Togami} requires the initial values for the linear filters $\mathbf{\underline{H}}(f)$ and $\mathbf{\underline{G}}(f)$, and for the PSDs of the reverberant near-end speech $v_{s}(n,f) = \frac{1}{M} \rVert \mathbf{s}(n,f) \rVert^2 $ and the noise signal $v_{b}(n,f) = \frac{1}{M} \rVert \mathbf{b}(n,f) \rVert^2 $. We initialize $\mathbf{\underline{H}}(f)$ and $\mathbf{\underline{G}}(f)$ by applying SpeexDSP and WPE on $\mathbf{d}(n,f)$, respectively, with the same hyperparameters as above. Since the authors did not specify how to initialize the PSDs \cite{togami_simultaneous_2014}, we estimate them using an NN similar to $\text{NN}_0$ where the type-I input for $|\widetilde{e}_\text{l}(n,f)|$ is replaced by $|\widetilde{d}_\text{l}(n,f)|$ obtained similarly to (\ref{eq:input_type_I}) from the corresponding multichannel signal $\mathbf{\widehat{d}}_\text{l}(n,f) = \sum_{l=\Delta}^{\Delta+L-1} \mathbf{G}(l,f) \mathbf{d}(n-l,f)$ (see Fig.\ \ref{fig:togami}). All the SCMs are initialized to $\mathbf{I}_M$. We perform $I=3$ iterations of \emph{Togami}'s EM algorithm using the same STFT hyperparameters and values of $K$, $L$ and $\Delta$ as for our approach. 

%\subsubsection{Hyperparameters of the cascade approach}
\subsubsection{Hyperparameters of Cascade}
We compute and fix the linear filters to $\mathbf{\underline{H}}(f) = \mathbf{\underline{H}}_0(f)$ and $\mathbf{\underline{G}}(f) = \mathbf{\underline{G}}_0(f)$ with the same hyperparameters as \emph{NN-joint}. Using $\mathbf{\underline{H}}_0(f)$ for echo cancellation is particularly efficient in time-varying conditions (see Section \ref{sec:echo_reduc}). The NN architecture and inputs are identical to those in \emph{NN-joint}, and the ground truth PSDs are computed using the same procedure where the linear filters are fixed to $\mathbf{\underline{H}}(f)= \mathbf{\underline{H}}_0(f)$ and $\mathbf{\underline{G}}(f) = \mathbf{\underline{G}}_0(f)$ (see Section \ref{sec:targets}). Note that the \mbox{type-I} inputs for $|\widetilde{y}(n,f)|$, $|\widetilde{e}(n,f)|$, $|\widetilde{e}_\text{l}(n,f)|$ and $|\widetilde{r}(n,f)|$ remain fixed over the EM iterations because of the fixed linear filters. %The NNs are trained using the training and validation sets, where we applied successively SpeexDSP and WPE. 

%\subsubsection{Hyperparameters of the parallel of the NN-supported BCA algorithm}
\subsubsection{\color{black}Hyperparameters of NN-parallel}
\textcolor{black}{We compute the linear filters $\mathbf{\underline{H}}(f)$ and $\mathbf{\underline{G}}(f)$ with the same hyperparameters as \emph{NN-joint}. The NN architecture and inputs are identical to those in \emph{NN-joint}, except for the type-I input for $|\widetilde{e}_\text{l}(n,f)|$ which is replaced by $|\widetilde{d}_\text{l}(n,f)|$ (see Section \ref{sec:togami_hpp}). The ground truth PSDs are computed using the same procedure as \emph{NN-joint} but where the linear filters are applied in parallel \cite{carbajal_supporting_2019}. We initialize $\mathbf{\underline{H}}(f)$ and $\mathbf{\underline{G}}(f)$ similarly to \emph{Togami}.}

%\subsubsection{Hyperparameters of the NN-supported cascade approach}
\subsubsection{\color{black}Hyperparameters of NN-cascade}

\textcolor{black}{All the filters are computed with the same hyperparameters as \emph{Cascade}. For echo cancellation, we compute $\mathbf{\underline{H}}_0(f)$ by applying the echo-only variant of \emph{NN-joint}. To estimate $\mathbf{\underline{H}}_0(f)$, the NN architecture and inputs are identical to those in \emph{NN-joint}, without the type-I inputs $|\widetilde{e}_\text{l}(n,f)|$ and $|\widetilde{r}(n,f)|$ related to dereverberation. We initialize the echo-only variant for estimating $\mathbf{\underline{H}}_0(f)$ by applying SpeexDSP on $\mathbf{d}(n,f)$ with the same hyperparameters as above. The ground truth PSDs of the echo-only variant for estimating $\mathbf{\underline{H}}_0(f)$ are computed using the same procedure as \emph{NN-joint} without linear dereverberation. At test time, we perform $I=3$ iterations of the echo-only variant for estimating $\mathbf{\underline{H}}_0(f)$ with $1$ spatial and $1$ spectral update for each iteration $i$.}

\subsubsection{Regularization}
% TODO : also for Togami et Nugraha
In order to avoid numerical instabilities and ill-conditioned matrices, we add a regularization scalar $\epsilon$ to the denominator in (\ref{eq:weighted_Rc_wiener}) and a regularization matrix $\epsilon \mathbf{I}$ to the matrix to be inverted in (\ref{eq:wiener}), (\ref{eq:update_h}) and (\ref{eq:update_g}). We also regularize the training loss in (\ref{eq:kullback}) similarly to Nugraha et al. \cite{nugraha}. We regularize likewise the four baseline approaches. The regularization hyperparameter is fixed to $\epsilon=10^{-5}$. 

\section{Results and discussion \label{sec:approach_comp}}
\textcolor{black}{In this section, \emph{NN-joint} is compared to \emph{Togami}, \emph{Cascade}, \emph{NN-parallel} and \emph{NN-cascade}. First, we investigate the influence of NN inputs on the performance of \emph{NN-joint}.} Secondly, we analyze the results of the five approaches in time-invariant conditions. Finally, we discuss their results in time-varying conditions and we compare their computation time. Audio examples are provided online\footnote{\url{https://guillaumecarbajal.github.io/}}.

\subsection{\color{black}Analysis of NN inputs \label{sec:dnn_inputs}}

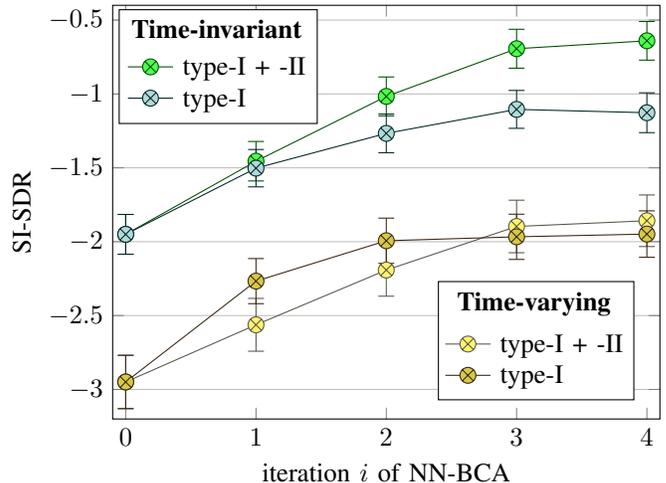
\begin{figure}[t]
\centering
\color{black}
%\subfloat[{Time-invariant conditions} \label{fig:type-II_inputs_time-invariant}]{
\begin{tikzpicture}
\begin{axis}[
    ylabel={SI-SDR},
    xlabel={iteration $i$ of  NN-BCA},
    xmin=0, xmax=4,
    ymin=-3.2, ymax=-0.4,
    xtick={0,1,...,4},
    ytick={-3,-2.5,...,0},
    %legend pos=north west,
    legend style={at={(0.28,1)},anchor=north},
    ymajorgrids=true,
	width=\columnwidth,
    height=0.8\columnwidth,
    enlarge x limits={0.025},
    enlarge y limits = 0.0,
    legend cell align={left}
]
\addplot+[
    green!30!black,
    mark=otimes*,
    mark size=3pt,
    mark options={fill=green!70!white},
    error bars/.cd,
    y dir=both,
    y explicit
    ]
    coordinates {
    (0,-1.95014)+- (0.0, 0.1346108187)
    (1,-1.45523)+- (0.0, 0.1336597032)
    (2,-1.016681)+- (0.0, 0.1317667339)
    (3,-0.693912)+- (0.0, 0.1316438401)
    (4,-0.64022)+- (0.0, 0.1307144135)
    };
\label{p1}    
\addplot+[
    teal!30!black,
    mark=otimes*,
    mark size=3pt,    
    mark options={fill=teal!30!white},
    error bars/.cd,
    y dir=both,
    y explicit
    ]
    coordinates {
    (0,-1.95014)+- (0.0, 0.1346108187)
    (1,-1.502426)+- (0.0, 0.1264414719)
    (2,-1.26691)+- (0.0, 0.1307525155)
    (3,-1.104424)+- (0.0, 0.1282408566)
    (4,-1.12708)+- (0.0, 0.1348206438)
    };
\label{p2}    
\addplot+[
    yellow!20!black,
    mark=otimes*,
    mark size=3pt,    
    mark options={fill=yellow!70!white},
    error bars/.cd,
    y dir=both,
    y explicit
    ]
    coordinates {
    (0,-2.94936)+- (0.0, 0.1805999419)
    (1,-2.5626)+- (0.0, 0.1787119002)
    (2,-2.192043)+- (0.0, 0.1769370188)
    (3,-1.89688)+- (0.0, 0.1777903717)
    (4,-1.8582)+- (0.0, 0.1745624013)
    };
\label{p3}      
\addplot+[
    brown!30!black,
    mark=otimes*,
    mark size=3pt,    
    mark options={fill=yellow!80!black},
    error bars/.cd,
    y dir=both,
    y explicit
    ]
    coordinates {
    (0,-2.94936)+- (0.0, 0.1805999419)
    (1,-2.26686)+- (0.0, 0.1526103684)
    (2,-1.993532)+- (0.0, 0.1530450154)
    (3,-1.96694)+- (0.0, 0.1530450154)
    (4,-1.94838)+- (0.0, 0.1570494282)
    }; 
\label{p4}      

    % Draw first "Legend" node using a left justified shortstack, position using relative axis coordinates
   % place legend
   \node [fit={(-0.05,-0.43)(1.47,-1.2)},draw,inner sep=0pt,fill=white] {};
   \node [above left] (M) at (rel axis cs:0.37,0.72) {\shortstack[l]{
\ref{p1} type-I + -II\\
\ref{p2} type-I}};   
   % Add title
   \node [above,font=\bfseries] (MT) at (M.north) {Time-invariant};

    % Draw first "Legend" node using a left justified shortstack, position using relative axis coordinates
   % place legend
   \node [fit={(2.4,-2.27)(3.9,-3.07)},draw,inner sep=0pt,fill=white] {};
   \node [above left] (L) at (rel axis cs:0.95,0.05) {\shortstack[l]{
\ref{p3} type-I + -II\\
\ref{p4} type-I}};   
   % Add title
   \node [above,font=\bfseries] (LT) at (L.north) {Time-varying};
   % if needed, add frame
   %\node [fit=(L)(LT),draw,inner sep=0pt,fill=white,on layer={axis background}] {};    
%\node [draw,fill=white] at (rel axis cs: 0.26,.92) {\shortstack[l]{
%\ref{p1} time-invariant I+II\\
%\ref{p2} time-invariant I}};   

%\node [draw,fill=white] at (rel axis cs: 0.73,.19) {\shortstack[l]{
%\ref{p3} time-varying I+II \ \\
%\ref{p4} time-varying I}};   
    %\legend{time-invariant I+II, time-invariant I, time-varying I+II, time-varying I}
\end{axis}
\end{tikzpicture}
\iffalse
\\
\subfloat[{Time-varying conditions}]{
\begin{tikzpicture}
\begin{axis}[
    ylabel={SI-SDR},
    xlabel={iteration $i$ of  NN-BCA},
    xmin=0, xmax=4,
    ymin=-3.2, ymax=0,
    xtick={0,1,...,4},
    ytick={-3,-2.5,...,0},
    legend pos=north west,
    ymajorgrids=true,
    %grid style=dashed,
    enlarge x limits={0.025},
    enlarge y limits = 0.0,
    legend cell align={left}
]
\addplot+[
    green!30!black,
    mark=otimes*,
    mark size=3pt,    
    mark options={fill=green!70!white},
    error bars/.cd,
    y dir=both,
    y explicit
    ]
    coordinates {
    (0,-2.94936)+- (0.0, 0.1805999419)
    (1,-2.5626)+- (0.0, 0.1787119002)
    (2,-2.192043)+- (0.0, 0.1769370188)
    (3,-1.89688)+- (0.0, 0.1777903717)
    (4,-1.8582)+- (0.0, 0.1745624013)
    };
\addplot+[
    teal!30!black,
    mark=otimes*,
    mark size=3pt,    
    mark options={fill=teal!30!white},
    error bars/.cd,
    y dir=both,
    y explicit
    ]
    coordinates {
    (0,-2.94936)+- (0.0, 0.1805999419)
    (1,-2.26686)+- (0.0, 0.1526103684)
    (2,-1.993532)+- (0.0, 0.1530450154)
    (3,-1.96694)+- (0.0, 0.1530450154)
    (4,-1.94838)+- (0.0, 0.1570494282)
    };
    \legend{type-I + type-II, type-I}
\end{axis}
\end{tikzpicture}
}
\fi
\caption{\color{black}Average overall distorsion results of \emph{NN-joint} (in dB) w.r.t. NN inputs. \label{fig:type-II_inputs}}
\end{figure}

\textcolor{black}{Fig.\ \ref{fig:type-II_inputs} shows the average SI-SDR of two NN input configurations of \emph{NN-joint} 1) both type-I and type-II inputs are used, 2) only type-I inputs are used. In time-invariant conditions, configuration 1) outperforms configuration 2) in terms of SI-SDR starting from $2$ iterations of the NN-supported BCA algorithm. This confirms that the type-II inputs improve the performance in source separation \cite{nugraha}. Note that for iteration $i=0$, the two configurations are the same as type-II inputs are not available at initialization (see Fig.\ \ref{fig:lstm}).}
\par
\textcolor{black}{In time-varying conditions, the two configurations perform similarly in terms of SI-SDR, except for iteration $i=1$. Indeed, the type-II inputs are computed with fixed SCMs $\mathbf{R}_c(f)$ while the spatial properties of target $\mathbf{s}_\text{e}$ and the near-end residual reverberation $\mathbf{s}_\text{r}$ vary over time. As a result, the type-II inputs do not improve NN estimation in configuration 1).}
%In addition the SI-SDR is lower than in time-invariant conditions since the spatial properties of target $\mathbf{s}_\text{e}$ and the near-end residual reverberation $\mathbf{s}_\text{r}$ vary over time while their SCMs $\mathbf{R}_c(f)$ remain fixed. In addition, 

\subsection{Time-invariant conditions}
%The proposed approach outperforms the cascade approaches by roughly $1$~dB in terms of SI-SDR,  $4$~dB in terms of SER, $4.5$~dB in terms of ERLE, $0.7$~dB in terms of SNR and $1$~dB in terms of SI-SAR. The proposed approach is outperformed by Togami et al.'s approach in terms of SER, ERLE, ELR and SNR, but outperforms Togami et al.'s approach  in terms of SI-SDR and SI-SAR. 
%To sum up, Togami et al.'s approach reduces greatly echo, reverberation and noise but at the cost of significant degradation of the target, while the proposed and cascade approaches achieve satisfactory reduction while preserving the target.

\subsubsection{Average performance}
\begin{figure}[t]
\centering
\color{black}
%\subfloat[{Time-invariant conditions} \label{fig:average_results_time-invariant}]{
\begin{tikzpicture}
\begin{axis}[
name=ax1,
%legend pos=outer north east,
legend columns=2,
transpose legend,
legend style={at={(1.44,1.24)},anchor=north},
enlargelimits={abs=0.5},
ybar=0pt,
bar width=0.17,
xtick={0.5,1.5,...,6.5},
xticklabels={\textbf{SI-SDR}},
x tick label as interval,
ymajorgrids = true,
ytick={-20,-18,...,4},
width=0.34\columnwidth,
height=.8\columnwidth,
ymin=-10,ymax=4.2,
enlarge y limits = 0.0,
legend cell align={left}
]

%\addplot+[blue!30!black,fill=LIGHTBLUE!40!white,mark=none,
\addplot+[blue,fill=blue!30!white,mark=none,
error bars/.cd,
y dir=both,y explicit]
coordinates {
	(1,-4.540174) +- (0.0, 0.1502523616)
    };
\addplot+[red!30!black,fill=red!35!white,mark=none,
error bars/.cd,
y dir=both,y explicit]
coordinates {
    (1,-1.706946) +- (0.0, 0.140545831)
    };
\addplot+[brown!30!black,fill=brown!30!white,mark=none,
error bars/.cd,
y dir=both,y explicit]
coordinates {
    (1,-0.763222) +- (0.0, 0.1292936753)
    };
\addplot+[violet!30!black,fill=violet!30!white,mark=none,
error bars/.cd,
y dir=both,y explicit]
coordinates {
    (1,-0.745184) +- (0.0, 0.1329036996)
    };
\addplot+[green!30!black,fill=green!70!white,mark=none,
error bars/.cd,
y dir=both,y explicit]
coordinates {
    (1,-0.693912) +- (0.0, 0.1316438401)
    };
%\legend{Togami, Cascade, NN-cascade, NN-parallel, NN-joint}
%\draw ({rel axis cs:0,0}|-{axis cs:0,0}) -- ({rel axis cs:1,0}|-{axis cs:0,0});
\end{axis}
\begin{axis}[
name=ax2,
at={(ax1.south east)},
xshift=.8cm,
legend columns=3,
legend style={at={(0.4,1.2)},anchor=north},
enlargelimits={abs=0.5},
ybar=0pt,
bar width=0.17,
xtick={0.5,1.5,...,6.5},
xticklabels={\textcolor{RED}{ERLE}},
x tick label as interval,
ymajorgrids = true,
ytick={0,20,...,100},
width=0.34\columnwidth,
height=.8\columnwidth,
ymin=0,ymax=100,
enlarge y limits = 0.0,
legend cell align={left}
]
\addplot+[blue,fill=blue!30!white,mark=none,
error bars/.cd,
y dir=both,y explicit]
coordinates {
	(1,95.255328) +- (0.0, 0.1976093023)
    };
\addplot+[red!30!black,fill=red!35!white,mark=none,
error bars/.cd,
y dir=both,y explicit]
coordinates {
    (1,75.92209) +- (0.0, 0.3275883752)
    };
\addplot+[brown!30!black,fill=brown!30!white,mark=none,
error bars/.cd,
y dir=both,y explicit]
coordinates {
    (1,71.426669) +- (0.0, 0.3056921144)
    };
\addplot+[violet!30!black,fill=violet!30!white,mark=none,
error bars/.cd,
y dir=both,y explicit]
coordinates {
    (1,78.032836) +- (0.0, 0.3823908407)
    };
\addplot+[green!30!black,fill=green!70!white,mark=none,
error bars/.cd,
y dir=both,y explicit]
coordinates {
    (1,82.00475) +- (0.0, 0.3530566755)
    };
%\legend{Valin, Schwarz, Lee, Lee + $|\widehat{y}|$, Lee + FSP, Prop.}
%\draw ({rel axis cs:0,0}|-{axis cs:0,0}) -- ({rel axis cs:1,0}|-{axis cs:0,0});
\end{axis}
\begin{axis}[
name=ax3,
at={(ax2.south east)},
xshift=.8cm,
%legend pos=north west,
%legend columns=2,
transpose legend,
legend style={at={(2.4,0.7)},anchor=north},
enlargelimits={abs=0.5},
ybar=0pt,
bar width=0.17,
xtick={0.5,1.5,...,6.5},
xticklabels={\textcolor{RED}{SER}},
x tick label as interval,
ymajorgrids = true,
ytick={0,10,...,50},
width=.34\columnwidth,
height=.8\columnwidth,
xmin=1,
ymin=0,ymax=40,
enlarge y limits = 0.0,
legend cell align={left}
]
\addplot+[blue,fill=blue!30!white,mark=none,
error bars/.cd,
y dir=both,y explicit]
coordinates {
	(1,36.931193) +- (0.0, 0.4786781323)
    };
\addplot+[red!30!black,fill=red!35!white,mark=none,
error bars/.cd,
y dir=both,y explicit]
coordinates {
    (1,31.787629) +- (0.0, 0.3121819514)
    };
\addplot+[brown!30!black,fill=brown!30!white,mark=none,
error bars/.cd,
y dir=both,y explicit]
coordinates {
    (1,31.00181) +- (0.0, 0.3187257174)
    };
\addplot+[violet!30!black,fill=violet!30!white,mark=none,
error bars/.cd,
y dir=both,y explicit]
coordinates {
    (1,35.661762) +- (0.0, 0.317697284)
    };
\addplot+[green!30!black,fill=green!70!white,mark=none,
error bars/.cd,
y dir=both,y explicit]
coordinates {
    (1,38.89151) +- (0.0, 0.3298402073)
    };
\legend{Togami, Cascade, NN-cascade, NN-parallel, NN-joint}
%\draw ({rel axis cs:0,0}|-{axis cs:0,0}) -- ({rel axis cs:1,0}|-{axis cs:0,0});
\end{axis}
\begin{axis}[
name=ax4,
at={(ax1.south west)},
yshift=-6.5cm,
%legend pos=north west,
legend columns=2,
transpose legend,
legend style={at={(2.0,-0.15)},anchor=north},
enlargelimits={abs=0.5},
ybar=0pt,
bar width=0.17,
xtick={0.5,1.5,...,6.5},
xticklabels={\textcolor{GREEN}{ELR}},
x tick label as interval,
ymajorgrids = true,
ytick={0,2,...,16},
width=.34\columnwidth,
height=.8\columnwidth,
xmin=1,
ymin=0,ymax=15,
enlarge y limits = 0.0,
legend cell align={left}
]

\addplot+[blue,fill=blue!30!white,mark=none,
error bars/.cd,
y dir=both,y explicit]
coordinates {
	(1,12.764978) +- (0.0, 0.1588297815)
    };
\addplot+[red!30!black,fill=red!35!white,mark=none,
error bars/.cd,
y dir=both,y explicit]
coordinates {
    (1,7.104763) +- (0.0, 0.1034780894)
    };
\addplot+[brown!30!black,fill=brown!30!white,mark=none,
error bars/.cd,
y dir=both,y explicit]
coordinates {
    (1,8.339045) +- (0.0, 0.1067255578)
    };
\addplot+[violet!30!black,fill=violet!30!white,mark=none,
error bars/.cd,
y dir=both,y explicit]
coordinates {
    (1,7.888524) +- (0.0, 0.1000849246)
    };
\addplot+[green!30!black,fill=green!70!white,mark=none,
error bars/.cd,
y dir=both,y explicit]
coordinates {
    (1,8.09435) +- (0.0, 0.09842496408)
    };
%\legend{Togami, Cascade, NN-cascade, NN-parallel, NN-joint}
%\draw ({rel axis cs:0,0}|-{axis cs:0,0}) -- ({rel axis cs:1,0}|-{axis cs:0,0});
\end{axis}
\begin{axis}[
%legend pos=north west,
name=ax5,
at={(ax4.south east)},
xshift=0.8cm,
legend columns=3,
transpose legend,
legend style={at={(0.7	,1.0)},anchor=north},
enlargelimits={abs=0.5},
ybar=0pt,
bar width=0.17,
xtick={0.5,1.5,...,6.5},
xticklabels={\textcolor{ORANGE}{SNR}},
x tick label as interval,
ymajorgrids = true,
ytick={-5,0,...,25},
width=.34\columnwidth,
height=.8\columnwidth,
xmin=1,
ymin=0,ymax=25,
enlarge y limits = 0.0,
legend cell align={left}
]

\addplot+[blue,fill=blue!30!white,mark=none,
error bars/.cd,
y dir=both,y explicit]
coordinates {
	(1,21.391692) +- (0.0, 0.3023168575)
    };
\addplot+[red!30!black,fill=red!35!white,mark=none,
error bars/.cd,
y dir=both,y explicit]
coordinates {
    (1,13.858129) +- (0.0, 0.2247932214)        
    };
\addplot+[brown!30!black,fill=brown!30!white,mark=none,
error bars/.cd,
y dir=both,y explicit]
coordinates {
    (1,17.182428) +- (0.0, 0.1822828928)        
    };
\addplot+[violet!30!black,fill=violet!30!white,mark=none,
error bars/.cd,
y dir=both,y explicit]
coordinates {
    (1,16.15711) +- (0.0, 0.2228241653)        
    };
\addplot+[green!30!black,fill=green!70!white,mark=none,
error bars/.cd,
y dir=both,y explicit]
coordinates {
    (1,15.33623) +- (0.0, 0.2243898085)        
    };
%\legend{Togami, Cascade, NN-echo, Parallel, Prop., Only type-I}
%\draw ({rel axis cs:0,0}|-{axis cs:0,0}) -- ({rel axis cs:1,0}|-{axis cs:0,0});
\end{axis}
\begin{axis}[
%legend pos=outer north east,
at={(ax5.south east)},
xshift=0.8cm,
transpose legend,
legend style={at={(2.7,0.5)},anchor=north},
enlargelimits={abs=0.5},
ybar=0pt,
bar width=0.17,
xtick={0.5,1.5,...,6.5},
xticklabels={SI-SAR},
x tick label as interval,
ymajorgrids = true,
ytick={-10,-8,...,4},
width=0.34\columnwidth,
height=.8\columnwidth,
ymin=-10,ymax=4.2,
enlarge y limits = 0.0,
legend cell align={left}
]

\addplot+[blue,fill=blue!30!white,mark=none,
error bars/.cd,
y dir=both,y explicit]
coordinates {
	(1,-4.327631) +- (0.0, 0.1494489356)
    };
\addplot+[red!30!black,fill=red!35!white,mark=none,
error bars/.cd,
y dir=both,y explicit]
coordinates {
    (1,-0.226954) +- (0.0, 0.1490582136)
    };
\addplot+[brown!30!black,fill=brown!30!white,mark=none,
error bars/.cd,
y dir=both,y explicit]
coordinates {
    (1,0.428846) +- (0.0, 0.1372846458)
    };
\addplot+[violet!30!black,fill=violet!30!white,mark=none,
error bars/.cd,
y dir=both,y explicit]
coordinates {
    (1,0.524212) +- (0.0, 0.140019788)
    };
\addplot+[green!30!black,fill=green!70!white,mark=none,
error bars/.cd,
y dir=both,y explicit]
coordinates {
    (1,0.64678) +- (0.0, 0.139324308)
    };
%\legend{Togami, Cascade, Prop. cascade, Prop. parallel, Prop.}
%\draw ({rel axis cs:0,0}|-{axis cs:0,0}) -- ({rel axis cs:1,0}|-{axis cs:0,0});
\end{axis}
\end{tikzpicture}
%}
\caption{\color{black}Average results (in dB) in time-invariant conditions \label{fig:average_results_time-invariant}}
\end{figure}
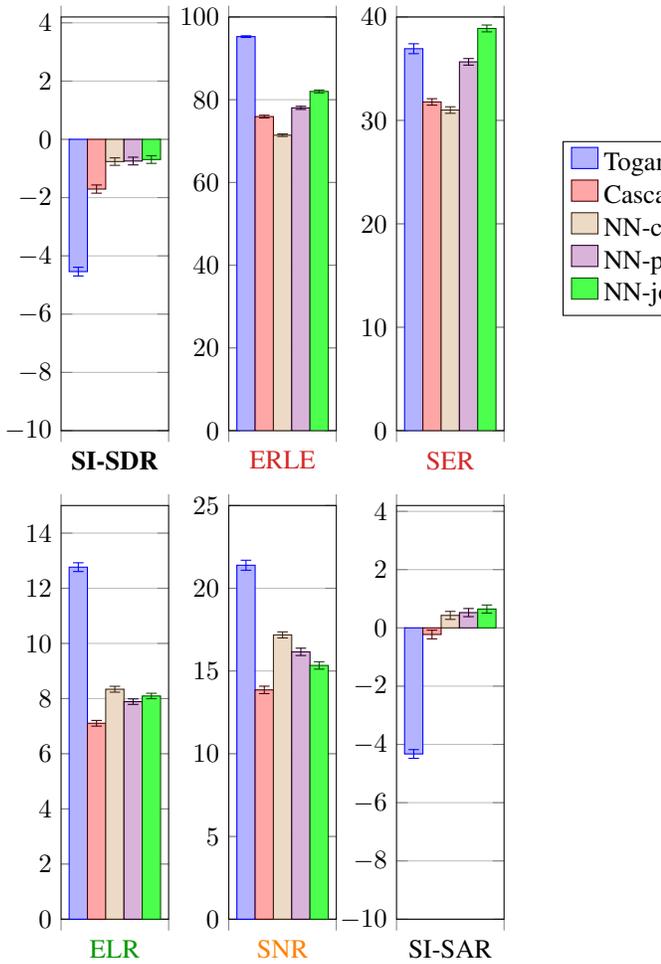

\begin{table}[b]
\color{black}
\centering
{\tabulinesep=1.3mm
\setlength{\tabcolsep}{2.5pt}
\begin{tabu}{|c||c|c|c|c|}
  \hline
\textbf{SI-SDR} & \textcolor{RED}{ERLE} &  \textcolor{RED}{SER}  & \textcolor{GREEN}{ELR} &  \textcolor{ORANGE}{SNR}  \\
\hline
   $\mathbf{-16.2 \pm 0.2}$ & $0.0$ & $-25.8 \pm 0.3$ & $-1.1\pm 0.1$ & $-2.6 \pm 0.2$ \\
  \hline
\end{tabu}}
\caption{\label{tab:input_metrics} \color{black} Metrics (in dB) related to the mixture signal $\mathbf{d}$ in the test set. The metrics are computed using the decomposition in \eqref{eq:mixture}. ERLE $=0$~dB since there is no echo reduction. The SI-SAR is not computed since there is no artifacts in the unprocessed target $\mathbf{s}_\text{e}$.}
\end{table}
\textcolor{black}{Table \ref{tab:input_metrics} shows the metrics related to the mixture $\mathbf{d}$. Fig.\ \ref{fig:average_results_time-invariant} shows the average results in time-invariant conditions. All approaches have a negative SI-SDR, which is caused by the challenging test set conditions.}
\par
%This degradation may be caused by the interfering application of the linear filters $\mathbf{\underline{H}}(f)$ and $\mathbf{\underline{G}}(f)$ and the probabilistic modeling of the echo path $\{\mathbf{a}_y(n,f)\}_{n,f}$ and the inverse of the near-end RIR $\{\mathbf{a}_s(n,f)\}_{n,f}$ (see Section \ref{sec:togami}). Although their approach provides much greater ERLE, this degradation mitigates their performance in SER. 
\textcolor{black}{\emph{NN-joint} outperforms \emph{Togami} by $3.8$~dB in terms of SI-SDR. \emph{NN-parallel} provides information about this difference in performance since it applies the linear filters $\mathbf{\underline{H}}(f)$ and $\mathbf{\underline{G}}(f)$ in the same order as in \emph{Togami}, but uses a similar signal model and optimization algorithm as \emph{NN-joint} (see Section \ref{sec:togami}). \emph{NN-parallel} also outperforms \emph{Togami} by $3.8$~dB in terms of SI-SDR. Therefore our proposed signal model and optimization algorithm explain the SI-SDR difference with \emph{Togami}. Although the parallel variant achieves lower reduction of echo, reverberation and noise than \emph{Togami}, it introduces lower degradation in the target $\mathbf{s}_\text{e}$. Regarding \emph{NN-joint} and \emph{NN-parallel}, applying the linear filters $\mathbf{\underline{H}}(f)$ and $\mathbf{\underline{G}}(f)$ one after another only modifies the distribution of the overall distortion over the echo (greater reduction), reverberation (greater reduction) and noise (lower reduction).}
\par
\textcolor{black}{\emph{NN-joint} outperforms \emph{Cascade} $1.0$~dB in terms of SI-SDR. \emph{NN-cascade} provides information about this difference in performance since it also uses an NN-supported echo cancellation as in \emph{NN-joint}, but estimates each filter separately. \emph{NN-cascade} also outperforms \emph{Cascade} by $1.0$~dB in terms of SI-SDR. Therefore the proposed NN-supported echo cancellation in \emph{NN-joint} explains the SI-SDR difference with \emph{Cascade}. Regarding the optimization of the filters, jointly optimizing them modifies the distribution of the overall distortion over the echo (greater reduction), reverberation (lower  reduction) and noise (lower reduction).}
\par
\textcolor{black}{From informal listening tests, we can state that the estimated target speech $\mathbf{\widehat{s}}_\text{e}$ is often highly attenuated and distorted during double-talk for all approaches when $\text{SER} \leq -20$~dB. Regarding \emph{Togami}, a slight reverberation remains, but noise and echo seem completely removed. However, the estimated target speech $\mathbf{\widehat{s}}_\text{e}$ is much more attenuated and distorted than with the other approaches, especially during \emph{double-talk} where the estimated target speech $\mathbf{\widehat{s}}_\text{e}$ can never be heard. Regarding the other approaches, post-residual distortions seem louder than with \emph{Togami}, but a comparison between these approaches is difficult to make.}

% TODO : difference between NN-joint/NN-parallel very subtle. In double-talk, target speech is often attenuated, sometimes cut. Still reverb and noise and a bit of echo which looks like noise.

% TODO : cascade / NN-cascade / NN-joint no significant difference. sometimes a bit more reverb for NN-cascade.

% TODO : NN-cascade / NN-joint: no significant hearable difference. In some recordings,  More noise in NN-cascade, especially when the target speech is absent. a bit more echo in NN-cascade during double-talk.

%The joint and cascade NN-supported approaches performs similarly in terms of SI-SDR. Although the joint approach a lower dereverberation and a significantly lower noise reduction, it has significantly greater echo reduction while introducing the same amount of artifacts. By significantly improving echo reduction, the joint approach does not degrade performance in terms of the overall distortion.
\begin{figure}
\color{black}
\centering
%\subfloat[{Time-varying conditions} \label{fig:average_results_time-varying}]{
\begin{tikzpicture}
\begin{axis}[
name=ax1,
%legend pos=outer north east,
legend columns=2,
transpose legend,
legend style={at={(1.43,1.22)},anchor=north},
enlargelimits={abs=0.5},
ybar=0pt,
bar width=0.17,
xtick={0.5,1.5,...,6.5},
xticklabels={\textbf{SI-SDR}},
x tick label as interval,
ymajorgrids = true,
ytick={-10,-8,...,4},
width=0.34\columnwidth,
height=.8\columnwidth,
ymin=-10,ymax=4.2,
enlarge y limits = 0.0,
legend cell align={left}
]
\addplot+[blue,fill=blue!30!white,mark=none,
error bars/.cd,
y dir=both,y explicit]
coordinates {
	(1,-4.860847) +- (0.0, 0.1876210722)
    };
\addplot+[red!30!black,fill=red!35!white,mark=none,
error bars/.cd,
y dir=both,y explicit]
coordinates {
    (1,-2.826461) +- (0.0, 0.1906279631)
    };
\addplot+[brown!30!black,fill=brown!30!white,mark=none,
error bars/.cd,
y dir=both,y explicit]
coordinates {
    (1,-1.956366) +- (0.0, 0.1714729222)
    };
\addplot+[violet!30!black,fill=violet!30!white,mark=none,
error bars/.cd,
y dir=both,y explicit]
coordinates {
    (1,-2.111917) +- (0.0, 0.1798192194)
    };
\addplot+[green!30!black,fill=green!70!white,mark=none,
error bars/.cd,
y dir=both,y explicit]
coordinates {
    (1,-1.89688) +- (0.0, 0.1777903717)
    };
%\legend{Togami, Cascade, Prop. cascade, Prop. parallel, Prop.}
%\draw ({rel axis cs:0,0}|-{axis cs:0,0}) -- ({rel axis cs:1,0}|-{axis cs:0,0});
\end{axis}
\begin{axis}[
name=ax2,
at={(ax1.south east)},
xshift=.8cm,
legend columns=3,
legend style={at={(0.4,1.2)},anchor=north},
enlargelimits={abs=0.5},
ybar=0pt,
bar width=0.17,
xtick={0.5,1.5,...,6.5},
xticklabels={\textcolor{RED}{ERLE}},
x tick label as interval,
ymajorgrids = true,
ytick={0,20,...,100},
width=0.34\columnwidth,
height=.8\columnwidth,
ymin=0,ymax=100,
enlarge y limits = 0.0,
legend cell align={left}
]
\addplot+[blue,fill=blue!30!white,mark=none,
error bars/.cd,
y dir=both,y explicit]
coordinates {
	(1,96.101162) +- (0.0, 0.2685082187)
    };
\addplot+[red!30!black,fill=red!35!white,mark=none,
error bars/.cd,
y dir=both,y explicit]
coordinates {
    (1,86.685935) +- (0.0, 0.5732541717)        
    };
\addplot+[brown!30!black,fill=brown!30!white,mark=none,
error bars/.cd,
y dir=both,y explicit]
coordinates {
    (1,64.979571) +- (0.0, 0.3131790748)        
    };
\addplot+[violet!30!black,fill=violet!30!white,mark=none,
error bars/.cd,
y dir=both,y explicit]
coordinates {
    (1,80.308901) +- (0.0, 0.6504728921)        
    };
\addplot+[green!30!black,fill=green!70!white,mark=none,
error bars/.cd,
y dir=both,y explicit]
coordinates {
    (1,79.74027) +- (0.0, 0.5052203832)
    };
%\legend{Valin, Schwarz, Lee, Lee + $|\widehat{y}|$, Lee + FSP, Prop.}
%\draw ({rel axis cs:0,0}|-{axis cs:0,0}) -- ({rel axis cs:1,0}|-{axis cs:0,0});
\end{axis}
\begin{axis}[
name=ax3,
at={(ax2.south east)},
xshift=.8cm,
%legend pos=north west,
legend columns=3,
transpose legend,
legend style={at={(0.7	,1.0)},anchor=north},
enlargelimits={abs=0.5},
ybar=0pt,
bar width=0.17,
xtick={0.5,1.5,...,6.5},
xticklabels={\textcolor{RED}{SER}},
x tick label as interval,
ymajorgrids = true,
ytick={0,10,...,50},
width=.34\columnwidth,
height=.8\columnwidth,
xmin=1,
ymin=0,ymax=40,
enlarge y limits = 0.0,
legend cell align={left}
]
\addplot+[blue,fill=blue!30!white,mark=none,
error bars/.cd,
y dir=both,y explicit]
coordinates {
	(1,37.127782) +- (0.0, 0.6009061862)
    };
\addplot+[red!30!black,fill=red!35!white,mark=none,
error bars/.cd,
y dir=both,y explicit]
coordinates {
    (1,39.110248) +- (0.0, 0.4545698986)
    };
\addplot+[brown!30!black,fill=brown!30!white,mark=none,
error bars/.cd,
y dir=both,y explicit]
coordinates {
    (1,27.446077) +- (0.0, 0.3572037254)
    };
\addplot+[violet!30!black,fill=violet!30!white,mark=none,
error bars/.cd,
y dir=both,y explicit]
coordinates {
    (1,32.197307) +- (0.0, 0.3963710908)
    };
\addplot+[green!30!black,fill=green!70!white,mark=none,
error bars/.cd,
y dir=both,y explicit]
coordinates {
    (1,33.90567) +- (0.0, 0.381687516)
    };
%\legend{Togami, Cascade, Prop. cascade, Prop. parallel, Prop.}
%\draw ({rel axis cs:0,0}|-{axis cs:0,0}) -- ({rel axis cs:1,0}|-{axis cs:0,0});
\end{axis}
\end{tikzpicture}
%\hspace{3.93cm}
%\label{fig:average_results} 
%\caption{Average results (in dB).}
\caption{\color{black}Average results (in dB) in time-varying conditions. \label{fig:average_results_time-varying}}
\end{figure}
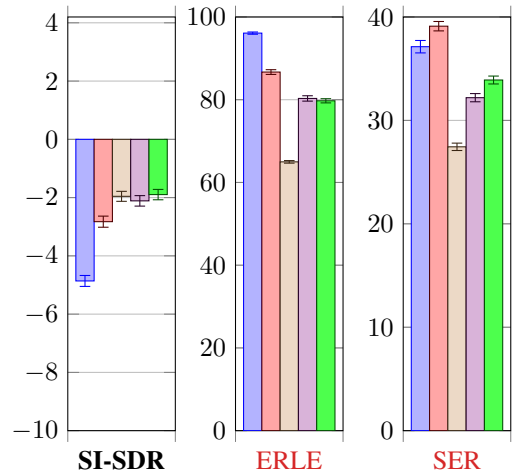

\subsubsection{Interactions of system components}
% TODO : SI-SDR positif en absence d'echo
% TODO : see figure 2 near-end talk
% TODO : see figure 3 double-talk
While the above results show the performance averaged over all periods (\textit{near-end talk}, \textit{far-end talk} and \textit{double-talk}), we need further performance analysis when only noise and reverberation are present, i.e. during \textit{near-end talk}, and when echo, reverberation and noise are present simultaneously, i.e.  during \textit{double-talk}, to investigate how the system components interact with each other. We discard the analysis of \textit{far-end talk} as the target $\mathbf{s}_\text{e}$ is absent in this scenario. %As we observe similar SI-SDR difference between the approaches in both time-invariant and time-varying conditions (see Fig.\ \ref{fig:average_results}), we analyze only in time-invariant conditions.
\par
\begin{figure}[t]
\centering
\color{black}
\hspace{-9pt}
\begin{tikzpicture}
\begin{axis}[
name=ax1,
%legend pos=outer north east,
legend columns=2,
transpose legend,
legend style={at={(1.7,1.2)},anchor=north},
enlargelimits={abs=0.5},
ybar=0pt,
bar width=0.17,
xtick={0.5,1.5,...,6.5},
xticklabels={\textbf{SI-SDR}},
x tick label as interval,
ymajorgrids = true,
ytick={-10,-8,...,4},
width=0.34\columnwidth,
height=.8\columnwidth,
ymin=-10,ymax=4.2,
enlarge y limits = 0.0,
legend cell align={left}
]

\addplot+[blue,fill=blue!30!white,mark=none,
error bars/.cd,
y dir=both,y explicit]
coordinates {
	(1,0.00655) +- (0.0, 0.0946840153)
    };
\addplot+[red!30!black,fill=red!35!white,mark=none,
error bars/.cd,
y dir=both,y explicit]
coordinates {
    (1,1.554376) +- (0.0, 0.107273231)
    };
\addplot+[brown!30!black,fill=brown!30!white,mark=none,
error bars/.cd,
y dir=both,y explicit]
coordinates {
    (1,2.2206) +- (0.0, 0.09982937693)
    };
\addplot+[violet!30!black,fill=violet!30!white,mark=none,
error bars/.cd,
y dir=both,y explicit]
coordinates {
    (1,1.742043) +- (0.0, 0.1329036996)
    };
\addplot+[green!30!black,fill=green!70!white,mark=none,
error bars/.cd,
y dir=both,y explicit]
coordinates {
    (1,1.659894) +- (0.0, 0.1073588141)
    };
%\legend{Togami, Cascade, Prop. cascade, Prop. parallel, Prop.}
%\draw ({rel axis cs:0,0}|-{axis cs:0,0}) -- ({rel axis cs:1,0}|-{axis cs:0,0});
\end{axis}
\begin{axis}[
name=ax3,
at={(ax1.south east)},
xshift=0.8cm,
%legend pos=north west,
legend columns=3,
transpose legend,
legend style={at={(0.7	,1.0)},anchor=north},
enlargelimits={abs=0.5},
ybar=0pt,
bar width=0.17,
xtick={0.5,1.5,...,6.5},
xticklabels={\textcolor{GREEN}{ELR}},
x tick label as interval,
ymajorgrids = true,
ytick={0,2,...,16},
width=.34\columnwidth,
height=.8\columnwidth,
xmin=1,
ymin=0,ymax=15,
enlarge y limits = 0.0,
legend cell align={left}
]

\addplot+[blue,fill=blue!30!white,mark=none,
error bars/.cd,
y dir=both,y explicit]
coordinates {
	(1,14.511214) +- (0.0, 0.149971989)
    };
\addplot+[red!30!black,fill=red!35!white,mark=none,
error bars/.cd,
y dir=both,y explicit]
coordinates {
    (1,7.618427) +- (0.0, 0.09720900992)
    };
\addplot+[brown!30!black,fill=brown!30!white,mark=none,
error bars/.cd,
y dir=both,y explicit]
coordinates {
    (1,8.73644) +- (0.0, 0.1049297488)
    };
\addplot+[violet!30!black,fill=violet!30!white,mark=none,
error bars/.cd,
y dir=both,y explicit]
coordinates {
    (1,8.226989) +- (0.0, 0.09684973674)
    };
\addplot+[green!30!black,fill=green!70!white,mark=none,
error bars/.cd,
y dir=both,y explicit]
coordinates {
    (1,8.35457) +- (0.0, 0.09671154348)
    };
%\legend{Togami, Cascade, NN-echo, Parallel, Prop., Only type-I}
%\draw ({rel axis cs:0,0}|-{axis cs:0,0}) -- ({rel axis cs:1,0}|-{axis cs:0,0});
\end{axis}
\begin{axis}[
name=ax4,
at={(ax3.south east)},
xshift=.8cm,
%legend pos=north west,
legend columns=3,
transpose legend,
legend style={at={(0.7	,1.0)},anchor=north},
enlargelimits={abs=0.5},
ybar=0pt,
bar width=0.17,
xtick={0.5,1.5,...,6.5},
xticklabels={\textcolor{ORANGE}{SNR}},
x tick label as interval,
ymajorgrids = true,
ytick={0,5,...,25},
width=.34\columnwidth,
height=.8\columnwidth,
xmin=1,
ymin=0,ymax=25,
enlarge y limits = 0.0,
legend cell align={left}
]

\addplot+[blue,fill=blue!30!white,mark=none,
error bars/.cd,
y dir=both,y explicit]
coordinates {
	(1,22.902153) +- (0.0, 0.2572662934)
    };
\addplot+[red!30!black,fill=red!35!white,mark=none,
error bars/.cd,
y dir=both,y explicit]
coordinates {
    (1,13.885252) +- (0.0, 0.1824891129)        
    };
\addplot+[brown!30!black,fill=brown!30!white,mark=none,
error bars/.cd,
y dir=both,y explicit]
coordinates {
    (1,17.254435) +- (0.0, 0.1621283067)        
    };
\addplot+[violet!30!black,fill=violet!30!white,mark=none,
error bars/.cd,
y dir=both,y explicit]
coordinates {
    (1,15.952748) +- (0.0, 0.1877867069)        
    };
\addplot+[green!30!black,fill=green!70!white,mark=none,
error bars/.cd,
y dir=both,y explicit]
coordinates {
    (1,14.59268) +- (0.0, 0.190502417)        
    };
%\legend{Togami, Cascade, NN-echo, Parallel, Prop., Only type-I}
%\draw ({rel axis cs:0,0}|-{axis cs:0,0}) -- ({rel axis cs:1,0}|-{axis cs:0,0});
\end{axis}
\begin{axis}[
%legend pos=outer north east,
at={(ax4.south east)},
xshift=.8cm,
legend columns=3,
transpose legend,
legend style={at={(0.2,.5)},anchor=north},
enlargelimits={abs=0.5},
ybar=0pt,
bar width=0.17,
xtick={0.5,1.5,...,6.5},
xticklabels={SI-SAR},
x tick label as interval,
ymajorgrids = true,
ytick={-10,-8,...,4},
width=0.34\columnwidth,
height=.8\columnwidth,
ymin=-10,ymax=4.2,
enlarge y limits = 0.0,
legend cell align={left}
]

\addplot+[blue,fill=blue!30!white,mark=none,
error bars/.cd,
y dir=both,y explicit]
coordinates {
	(1,0.290724) +- (0.0, 0.09218073887)
    };
\addplot+[red!30!black,fill=red!35!white,mark=none,
error bars/.cd,
y dir=both,y explicit]
coordinates {
    (1,3.789631) +- (0.0, 0.1173242337)
    };
\addplot+[brown!30!black,fill=brown!30!white,mark=none,
error bars/.cd,
y dir=both,y explicit]
coordinates {
    (1,3.947864) +- (0.0, 0.1082922855)
    };
\addplot+[violet!30!black,fill=violet!30!white,mark=none,
error bars/.cd,
y dir=both,y explicit]
coordinates {
    (1,3.496491) +- (0.0, 0.1179665759)
    };
\addplot+[green!30!black,fill=green!70!white,mark=none,
error bars/.cd,
y dir=both,y explicit]
coordinates {
    (1,3.54806) +- (0.0, 0.1156456325)
    };
%\legend{Togami, Cascade, NN-echo, Parallel, Prop., Only type-I}
%\draw ({rel axis cs:0,0}|-{axis cs:0,0}) -- ({rel axis cs:1,0}|-{axis cs:0,0});
\end{axis}
\end{tikzpicture}
%\label{fig:average_results} 
\caption{\color{black}Results (in dB) during near-end talk in time-invariant conditions. \label{fig:average_results_near-end_talk}}
\end{figure}
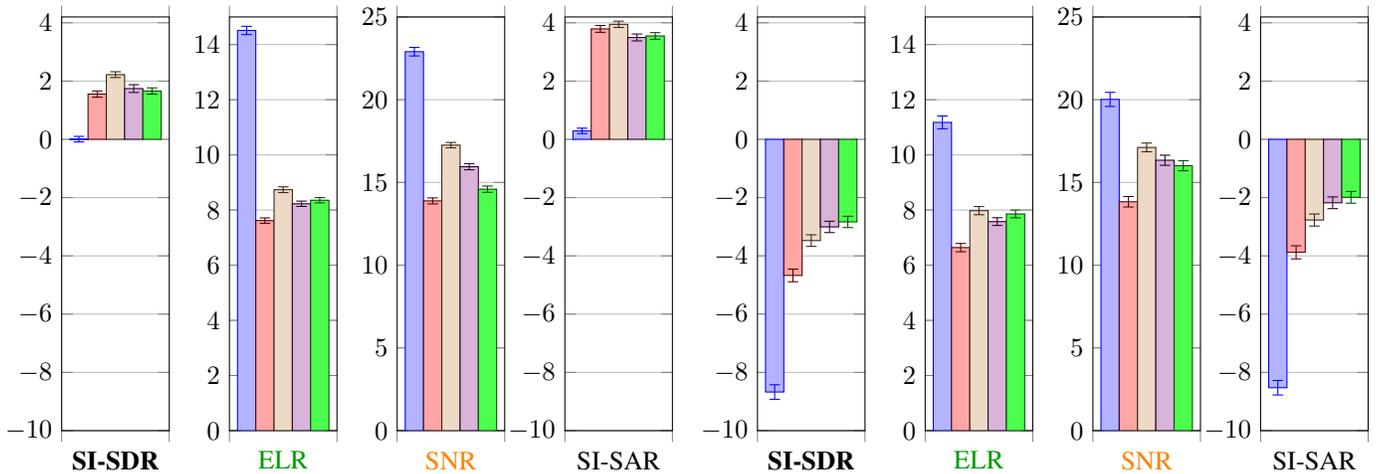

%The trend in performance between Togami et al.'s approach and the parallel variant, between the joint approach and the parallel variant, and between the cascade approach and the cascade NN-supported approach, is also similar to the results averaged over all periods. The joint 
%or close to $0$~dB.
%The SI-SAR is also positive for all approaches, i.e. there is a lower degradation of the target $\mathbf{s}_\text{e}$ during \textit{near-end talk}.
%The proposed approach performs similarly or slightly better than the cascade approach for all metrics.
%Togami et al.'s approach is outperformed by all other approaches in terms of SI-SDR. The joint approach, the parallel variant and the cascade approach perform similarly in terms of SI-SDR. However, the joint approach is outperformed by the cascade NN-supported approach by $0.6$~dB in terms of SI-SDR. This is due to lower dereverberation, noise reduction and a greater degradation of the target $\mathbf{s}_\text{e}$. Thus joint optimization slightly degrades performance during \textit{near-end talk}.
\textcolor{black}{Fig.\ \ref{fig:average_results_near-end_talk} shows the results during \textit{near-end talk}. SER and ERLE are not evaluated as echo is absent. All approaches have a positive SI-SDR. The SI-SAR is also positive for all approaches. The trend in performance between \emph{NN-joint}, \emph{NN-parallel} and \emph{Togami} is similar to the results averaged over all periods. \emph{NN-cascade} outperforms \emph{Cascade} by $+0.6$~dB in terms of SI-SDR. This is due to greater dereverberation and noise reduction with comparable degradation of the target $\mathbf{s}_\text{e}$. This might be due to the performance before post-filtering \cite[Sec. 8.1]{carbajal_supporting_2019}: linear dereverberation in \emph{NN-cascade} also achieves greater dereverberation and noise reduction than \emph{Cascade}. This results from echo cancellation: since the dereverberation filter $\mathbf{\underline{G}}(f)$ is time-invariant, its performance during \emph{near-end talk} is also affected by echo cancellation during \emph{double-talk}. In \emph{NN-cascade}, the NN-supported echo cancellation achieves greater echo reduction than Valin's echo cancellation in \emph{Cascade}. As a result, linear dereverberation in \emph{NN-cascade} is able to achieve greater reduction of the other distortion signals, i.e. reverberation and noise.}
%Hence the greater performance of  \emph{NN-cascade}  in SI-SDR.
\par
% Therefore, the proposed NN-supported echo cancellation in \emph{NN-joint} improves the SI-SDR compared to \emph{Cascade} as it might detect \emph{near-end talk} periods more precisely than Valin's echo cancellation in \emph{Cascade} \cite{valin_adjusting_2007}. 
%the three filters $\mathbf{\underline{H}}(f)$, $\mathbf{\underline{G}}(f)$ and $\mathbf{W}_{s_\text{e}}(n,f)$ 
\textcolor{black}{\emph{NN-cascade} also outperforms \emph{NN-joint} in terms of SI-SDR during \emph{near-end talk}.  Indeed, joint estimation of the filters in \emph{NN-joint} implies a performance compromise during all periods in order to reduce all the distortion signals. In the case of \emph{NN-cascade}, there is no performance compromise as the filters are estimated separately. Thus \emph{NN-cascade} might better perform when one distortion source is absent. Hence the greater performance of \emph{NN-cascade} during \emph{near-end talk} where echo is absent. All in all, \emph{NN-joint} does not improve performance when only reverberation and noise are present, but does not degrade it either compared to \emph{Cascade}.}
%Thus the estimation problem during \emph{near-end talk} reduces to the estimation of only the two filters $\mathbf{\underline{G}}(f)$ and $\mathbf{W}_{s_\text{e}}(n,f)$, 

% In the case of \emph{NN-cascade}, there is no performance compromise between the filters. 
%However, joint optimization of the filters in \emph{NN-joint} reduces this improvement in SI-SDR. 

%As separately estimating the filters in \emph{NN-cascade} does not make any performance compromise. 
% In \emph{NN-joint} and \emph{NN-parallel}, echo cancellation is performed by modeling the target $\mathbf{s}_\text{e}$ and the residual reverberation $\mathbf{s}_\text{r}$. In \emph{NN-cascade}, echo cancellation is performed by modeling the reverberant near-end target $\mathbf{s}$. This seems to be more accurate to detect the change of periods (\textit{near-end talk}, \textit{double-talk}) since \emph{NN-cascade} introduces lower degradation of the target $\mathbf{s}_\text{e}$ than \emph{NN-joint} and \emph{NN-parallel}. 
 %The proposed NN-supported echo cancellation in our joint approach might detect \emph{near-end talk} periods more precisely than Valin's echo cancellation, but the joint optimization mitigates this detection.
%the echo filter $\mathbf{\underline{H}}(f)$ 
%In our joint approach and the parallel variant, the echo cancellation filter $\mathbf{\underline{H}}(f)$ is estimated using the target $\mathbf{s}_\text{e}$ and the residual reverberation $\mathbf{s}_\text{r}$.
 \par
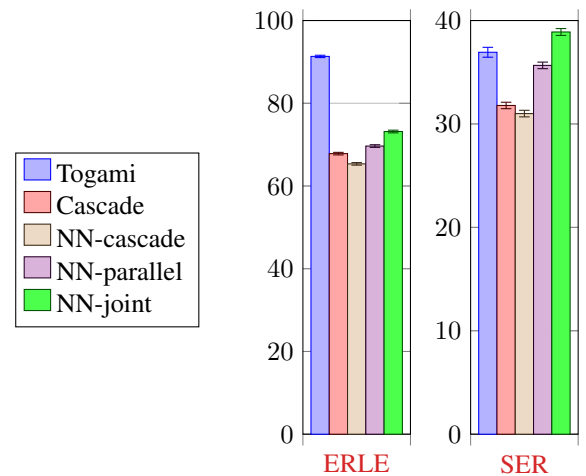
\begin{figure}[t]
\hspace{-9pt}
\color{black}
\begin{tikzpicture}
\begin{axis}[
name=ax1,
%legend pos=outer north east,
%legend columns=3,
transpose legend,
legend style={at={(1.3,-.5)},anchor=north},
enlargelimits={abs=0.5},
ybar=0pt,
bar width=0.17,
xtick={0.5,1.5,...,6.5},
xticklabels={\textbf{SI-SDR}},
x tick label as interval,
ymajorgrids = true,
ytick={-10,-8,...,4},
width=0.34\columnwidth,
height=.8\columnwidth,
ymin=-10,ymax=4.2,
enlarge y limits = 0.0,
legend cell align={left}
]

\addplot+[error bars/.cd,
y dir=both,y explicit]
coordinates {
	(1,-8.67356) +- (0.0, 0.2490054724)
    };
\addplot+[red!30!black,fill=red!35!white,mark=none,
error bars/.cd,
y dir=both,y explicit]
coordinates {
    (1,-4.671784) +- (0.0, 0.2165093633)
    };
\addplot+[brown!30!black,fill=brown!30!white,mark=none,
error bars/.cd,
y dir=both,y explicit]
coordinates {
    (1,-3.475786) +- (0.0, 0.1988559433)
    };
\addplot+[violet!30!black,fill=violet!30!white,mark=none,
error bars/.cd,
y dir=both,y explicit]
coordinates {
    (1,-3.006298) +- (0.0, 0.1940568122)
    };
\addplot+[green!30!black,fill=green!70!white,mark=none,
error bars/.cd,
y dir=both,y explicit]
coordinates {
    (1,-2.833735) +- (0.0, 0.1946256468)
    };
\legend{Togami, Cascade, NN-cascade, NN-parallel, NN-joint}
%\draw ({rel axis cs:0,0}|-{axis cs:0,0}) -- ({rel axis cs:1,0}|-{axis cs:0,0});
\end{axis}
\begin{axis}[
name=ax4,
at={(ax1.south east)},
xshift=.8cm,
%legend pos=north west,
legend columns=3,
transpose legend,
legend style={at={(0.7	,1.0)},anchor=north},
enlargelimits={abs=0.5},
ybar=0pt,
bar width=0.17,
xtick={0.5,1.5,...,6.5},
xticklabels={\textcolor{GREEN}{ELR}},
x tick label as interval,
ymajorgrids = true,
ytick={0,2,...,16},
width=.34\columnwidth,
height=.8\columnwidth,
xmin=1,
ymin=0,ymax=15,
enlarge y limits = 0.0,
legend cell align={left}
]
\addplot+[error bars/.cd,
y dir=both,y explicit]
coordinates {
	(1,11.17749) +- (0.0, 0.2312293931)
    };
\addplot+[red!30!black,fill=red!35!white,mark=none,
error bars/.cd,
y dir=both,y explicit]
coordinates {
    (1,6.637797) +- (0.0, 0.1509624669)
    };
\addplot+[brown!30!black,fill=brown!30!white,mark=none,
error bars/.cd,
y dir=both,y explicit]
coordinates {
    (1,7.977777) +- (0.0, 0.1483071339)
    };
\addplot+[violet!30!black,fill=violet!30!white,mark=none,
error bars/.cd,
y dir=both,y explicit]
coordinates {
    (1,7.580828) +- (0.0, 0.139637976)
    };
\addplot+[green!30!black,fill=green!70!white,mark=none,
error bars/.cd,
y dir=both,y explicit]
coordinates {
    (1,7.85779) +- (0.0, 0.1404099238)
    };
%\legend{Togami, Cascade, NN-echo, Parallel, Prop., Only type-I}
%\draw ({rel axis cs:0,0}|-{axis cs:0,0}) -- ({rel axis cs:1,0}|-{axis cs:0,0});
\end{axis}
\begin{axis}[
%legend pos=north west,
name=ax5,
at={(ax4.south east)},
xshift=.8cm,
legend columns=3,
transpose legend,
legend style={at={(0.7	,1.0)},anchor=north},
enlargelimits={abs=0.5},
ybar=0pt,
bar width=0.17,
xtick={0.5,1.5,...,6.5},
xticklabels={\textcolor{ORANGE}{SNR}},
x tick label as interval,
ymajorgrids = true,
ytick={0,5,...,25},
width=.34\columnwidth,
height=.8\columnwidth,
xmin=1,
ymin=0,ymax=25,
enlarge y limits = 0.0,
legend cell align={left}
]

\addplot+[error bars/.cd,
y dir=both,y explicit]
coordinates {
	(1,20.018547) +- (0.0, 0.4274915529)
    };
\addplot+[red!30!black,fill=red!35!white,mark=none,
error bars/.cd,
y dir=both,y explicit]
coordinates {
    (1,13.833471) +- (0.0, 0.3160263326)        
    };
\addplot+[brown!30!black,fill=brown!30!white,mark=none,
error bars/.cd,
y dir=both,y explicit]
coordinates {
    (1,17.116968) +- (0.0, 0.2629375808)        
    };
\addplot+[violet!30!black,fill=violet!30!white,mark=none,
error bars/.cd,
y dir=both,y explicit]
coordinates {
    (1,16.342895) +- (0.0, 0.3056987676)        
    };
\addplot+[green!30!black,fill=green!70!white,mark=none,
error bars/.cd,
y dir=both,y explicit]
coordinates {
    (1,16.01218) +- (0.0, 0.3062740795)        
    };
%\legend{Togami, Cascade, NN-echo, Parallel, Prop., Only type-I}
%\draw ({rel axis cs:0,0}|-{axis cs:0,0}) -- ({rel axis cs:1,0}|-{axis cs:0,0});
\end{axis}
\begin{axis}[
%legend pos=outer north east,
at={(ax5.south east)},
xshift=.8cm,
legend columns=3,
transpose legend,
legend style={at={(0.2,.5)},anchor=north},
enlargelimits={abs=0.5},
ybar=0pt,
bar width=0.17,
xtick={0.5,1.5,...,6.5},
xticklabels={SI-SAR},
x tick label as interval,
ymajorgrids = true,
ytick={-10,-8,...,4},
width=0.34\columnwidth,
height=.8\columnwidth,
ymin=-10,ymax=4.2,
enlarge y limits = 0.0,
legend cell align={left}
]

\addplot+[error bars/.cd,
y dir=both,y explicit]
coordinates {
	(1,-8.526135) +- (0.0, 0.2493777882)
    };
\addplot+[red!30!black,fill=red!35!white,mark=none,
error bars/.cd,
y dir=both,y explicit]
coordinates {
    (1,-3.878394) +- (0.0, 0.2247001058)
    };
\addplot+[brown!30!black,fill=brown!30!white,mark=none,
error bars/.cd,
y dir=both,y explicit]
coordinates {
    (1,-2.770261) +- (0.0, 0.2067798835)
    };
\addplot+[violet!30!black,fill=violet!30!white,mark=none,
error bars/.cd,
y dir=both,y explicit]
coordinates {
    (1,-2.177859) +- (0.0, 0.2016241464)
    };
\addplot+[green!30!black,fill=green!70!white,mark=none,
error bars/.cd,
y dir=both,y explicit]
coordinates {
    (1,-1.99076) +- (0.0, 0.2023442461)
    };

%\legend{Togami, Cascade, NN-echo, Parallel, Prop., Only type-I}
%\draw ({rel axis cs:0,0}|-{axis cs:0,0}) -- ({rel axis cs:1,0}|-{axis cs:0,0});
\end{axis}
\begin{axis}[
%legend pos=outer north east,
name=ax2,
at={(ax5.south west)},
yshift=-6.5cm,
legend columns=3,
legend style={at={(0.4,1.2)},anchor=north},
enlargelimits={abs=0.5},
ybar=0pt,
bar width=0.17,
xtick={0.5,1.5,...,6.5},
xticklabels={\textcolor{RED}{ERLE}},
x tick label as interval,
ymajorgrids = true,
ytick={0,20,...,100},
width=0.34\columnwidth,
height=.8\columnwidth,
ymin=0,ymax=100,
enlarge y limits = 0.0,
legend cell align={left}
]

\addplot+[error bars/.cd,
y dir=both,y explicit]
coordinates {
	(1,91.326195) +- (0.0, 0.2853266773)
    };
\addplot+[red!30!black,fill=red!35!white,mark=none,
error bars/.cd,
y dir=both,y explicit]
coordinates {
    (1,67.82373) +- (0.0, 0.3099875947)
    };
\addplot+[brown!30!black,fill=brown!30!white,mark=none,
error bars/.cd,
y dir=both,y explicit]
coordinates {
    (1,65.349505) +- (0.0, 0.3077181185)
    };
\addplot+[violet!30!black,fill=violet!30!white,mark=none,
error bars/.cd,
y dir=both,y explicit]
coordinates {
    (1,69.655746) +- (0.0, 0.3205888498)
    };
\addplot+[green!30!black,fill=green!70!white,mark=none,
error bars/.cd,
y dir=both,y explicit]
coordinates {
    (1,73.16822) +- (0.0, 0.3286080451)
    };
%\legend{Valin, Schwarz, Lee, Lee + $|\widehat{y}|$, Lee + FSP, Prop.}
%\draw ({rel axis cs:0,0}|-{axis cs:0,0}) -- ({rel axis cs:1,0}|-{axis cs:0,0});
\end{axis}
\begin{axis}[
name=ax3,
at={(ax2.south east)},
xshift=.8cm,
%legend pos=north west,
legend columns=3,
transpose legend,
legend style={at={(0.7	,1.0)},anchor=north},
enlargelimits={abs=0.5},
ybar=0pt,
bar width=0.17,
xtick={0.5,1.5,...,6.5},
xticklabels={\textcolor{RED}{SER}},
x tick label as interval,
ymajorgrids = true,
ytick={-10,0,...,50},
width=.34\columnwidth,
height=.8\columnwidth,
xmin=1,
ymin=0,ymax=40,
enlarge y limits = 0.0,
legend cell align={left}
]
\addplot+[error bars/.cd,
y dir=both,y explicit]
coordinates {
	(1,36.931193) +- (0.0, 0.4786781323)
    };
\addplot+[red!30!black,fill=red!35!white,mark=none,
error bars/.cd,
y dir=both,y explicit]
coordinates {
    (1,31.787629) +- (0.0, 0.3121819514)
    };
\addplot+[brown!30!black,fill=brown!30!white,mark=none,
error bars/.cd,
y dir=both,y explicit]
coordinates {
    (1,31.00181) +- (0.0, 0.3187257174)
    };
\addplot+[violet!30!black,fill=violet!30!white,mark=none,
error bars/.cd,
y dir=both,y explicit]
coordinates {
    (1,35.661762) +- (0.0, 0.317697284)
    };
\addplot+[green!30!black,fill=green!70!white,mark=none,
error bars/.cd,
y dir=both,y explicit]
coordinates {
    (1,38.89151) +- (0.0, 0.3298402073)
    };
%\legend{Togami, Cascade, NN-echo, Parallel, Prop., Only type-I}
%\draw ({rel axis cs:0,0}|-{axis cs:0,0}) -- ({rel axis cs:1,0}|-{axis cs:0,0});
\end{axis}
\end{tikzpicture}
\caption{\color{black}Results (in dB) during double-talk in time-invariant conditions. \label{fig:average_results_double-talk}}
\end{figure}

 %TODO : The proposed approach performs similarly or slightly better than the cascade approach for all metrics. 
%TODO :  L’optimisation conjointe améliore (nettement) les performances de double-talk (prop. cascade vs. prop)
\textcolor{black}{Fig.\ \ref{fig:average_results_double-talk} shows the results during \textit{double-talk}. The trend in performance between \emph{NN-joint}, \emph{NN-parallel} and \emph{Togami} is similar to the results averaged over all periods. \emph{NN-cascade} outperforms \emph{Cascade} by $1.2$~dB in terms of SI-SDR. \emph{NN-joint} outperforms \emph{NN-cascade} by $0.6$~dB in terms of SI-SDR. Therefore, during \emph{double-talk}, both joint optimization of the filters and the proposed NN-supported echo cancellation in \emph{NN-joint} explain the SI-SDR improvement between \emph{NN-joint} and \emph{Cascade}. Although \emph{NN-joint} achieves lower dereverberation and noise reduction than \emph{NN-cascade}, it achieves greater echo reduction with lower degradation of the target $\mathbf{s}_\text{e}$. As a result, \emph{NN-joint} improves performance when echo, reverberation and noise are present simultaneously.}
\par
% TODO : thus robustness is improved (drop in difference)
\textcolor{black}{From \textit{near-end talk} to \textit{double-talk}, the SI-SDR decreases by $4.5$~dB for \emph{NN-joint}, by $5.7$~dB for \emph{NN-cascade} and by $6.2$~dB for \emph{Cascade}. We conclude that \emph{NN-joint} improves robustness in terms of SI-SDR when echo, reverberation and noise are present simultaneously, while not degrading performance when only reverberation and noise are present.}

%As the proposed approach still outperforms the cascade approach in terms of SER but with a reduced difference than in time-invariant conditions,
%Fig.\ \ref{fig:average_results_time-varying} shows the average results in time-varying conditions. SI-SDR decreases for all approaches. The proposed approach still outperforms the cascade approach in terms of SI-SDR by $1$~dB. However, there is lower SER difference in these conditions. The performance in terms of SI-SDR is thus explained by greater SI-SAR difference. The lower SER difference is caused by comparable performance in ERLE. Indeed the SCMs of the target $\mathbf{s}_\text{e}$ and the near-end residual reverberation $\mathbf{s}_\text{r}$ are time-varying in these conditions. While the proposed approach is not designed to handle such case, the cascade approach uses Valin's approach is robust to time-varying conditions for echo reduction.
%The proposed approach still outperforms the cascade approach in terms of SI-SDR by $1$~dB. 
%even though SER difference between the proposed and cascade approaches is lower, due to comparable performance in ERLE.
% Indeed the SCMs of the target $\mathbf{s}_\text{e}$ and the near-end residual reverberation $\mathbf{s}_\text{r}$ are time-varying in these conditions. While the proposed approach is not designed to handle such case, 

%\subsection{Further analysis}
\subsection{Time-varying conditions}

% the approaches is similar to the results averaged over all periods. However, the proposed approach has greater difference in terms of SI-SDR due greater SI-SAR difference, i.e. lower degradation of the target $\mathbf{s}_\text{e}$ during \textit{double-talk}.
%since the spatial properties of target $\mathbf{s}_\text{e}$ and the near-end residual reverberation $\mathbf{s}_\text{r}$ vary over time while their SCMs $\mathbf{R}_c(f)$ remain fixed.
\textcolor{black}{Fig.\ \ref{fig:average_results_time-varying} shows the average results in time-varying conditions. As the trend in ELR, SNR and SAR is similar to the average results in time-invariant conditions for all approaches, we discard these metrics from the analysis and we provide them in the supporting document \cite[Sec. 8.2]{carbajal_supporting_2019}.} The SI-SDR is lower than in time-invariant conditions for all approaches (see Fig.\ \ref{fig:average_results_time-invariant}) since the spatial properties of target $\mathbf{s}_\text{e}$ and the near-end residual reverberation $\mathbf{s}_\text{r}$ vary over time while their SCMs $\mathbf{R}_c(f)$ remain fixed. \textcolor{black}{This also explains the drop in SI-SDR for the two NN input configurations of \emph{NN-joint} in time-varying conditions (see Fig.\ \ref{fig:type-II_inputs}).} \textcolor{black}{Informal listening tests provide the same observations as in time-invariant conditions.}
%In addition the SI-SDR is lower than in time-invariant conditions since the spatial properties of target $\mathbf{s}_\text{e}$ and the near-end residual reverberation $\mathbf{s}_\text{r}$ vary over time while their SCMs $\mathbf{R}_c(f)$ remain fixed. In addition, 
\par
%The trend in SI-SDR between \emph{NN-joint}, \emph{NN-cascade} and \emph{Cascade} is also similar to the average SI-SDR in time-invariant conditions.
%However, regarding \emph{NN-joint} and \emph{NN-cascade}, the distribution of the overall distortion is changed: jointly optimizing the filters improves the reduction of echo, reverberation and noise, at a cost of greater degradation of the target $\mathbf{s}_\text{e}$.
\textcolor{black}{The trend in performance between \emph{NN-joint}, \emph{NN-parallel} and \emph{Togami} is similar to the average performance in time-invariant conditions. The trend in SI-SDR between \emph{NN-joint}, \emph{NN-cascade} and \emph{Cascade} is also similar to the average SI-SDR in time-invariant conditions. However, \emph{Cascade} achieves here the greatest echo reduction. \emph{Cascade} also systematically achieves the greatest echo reduction after echo cancellation, and also after echo cancellation and dereveberation \cite[Sec. 8.1]{carbajal_supporting_2019}. This is explained by Valin's adaptive approach for echo cancellation which is designed for time-varying conditions \cite{valin_adjusting_2007} (see Section \ref{sec:echo_reduc}).}
%Although the proposed approach is not designed for time-varying conditions, it still outperforms the $2$ others approaches in terms of SI-SDR. Additionally, it still outperforms the cascade approach in terms of SER, even though the cascade approach uses Valin's adaptive approach for echo cancellation which is designed for time-varying conditions. 
%\vspace{-.2cm}

%TODO : togami always much more attenuated and distorted. no target speech during double talk. however no noise, echo. A bit of reverb.

%TODO : difference between NN-joint/NN-parallel very subtle. In double-talk, target speech is often attenuated, sometimes cut. Still reverb and noise and a bit of echo which looks like noise.

%TODO : cascade / NN-cascade / NN-joint: NN-cascade has more residual echo during far-end talk than cascade and NN-joint.

\subsection{Computation time}
\begin{table}[t]
\color{black}
\centering
{\tabulinesep=1.3mm
\begin{tabu}{|c|c|c|c|}
  \hline
\textbf{Togami} & {NN-cascade} &  {NN-parallel}  & {NN-joint} \\
\hline
   $\mathbf{-42 \%}$ & $+56 \%$ & $+7 \%$ & $+16\%$ \\
  \hline
\end{tabu}}
\caption{\label{tab:computation_time} \color{black} Computation time of the approaches compared to \emph{Cascade} (in percentage). }
\end{table}
%We discard the initialization as it is the same for all $3$ approaches (see Section \ref{sec:hyperparameters}). With a $2.7$ GHz Intel Core i5 CPU, computing the target $\mathbf{\widehat{s}}_e$ for a $8$~s utterance took $135.6 \pm 1.2$~s for the proposed approach, $121.6 \pm 1.4$~s for the cascade approach and $68.9 \pm 0.3$~s for Togami et al.'s approach. The proposed approach is slower than the cascade approach due to the update of the linear filters $\mathbf{\underline{H}}(f)$ and $\mathbf{\underline{G}}(f)$ at each iteration of the BCA algorithm. Both approaches are much slower than Togami et al.'s approach due to the NN update at each \mbox{NN-EM} iteration.

\textcolor{black}{We discard the initialization as it is the same for all $5$ approaches (see Section \ref{sec:hyperparameters}). We compute the target $\mathbf{\widehat{s}}_e$ for a $8$~s utterance with a $2.7$ GHz Intel Core i5 CPU. Table \ref{tab:computation_time} shows the computation time of the approaches compared to the cascade approach. \emph{NN-joint} is much faster than \emph{NN-cascade}. Therefore joint optimization of the filters significantly reduces the computation time. In addition, \emph{NN-parallel} is slightly faster than \emph{NN-joint}. Since \emph{Cascade} is one of the approaches implemented in today's industrial devices, we conclude that both \emph{NN-joint} and \emph{NN-parallel} could be implemented in real time.}

\section{Conclusion}

We proposed an NN-supported BCA algorithm for joint multichannel reduction of acoustic echo, reverberation and noise. The approach jointly models the spectra of the target and residual signals after echo cancellation and dereverberation with an NN. We evaluated our system on real recordings of acoustic echo, reverberation and noise acquired with a smart speaker in various situations.  When echo, reverberation and noise are present simultaneously, the proposed approach outperforms the cascade approach and Togami et al.'s joint reduction approach in terms of overall distortion reduction while not degrading performance when only reverberation and noise are present. Future work will focus on a recursive version of the approach in order to better handle time-varying conditions.

%In order to enhance the reverberated signal in such conditions, algorithms must be able to update their parameters in an online fashion. 
%Future work will be done on end-to-end multichannel to reduce the complexity. In ASR, NN have been used in end-to-end multichannel MMSE-based approach to predict all parameters, \cite{sainath_multichannel_2017, heymann_beamnet_2017}. 

% or


%%%%%%%%%%%%%%%
%%% Update 08-04-2020
%%%%%%%%%%%%%%%

%\section{}

%\section*{Acknowledgment}


\ifCLASSOPTIONcaptionsoff
  \newpage
\fi



%\bibliographystyle{IEEEtran}

\bibliographystyle{ieeetran}
\bibliography{refs}

\begin{IEEEbiography}[{\includegraphics[width=1in,height=1.25in,clip,keepaspectratio]{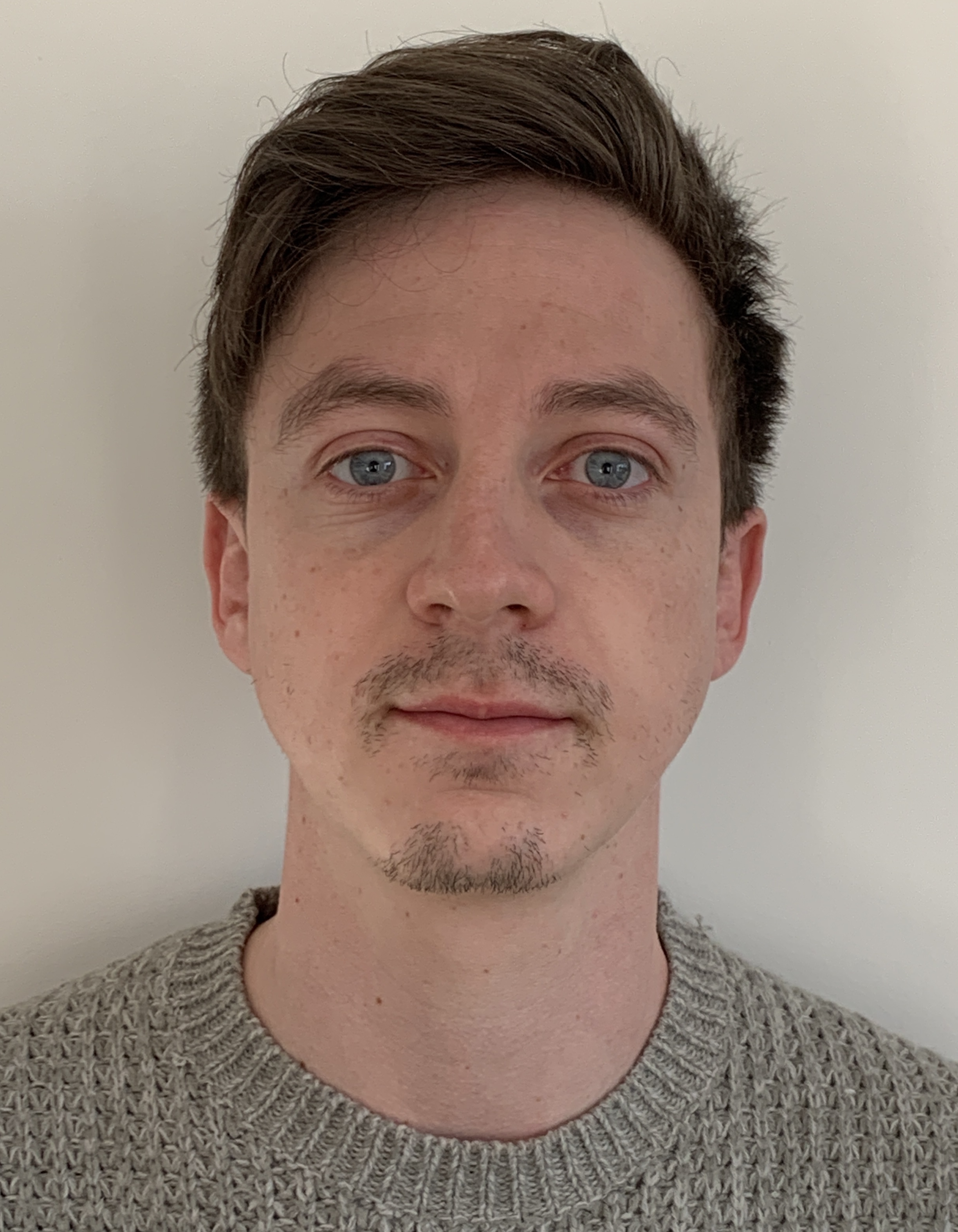}}]{Guillaume Carbajal}
received the M.Sc. degree in machine learning from the Université Paris-Saclay (Paris, France) in 2016
and the State Engineering from Ecole Nationale des Ponts et Chaussées (Paris, France) in 2017.
He worked as an audio research engineer at Invoxia (Issy-les-Moulineaux, France) from 2017 to 2020 
and obtained the Ph.D. degree in informatics from the Université de Lorraine and Inria Nancy–Grand-Est (Nancy, France) in 2020. 
He is currently a postdoctoral researcher at the computer science department of the Universität Hamburg (Hamburg, Germany).
His research interests include speech enhancement, audio-visual signal processing and machine learning.
\end{IEEEbiography}

\begin{IEEEbiography}[{\includegraphics[width=1in,height=1.25in,clip,keepaspectratio]{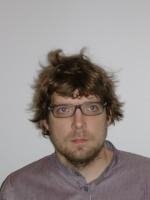}}]{Romain Serizel}
received the M.Eng. degree in Automatic System Engineering from ENSEM (Nancy, France) in 2005 and the M.Sc. degree in Signal Processing from Universit\'e Rennes 1 (Rennes, France) in 2006.
He received the Ph.D. degree in Engineering Sciences from the KU Leuven (Leuven, Belgium) in June 2011. 
From 2011 till 2016 he was a postdoctoral researcher at KU Leuven (Leuven, Belgium), FBK (Trento, Italy), and at Télécom ParisTech (Paris, France).
He is now an Associate Professor with Université de Lorraine (Nancy, France) doing research on robust speech communications and ambient sound detection and classification. Since 2019, he is coordinating the DCASE challenge series together with Annamaria Mesaros.
\end{IEEEbiography}

\begin{IEEEbiography}[{\includegraphics[width=1in,height=1.25in,clip,keepaspectratio]{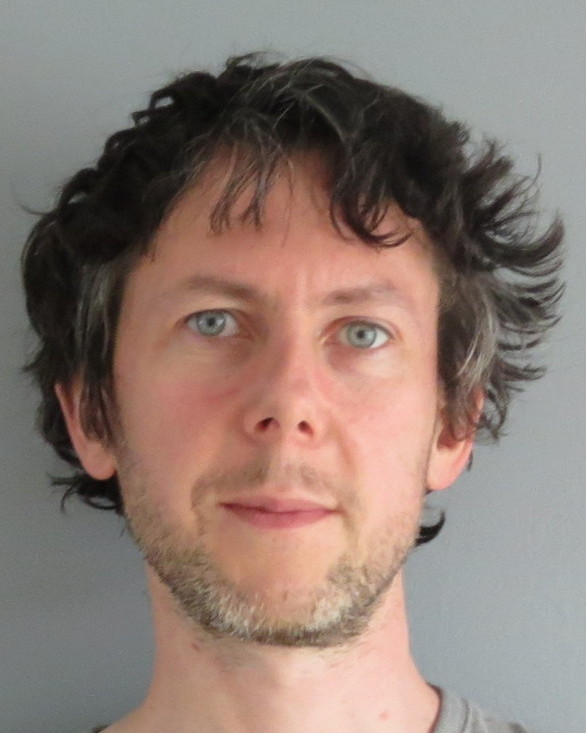}}]{Emmanuel Vincent}
    is a Senior Research Scientist with Inria (Nancy,
France). He received the Ph.D. degree in music signal processing from
the Institut de Recherche et Coordination Acoustique/Musique (Ircam,
Paris, France) in 2004 and worked as a Research Assistant with the
Centre for Digital Music at Queen Mary, University of London (United
Kingdom), from 2004 to 2006. His research focuses on statistical machine
learning for speech and audio signal processing, with application to
audio source localization and separation, speech enhancement, and robust
speech and speaker recognition. He is a founder of the series of CHiME,
SiSEC and VoicePrivacy challenges. He was an associate editor for IEEE
TRANSACTIONS ON AUDIO, SPEECH, AND LANGUAGE PROCESSING.
\end{IEEEbiography}

\begin{IEEEbiography}[{\includegraphics[width=1in,height=1.25in,clip,keepaspectratio]{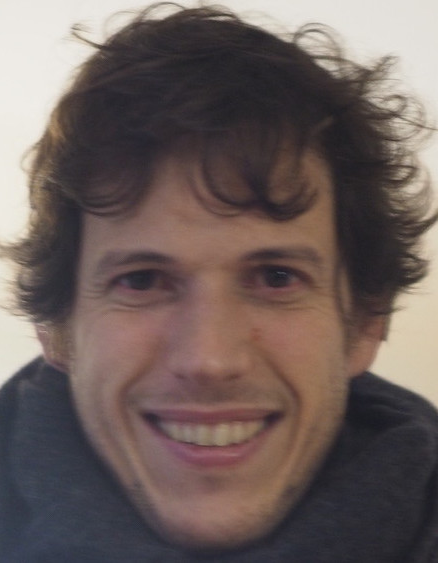}}]{Eric Humbert}
    received the State Engineering from Centrale-Supélec (Paris, France) in 2002
and the M.Sc. degree in acoustics, computer science and signal processing applied to music (ATIAM) from the Université Pierre et Marie Curie (Paris, France) in 2003.
He worked as an audio software engineer at Inventel (Paris, France) from 2003 to 2005,
at Thomson (Paris, France) from 2005 to 2009
and at Technicolor from 2009 to 2010.
He is currently the AI director at Invoxia (Paris, France).
\end{IEEEbiography}


%\newpage



%\vfill


\end{document}